\documentclass[twocolumn]{aastex63}

\newcommand{\Msun}{$M_{\odot}$}
\newcommand{\HI}{\hbox{{\rm H}\kern 0.2em{\sc i}}}

\received{xxxx, 2023}
\revised{xxxx, 2023}
\accepted{xxxx, 2023}
\shorttitle{Molecular gas of group galaxies}
\shortauthors{Lee et al.}
\graphicspath{{./}{figures/}}

\begin{document}

\title{ALMA/ACA CO Survey of the IC~1459 and NGC~4636 Groups: Environmental Effects on the Molecular Gas of Group Galaxies}

\correspondingauthor{Bumhyun Lee, Jing Wang}
\email{bhlee301@gmail.com, jwang\_astro@pku.edu.cn}

\author[0000-0002-3810-1806]{Bumhyun Lee}
\affiliation{Kavli Institute for Astronomy and Astrophysics, Peking University, Beijing 100871, People's Republic of China}
\affiliation{Korea Astronomy and Space Science Institute, 776 Daedeokdae-ro, Daejeon 34055, Republic of Korea}
\affiliation{Department of Astronomy, Yonsei University, 50 Yonsei-ro, Seodaemun-gu, Seoul 03722, Republic of Korea}

\author[0000-0002-6593-8820]{Jing Wang}
\affiliation{Kavli Institute for Astronomy and Astrophysics, Peking University, Beijing 100871, People's Republic of China}

\author[0000-0003-1440-8552]{Aeree Chung}
\affiliation{Department of Astronomy, Yonsei University, 50 Yonsei-ro, Seodaemun-gu, Seoul 03722, Republic of Korea}

\author[0000-0001-6947-5846]{Luis C. Ho}
\affiliation{Kavli Institute for Astronomy and Astrophysics, Peking University, Beijing 100871, People's Republic of China}
\affiliation{Department of Astronomy, School of Physics, Peking University, Beijing 100871, People's Republic of China}

\author[0000-0003-4956-5742]{Ran Wang}
\affiliation{Kavli Institute for Astronomy and Astrophysics, Peking University, Beijing 100871, People's Republic of China}
\affiliation{Department of Astronomy, School of Physics, Peking University, Beijing 100871, People's Republic of China}


\author[0000-0003-2475-7983]{Tomonari Michiyama}
\affiliation{Department of Earth and Space Science, Osaka University, 1-1 Machikaneyama, Toyonaka, Osaka 560-0043, Japan}
\affiliation{National Astronomical Observatory of Japan, 2-21-1 Osawa, Mitaka, Tokyo 181-8588, Japan}

\author[0000-0002-8136-8127]{Juan Molina}
\affiliation{Kavli Institute for Astronomy and Astrophysics, Peking University, Beijing 100871, People's Republic of China}

\author[0000-0003-1647-3286]{Yongjung Kim}
\affiliation{Department of Astronomy and Atmospheric Sciences, College of Natural Sciences, Kyungpook National University, Daegu 41566, Republic of Korea}
\affiliation{Kavli Institute for Astronomy and Astrophysics, Peking University, Beijing 100871, People's Republic of China}

\author[0000-0003-2015-777X]{Li Shao}
\affiliation{National Astronomical Observatories, Chinese Academy of Sciences, 20A Datun Road, Chaoyang District, Beijing, People’s Republic of China}

\author{Virginia Kilborn}
\affiliation{Centre for Astrophysics and Supercomputing, Swinburne University of Technology, P.O. Box 218, Hawthorn, VIC 3122, Australia}
\affiliation{ARC Centre of Excellence for All-Sky Astrophysics in 3 Dimensions (ASTRO 3D), Australia}

\author[0000-0002-9663-3384]{Shun Wang}
\affiliation{Kavli Institute for Astronomy and Astrophysics, Peking University, Beijing 100871, People's Republic of China}
\affiliation{Department of Astronomy, School of Physics, Peking University, Beijing 100871, People's Republic of China}

\author[0000-0002-4250-2709]{Xuchen Lin}
\affiliation{Department of Astronomy, School of Physics, Peking University, Beijing 100871, People's Republic of China}

\author[0000-0001-5717-3736]{Dawoon E. Kim}
\affiliation{Department of Astronomy, Yonsei University, 50 Yonsei-ro, Seodaemun-gu, Seoul 03722, Republic of Korea}
\affiliation{INAF - Istituto di Astrofisica e Planetologia Spaziali, Via del Fosso del Cavaliere 100, 00133 Roma, Italy}
\affiliation{Università di Roma La Sapienza, Dipartimento di Fisica, Piazzale Aldo Moro 5, 00185 Roma, Italy}
\affiliation{Università di Roma Tor Vergata, Dipartimento di Fisica, Via della Ricerca Scientifica 1, 00133 Roma, Italy}

\author[0000-0002-7625-562X]{Barbara Catinella}
\affiliation{International Centre for Radio Astronomy Research (ICRAR), The University of Western Australia, 35 Stirling Highway, Crawley WA 6009, Australia}
\affiliation{ARC Centre of Excellence for All-Sky Astrophysics in 3 Dimensions (ASTRO 3D), Australia}

\author[0000-0002-7422-9823]{Luca Cortese}
\affiliation{International Centre for Radio Astronomy Research (ICRAR), The University of Western Australia, 35 Stirling Highway, Crawley WA 6009, Australia}
\affiliation{ARC Centre of Excellence for All-Sky Astrophysics in 3 Dimensions (ASTRO 3D), Australia}

\author[0000-0003-3523-7633]{Nathan Deg}
\affiliation{Department of Physics, Engineering Physics, and Astronomy, Queen’s University, Kingston, ON, K7L 3N6, Canada}

\author[0000-0002-9214-8613]{Helga Dénes}
\affiliation{ASTRON - The Netherlands Institute for Radio Astronomy, 7991 PD Dwingeloo, The Netherlands}

\author{Ahmed Elagali}
\affiliation{Telethon Kids Institute, Perth Children’s Hospital, Perth, Australia}

\author[0000-0002-0196-5248]{Bi-Qing For}
\affiliation{International Centre for Radio Astronomy Research (ICRAR), The University of Western Australia, 35 Stirling Highway, Crawley WA 6009, Australia}
\affiliation{ARC Centre of Excellence for All-Sky Astrophysics in 3 Dimensions (ASTRO 3D), Australia}

\author[0000-0002-7573-555X]{Dane Kleiner}
\affiliation{INAF – Osservatorio Astronomico di Cagliari, Via della Scienza 5, 09047 Selargius, CA, Italy}

\author[0000-0003-4351-993X]{B\"{a}rbel S. Koribalski}
\affiliation{CSIRO Space and Astronomy, Australia Telescope National Facility, P.O. Box 76, NSW 1710, Australia}
\affiliation{School of Science, Western Sydney University, Locked Bag 1797, Penrith, NSW 2751, Australia}

\author[0000-0003-4844-8659]{Karen Lee-Waddell}
\affiliation{International Centre for Radio Astronomy Research (ICRAR), The University of Western Australia, 35 Stirling Highway, Crawley WA 6009, Australia} 
\affiliation{CSIRO Space and Astronomy, Australia Telescope National Facility, PO Box 1130, Bentley, WA 6102, Australia}

\author[0000-0001-8496-4306]{Jonghwan Rhee}
\affiliation{International Centre for Radio Astronomy Research (ICRAR), The University of Western Australia, 35 Stirling Highway, Crawley WA 6009, Australia}
\affiliation{ARC Centre of Excellence for All-Sky Astrophysics in 3 Dimensions (ASTRO 3D), Australia}

\author[0000-0002-0956-7949]{Kristine Spekkens}
\affiliation{Royal Military College of Canada, PO Box 17000, Station Forces, Kingston, Ontario, K7K7B4, Canada}

\author[0000-0002-5300-2486]{Tobias Westmeier}
\affiliation{International Centre for Radio Astronomy Research (ICRAR), The University of Western Australia, 35 Stirling Highway, Crawley WA 6009, Australia}

\author[0000-0003-4264-3509]{O. Ivy Wong}
\affiliation{CSIRO Space and Astronomy, Australia Telescope National Facility, PO Box 1130, Bentley, WA 6102, Australia}
\affiliation{International Centre for Radio Astronomy Research (ICRAR), The University of Western Australia, 35 Stirling Highway, Crawley WA 6009, Australia}
\affiliation{ARC Centre of Excellence for All-Sky Astrophysics in 3 Dimensions (ASTRO 3D), Australia}

\author{Frank Bigiel}
\affiliation{Argelander-Institut f\"ur Astronomie, Universit\"at Bonn, Auf dem H\"ugel 71, D-53121 Bonn, Germany}

\author[0000-0002-1128-6089]{Albert Bosma}
\affiliation{Aix Marseille Univ, CNRS, CNES, LAM, Marseille, France}

\author[0000-0002-4884-6756]{Benne W. Holwerda}
\affiliation{University of Louisville, Department of Physics and Astronomy, 102 Natural Science Building, 40292 KY Louisville, USA}

\author[0000-0002-9316-763X]{Jan M. van der Hulst}
\affiliation{Kapteyn Astronomical Institute, University of Groningen, The Netherlands}

\author[0000-0002-5820-4589]{Sambit Roychowdhury}
\affiliation{International Centre for Radio Astronomy Research (ICRAR), The University of Western Australia, 35 Stirling Highway, Crawley WA 6009, Australia}
\affiliation{ARC Centre of Excellence for All-Sky Astrophysics in 3 Dimensions (ASTRO 3D), Australia}

\author[0000-0003-0156-6180]{Lourdes Verdes-Montenegro}
\affiliation{Instituto de Astrofísica de Andalucía (CSIC), Spain}

\author[0000-0003-0101-1804]{Martin A. Zwaan}
\affiliation{European Southern Observatory, Karl-Schwarzschild-Str. 2, 85748 Garching b. München, Germany}




\begin{abstract}
We present new results of a $^{12}$CO($J$=1--0) imaging survey using the Atacama Compact Array (ACA) for 31 {\HI} detected galaxies in the IC~1459 and NGC~4636 groups. This is the first CO imaging survey for loose galaxy groups. We obtained well-resolved CO data ($\sim$0.7 $-$ 1.5 kpc) for a total of 16 galaxies in two environments. By comparing our ACA CO data with the {\HI} and UV data, we probe the impacts of the group environment on the cold gas components (CO and {\HI} gas) and star formation activity. We find that CO and/or {\HI} morphologies are disturbed in our group members, some of which show highly asymmetric CO distributions (e.g., IC~5264, NGC~7421, and NGC~7418). In comparison with isolated galaxies in the xCOLD GASS sample, our group galaxies tend to have low star formation rates and low H$_{2}$ gas fractions. Our findings suggest that the group environment can change the distribution of cold gas components, including the molecular gas, and star formation properties of galaxies. This is supporting evidence that preprocessing in the group-like environment can play an important role in galaxy evolution. 
\end{abstract}




\section{Introduction} \label{sec:intro}
Galaxy properties (e.g., morphology, star formation rate (SFR), and color) change with environment including the field, groups, and clusters. As the surrounding density increases, the fraction of early-type galaxies increases (i.e., the density-morphology relation; \citealt[][]{dressler1980,goto2003,houghton2015}). In particular, in the cluster environment, the galaxy population is dominated by red sequence galaxies with lower SFRs \citep[][]{lewis2002,gomez2003,hogg2004,Kauffmann2004}. Furthermore, this trend is also observed at the outskirts of galaxy clusters \citep{cybulski2014,jaffe2016,morokuma-matsui2021}.

In fact, in a hierarchical Universe, a significant fraction of cluster populations is accreted through the groups \citep[][]{mcgee2009,delucia2012}. This result implies that many galaxies are likely to be already affected by the environmental effects in groups (e.g., tidal interaction and ram pressure stripping (RPS); for a review see \citealt{cortese2021}), making them become red, passive, and gas deficient before they fall into a cluster. This process (known as “preprocessing”) can account for a significant fraction of the quenched galaxies at the outskirts of the cluster or beyond the virial radius of the cluster \citep[][]{haines2015,denes2016,jaffe2016,robert2017,jung2018,vulcani2018,dzudzar2019,seth2020,cortese2021,kleiner2021,morokuma-matsui2021,castignani2021}. Therefore, studying “preprocessing” (e.g., how significantly galaxies can be processed in the group environment before they enter a cluster) is important to understand galaxy evolution in the groups as well as the clusters.

In addition, at least half of all galaxies belong to galaxy groups in the local universe \citep[][]{eke2004,robotham2011}, indicating that the group environment is the common place where local galaxies evolve.

Although there are various external processes that can play a role in changing physical properties of galaxies, tidal interactions and merging events are more frequent, especially in the group environment, due to low-velocity dispersion of the group, which are thought to be the main mechanisms affecting group galaxies \citep[][]{zabludoff1998,bitsakis2014, alatalo2015,iodice2020,kleiner2021,s.wang2021}. Some groups are detected in X-ray, indicating the presence of a hot intragroup medium (IGrM) \citep[e.g.,][]{osmond2004}. However, the strength of ram pressure in groups is expected to be not as strong as the strength of ram pressure in clusters because the velocity of group galaxies relative to the IGrM is not as high and the density of IGrM is relatively low. Nevertheless, evidence of ram pressure stripping has been reported in galaxy groups \citep[][]{kantharia2005, rasmussen2006,westmeier2011,wolter2015,brown2017,roberts2021}.

The cold interstellar medium (ISM), mainly composed of atomic hydrogen ({\HI}) and molecular hydrogen (H$_{2}$), is one of the important baryonic components of a galaxy, as the fuel for star formation. Generally, the {\HI} gas disk of the star-forming field galaxies extends beyond the optical disk \citep[e.g.,][]{walter2008}. The large extent of the {\HI} disk and the low density of {\HI} gas make it more susceptible to environmental processes, such as tidal interactions \citep[][]{yun1994,saponara2018} and ram pressure stripping \citep[][]{chung2009,wang2020,wang2021}. In the group environment, galaxies are often reported with {\HI} deficiency, asymmetric {\HI} distribution, and a shrinking of {\HI} disk \citep[][]{denes2016,brown2017,for2019,leewaddell2019,for2021,kleiner2021,roychowdhury2022,s.wang2021}. 

In particular, the H$_{2}$ gas, which is mainly traced by carbon monoxide (CO), is known as the direct ingredient for star formation. Therefore, it is essential to understand how the group environmental processes affect the molecular gas of galaxies because a change of molecular gas properties by the environmental effects likely links to star formation activity and hence galaxy evolution. However, since the distribution of molecular gas is more compact within the stellar disk and its density is higher, it still remains unclear whether the molecular gas is also strongly affected by the group environmental processes. For cluster galaxies, recent studies on molecular gas have found evidence of the environmental effects on molecular gas of cluster galaxies \citep[][]{boselli2014b,lee2017,lee2018,jachym2019,zabel2019,moretti2020a,moretti2020b,cramer2020}. However, there are not many studies of the molecular gas in group galaxies \citep[e.g.,][]{alatalo2015}, and therefore we still lack a good understanding of the group environmental effects on the molecular gas.

In order to obtain a better understanding of the group environmental effects on the molecular gas and the star formation activity, we carried out a $^{12}$CO($J$=1--0) imaging survey of 31 galaxies in two loose groups (IC~1459 group and NGC~4636 group) using the Atacama Large Millimeter/submillimeter Array (ALMA)/the Atacama Compact Array (ACA) in Cycle 7. This is the first CO imaging survey for the loose groups. In particular, a loose group is an intermediate structure between compact groups and clusters. Loose groups host tens of galaxies over an area of $\sim$1 Mpc$^{2}$, with a median velocity dispersion of 165 km~s$^{-1}$ and a median virial mass of $\sim$1.9 $\times$ 10$^{13}h^{-1}$ {\Msun} \citep{tucker2000,pisano2004}. These are particular interesting objects in a study of structure formation in the hierarchical Universe.

However, previous CO imaging studies on group galaxies mainly focused on compact groups, showing violent interactions among group members \citep[e.g.,][]{alatalo2015}. Thus, the well-resolved CO imaging data of our survey can provide a unique opportunity to study the detailed molecular gas properties (e.g., CO distribution and velocity field) of group galaxies and is expected to show direct evidence for the group environmental effects on the molecular gas.

Recent high-resolution {\HI} imaging observations for the IC~1459 group and the NGC~4636 group (hereafter I1459G and N4636G, respectively) show that there are explicit signs (stripped {\HI} gas and asymmetric {\HI} morphology) of external perturbations \citep{serra2015,oosterloo2018,saponara2018,koribalski2020}. In addition, thanks to the proximity of these two groups (I1459G: 27.2 Mpc and N4636G: 13.6 Mpc) and the high spatial resolution of our ACA observations, individual group galaxies can be resolved on a kpc scale (I1459G: $\sim$1.5 kpc and N4636G: $\sim$0.7 kpc). Therefore, both the I1459G and N4636G are good laboratories to investigate the group environmental effects on molecular gas in detail. In particular, since the N4636G is falling into the Virgo cluster \citep{tully1984,nolthenius1993}, this makes the N4636G the best target to study the preprocessing in the group environment.

In this work, we present the results of our ACA CO survey of the two groups. In particular, we compare the CO images with existing {\HI} images for our group sample to investigate how these two phases of the cold ISM react to the group environmental processes, since the {\HI} gas and CO gas have different characteristics (e.g., the disk size and the density). The combination of CO and {\HI} images is possibly powerful for distinguishing between different stripping or quenching mechanisms.

In Section~\ref{sec:sample}, we describe the two galaxy groups and the sample selection along with the general properties of the target galaxies. Details of the ACA observations, data reduction, and ancillary data are described in Section~\ref{sec:obs}. We present CO properties and CO images of our sample in Section~\ref{sec:res}. In Section~\ref{sec:dis}, we investigate differences in global properties between the group sample and the extended CO Legacy Database for GASS\footnote{The Parkes Galactic All Sky Survey} (xCOLD GASS; \citealt{saintonge2017}) sample, and we discuss the impacts of the group environment on member galaxies. In Section~\ref{sec:sum}, we summarize our results and conclusions. Throughout this paper, we adopt distances of 27.2 Mpc and 13.6 Mpc to two groups, I1459G and N4636G, respectively \citep{brough2006}.

\section{Group sample} \label{sec:sample}
The Group Evolution Multiwavelength Study (GEMS) is a panchromatic survey of 60 nearby galaxy groups (the recession velocity of 1000~km~s$^{-1}$ $<$ $v_{\rm group}$ $<$ 3000~km~s$^{-1}$), which has been initiated to study environmental effects on galaxy evolution in the group environment \citep[][]{osmond2004, forbes2006}. The survey covers a broad range of wavelengths from radio \citep{kilborn2005,kilborn2009}, infrared \citep{brough2006}, optical \citep{miles2004}, and X-ray \citep{osmond2004} to investigate how strongly and frequently group members are influenced by various processes. Among 60 groups in the GEMS, in particular, both I1459G and N4636G were observed in the {\HI} emission, with the high spatial resolution of the Australian Square Kilometer Array Pathfinder (ASKAP, \citealt{serra2015a,koribalski2020}). The presence of high-resolution {\HI} images takes great advantage of verifying immediately the environmental effects on the cold gas disk (e.g., asymmetric {\HI} distribution; \citealt{serra2015a,for2019, leewaddell2019}). For these reasons, we selected these two groups to probe the group environmental effects on the cold gas and star formation activity of group galaxies.

Both I1459G and N4636G have many properties in common such as the detection of X-ray emission, the presence of a relatively large elliptical galaxy at the group center (i.e., bright group-centered galaxy; BGG) \citep[][]{osmond2004,brough2006}. However, their locations in the large scale structure are rather distinct. The I1459G is a relatively isolated group, while the N4636G falls into the Virgo cluster \citep{tully1984,nolthenius1993}. Therefore, the N4636G is an ideal laboratory to study “preprocessing” of group galaxies.

We describe the details of the properties of two groups as well as the sample selection for our ACA CO observations in the following sections.

\begin{figure*}[!htbp]
\begin{center}
\includegraphics[width=1.0\textwidth]{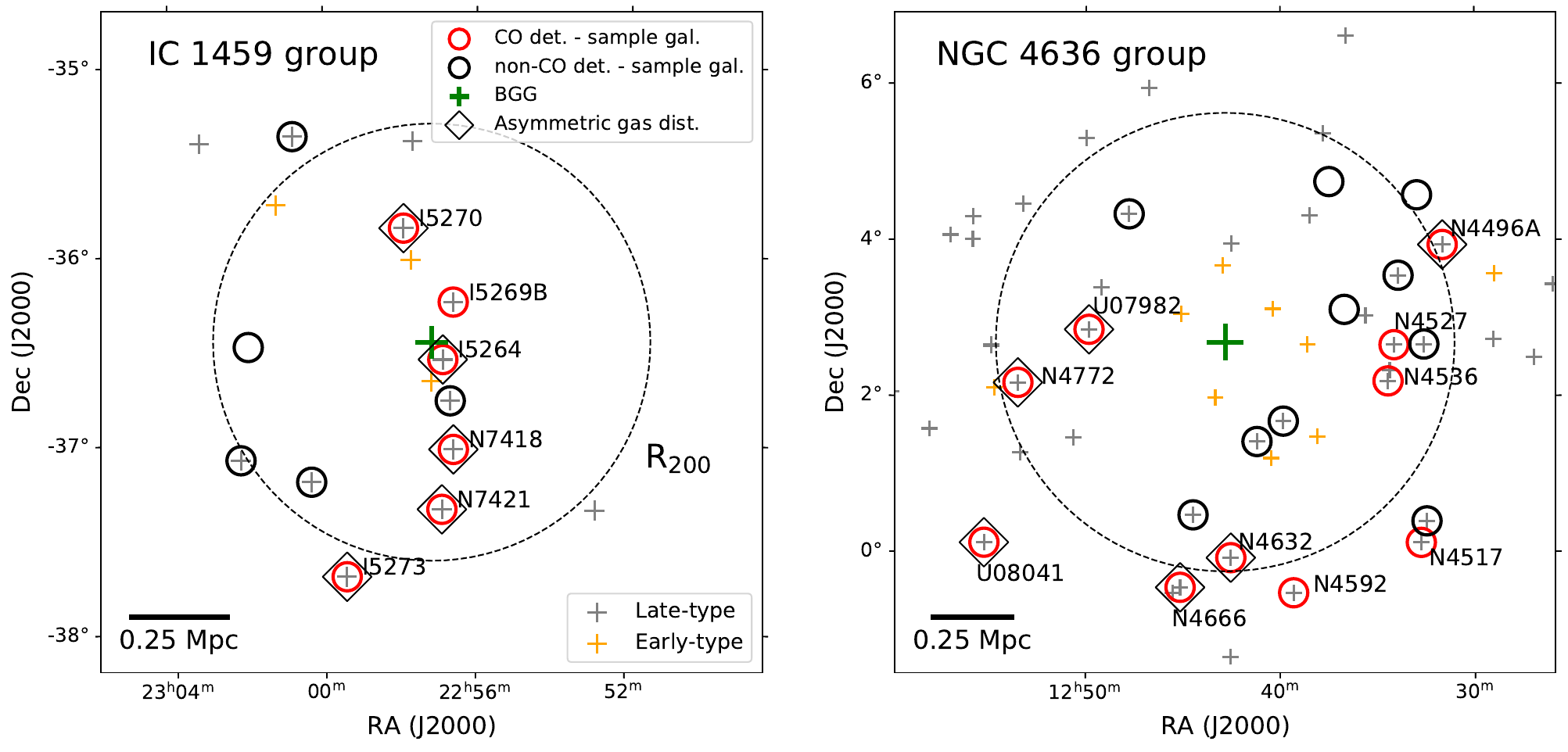}
\caption{Distribution of sample galaxies in the I1459G (left) and the N4636G (right). Our targets are indicated by open circles (red: detection of CO, black: nondetection of CO). Open diamonds display some of the samples, which display asymmetric {\HI} and/or CO distributions. Green large cross in each panel is the BGG. Small crosses (gray: late-type, yellow: early-type) represent galaxies taken from the Hyperleda database, which are also within 1.5 $\times$ $R_{200}$ (dashed circles, I1459G: 0.55 Mpc , N4636G: 0.70 Mpc) and $\pm$3 $\times$ $\sigma_{\rm group}$ (I1459G: 223 km~s$^{-1}$, N4636G: 284 km~s$^{-1}$) and brighter than the absolute magnitude of -15.5 (the faintest galaxy among small crosses in the I1459G) in the $B$-band. The bar in the bottom-left corner represents the physical scale of 0.25 Mpc. \label{fig:fig1}}
\end{center}
\end{figure*}

\begin{figure*}[!htbp]
\begin{center}
\includegraphics[width=1.0\textwidth]{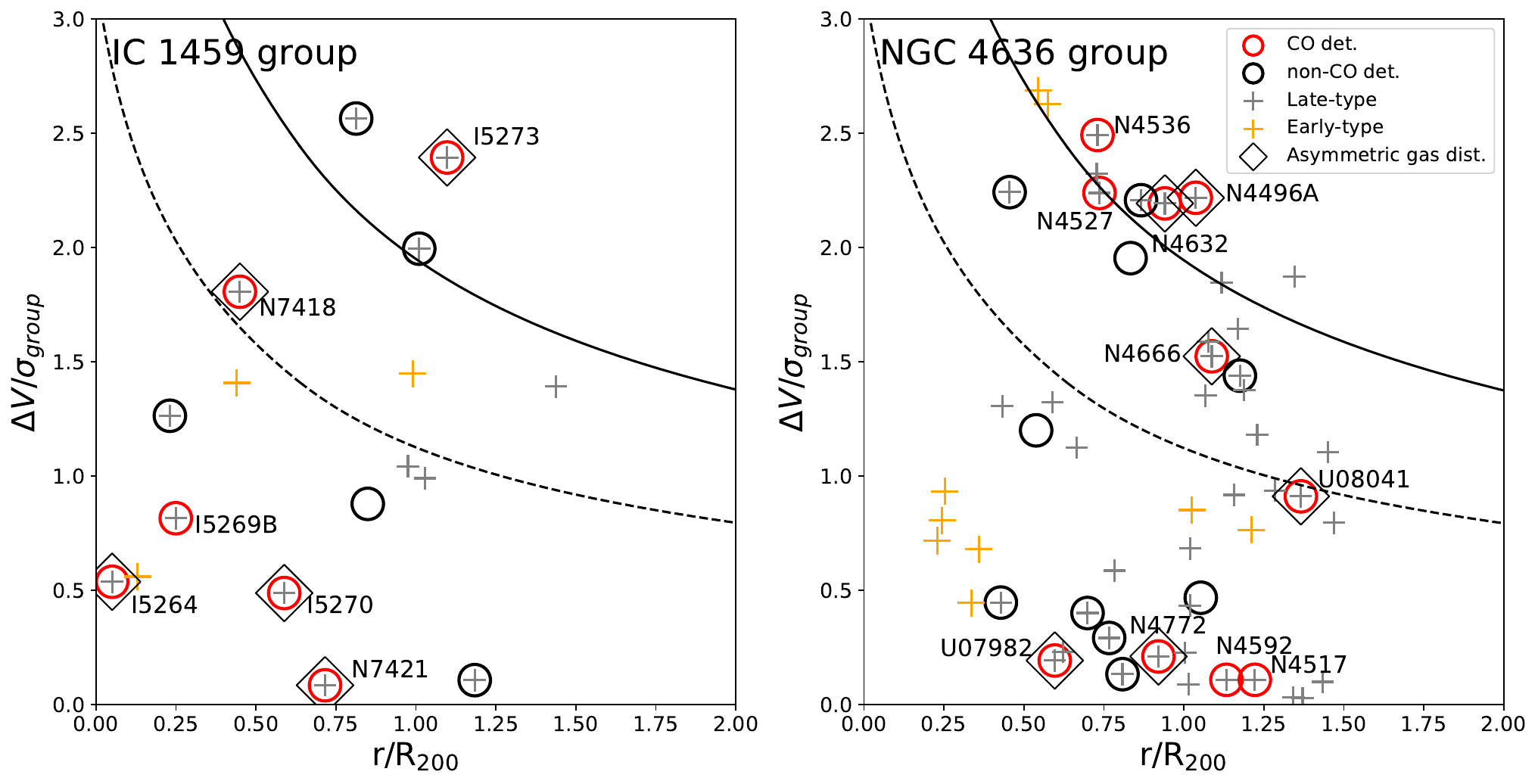}
\caption{Distribution of sample galaxies on the projected phase-space diagram (left: I1459G, right: N4636G). Symbols are the same as in Figure~\ref{fig:fig1}. The x-axis is a projected distance from the group center ($r/R_{200}$, normalized by $R_{200}$). The y-axis is a line of sight velocity with respect to the systemic velocity of the group ($\Delta V/\sigma_{\rm group}$, normalized by the group velocity dispersion). The dashed (solid) line indicates the averaged (the maximum) escape velocity as a function of the projected distance from the group center ($r/R_{200}$). \label{fig:fig2}}
\end{center}
\end{figure*}

\subsection{IC~1459 Group} \label{sec:ic1459g}
The I1459G is a loose galaxy group that includes $\sim$10 members identified by a friends-of-friends (FOF) algorithm \citep{brough2006}. Additional faint galaxies (e.g., dwarf and ultra-diffuse galaxies) are likely to be associated with the I1459G, based on previous {\HI} observations and recent deep optical imaging  \citep[e.g.,][]{kilborn2009,serra2015,forbes2020}.  

The BGG of the I1459G, with a stellar mass of $\sim$3~$\times$~10$^{11}$~{\Msun}, which we estimated using the the Wide-field Infrared Survey Explorer (WISE) data (see Section~\ref{subsec:add_data}), shows many peculiar features, such as a counter-rotating stellar core \citep{franx1988,prichard2019}, an irregular dust distribution \citep{forbes1994}, and the tidal tails and shells in the outskirts of the galaxy \citep{forbes1995,iodice2020}. All these features indicate that the BGG in the I1459G has experienced tidal interactions and/or merging events. X-ray observations also revealed that there is a hot IGrM at the center of the I1459G \citep{osmond2004}.

Many previous {\HI} imaging observations of the I1459G \citep[e.g.,][]{kilborn2009,serra2015,saponara2018,oosterloo2018} also suggest that galaxies in the I1459G are undergoing violent interactions (e.g., tidal interactions between galaxies). With the Australia Telescope Compact Array (ATCA), \cite{saponara2018} reported a substantial amount of {\HI} in clouds ($\sim$7.2$\times10^{8}$ {\Msun}) and in extended distributions near IC~1459 (BGG), IC~5264, and NGC~7418A, covering nearly half of the group region. In addition, \cite{oosterloo2018} found a long {\HI} tail across the I1459G with the Karoo Array Telescope (KAT-7). Recent {\HI} imaging data obtained from the ASKAP reveal that many of group galaxies show asymmetric and disturbed {\HI} morphologies \citep{serra2015}.

Under the assumption that the I1459G is in dynamical equilibrium \citep[e.g.,][]{balogh2007} and adopting the group velocity dispersion ($\sigma_{\rm group}$: 223 km~s$^{-1}$) taken from \cite{osmond2004}, we calculated the group virial radius ($R_{200}$: 0.55 Mpc) and the total group mass of $M_{200}$ (1.9 $\times$ 10$^{13}$ {\Msun}), where $R_{200}$ is the radius at which the system’s mean mass density is 200 times higher than the critical density of the Universe, and $M_{200}$ is the total mass enclosed within $R_{200}$.

\subsection{NGC~4636 Group} \label{sec:ngc4636g}
The N4636G is located towards the southeast region of the Virgo cluster and is falling into the cluster. In the group, there is an early-type BGG (NGC~4636) with a stellar mass of $\sim$3$\times$10$^{10}${\Msun} \citep{osullivan2018}.

The N4636G shows luminous and extended X-ray emission, indicating the existence of a hot IGrM \citep[e.g.,][]{osmond2004,baldi2009,ahoranta2016}. This X-ray emission is extended out to $\sim$0.3~Mpc from the group center \citep{matsushita1998}, suggesting that some group galaxies near the group center are possibly affected by ram pressure stripping. Interestingly, recent X-ray studies of NGC~4636 show that the morphology of X-ray emission near the BGG is complex, showing arm-like structures and bubbles related to previous active galactic nucleus (AGN) activity \citep[e.g.,][]{baldi2009}. The direction of the radio jets also coincides with the X-ray cavities \citep{giacintucci2011}.

The N4636G hosts at least 17 members identified by \cite {brough2006}. However, $\sim$100 systems, including low-mass dwarfs and massive spiral galaxies, could be associated with the N4636G, in a recent study \citep{lin2023}. Previous {\HI} observations of the N4636G found that there are many {\HI}-deficient galaxies, which implies that group members in the N4636G are affected by the environmental effects \citep{kilborn2009}.


Under the assumption that the N4636G is in dynamical equilibrium and adopting the group velocity dispersion ($\sigma_{\rm group}$: 284 km~s$^{-1}$) taken from \cite{osmond2004}, we calculated the total group mass of $M_{200}$ (3.9 $\times$ 10$^{13}$ {\Msun}) and the group virial radius ($R_{200}$: 0.70 Mpc).

\subsection{Sample Selection} \label{sec:selection}
The target galaxies of our ACA CO survey were selected from the GEMS-{\HI} survey that observed {\HI} gas of galaxies within a square region of 30 deg$^{2}$ surrounding the center of each group for a sample of 16 selected GEMS groups, with the Parkes radio telescope \citep{kilborn2009}. From this {\HI} survey, we selected a sample of 31 systems with {\HI} detections, which are located within 1.5 $\times$ $R_{200}$ and $\pm$3 $\times$ $\sigma_{\rm group}$ (group velocity dispersion), in these two groups (11 galaxies in the I1459G and 20 galaxies in the N4636G). Since the primary selection criterion is the presence of {\HI}, which implies that the sample galaxies may have a cold ISM with potentially molecular gas, our targets show a variety of morphological types from dwarfs to spirals (Table~\ref{tab:table1}), with a wide range of stellar masses (1.6 $\times$ 10$^{6}$ $-$ 4.9 $\times$ 10$^{10}$~{\Msun}; Table~\ref{tab:table1}) and {\HI} gas masses (3.1 $\times$ 10$^{7}$ $-$ 7.5 $\times$ 10$^{8}$ {\Msun}; Table~\ref{tab:table1}, \citealt{kilborn2009}).

Figure~\ref{fig:fig1} shows the locations of our sample in the I1459G and the N4636G (red circles: detection of CO, black circles: nondetection of CO). Their BGGs are shown as green large crosses. Small crosses (gray: late-type, yellow: early-type) also indicate galaxies taken from the Hyperleda database, which are also within 1.5 $\times$ $R_{200}$ and $\pm$3 $\times$ $\sigma_{\rm group}$ and brighter than the absolute magnitude of -15.5 (the faintest galaxy among small crosses in the I1459G) in the $B$-band. Our target galaxies of the I1459G are located around the BGG as well as in the outskirts of the group. On the other hand, many of our sample galaxies in the N4636G are located around its $R_{200}$.

Figure~\ref{fig:fig2} also displays the distributions of the sample galaxies on the projected phase-space diagram (PSD), which allows us to probe a projected distance from the group center ($r/R_{200}$, normalized by $R_{200}$) and a line of sight velocity with respect to the systemic velocity of the group ($\Delta V/\sigma_{\rm group}$, normalized by the group velocity dispersion), simultaneously. Some of our group targets, with high $\Delta V/\sigma_{\rm group}$ above the curve of the maximum escape velocity, may just pass through a group, but they can still be affected by all group environmental effects during this one fly-by trip.

The general properties (e.g., coordinates, stellar masses, {\HI} gas masses, star formation rates) of our sample are summarized in Table~\ref{tab:table1}.

\begin{deluxetable*}{lccccccccc}
\tabletypesize{\footnotesize}
\tablecaption{General Properties of Sample Galaxies in IC~1459 Group and NGC~4636 Group \label{tab:table1}}
\tablehead{
\multicolumn{1}{l}{Name} & \multicolumn{1}{c}{R.A.} & \multicolumn{1}{c}{Decl.} & \multicolumn{1}{c}{Type} & \multicolumn{1}{c}{Inc} & \multicolumn{1}{c}{PA} & \multicolumn{1}{c}{v (opt)} & \multicolumn{1}{c}{$M_{\rm\tiny {\HI}}$} & \multicolumn{1}{c}{log $M_\star$} & \multicolumn{1}{c}{SFR} \\
& (J2000)  & (J2000) &  & (deg) & (deg) & (km~s$^{-1}$) & ($\times$10$^8$\Msun) & (\Msun) & (\Msun~yr$^{-1}$) \\
(1) & (2) & (3) & (4) & (5) & (6) & (7) & (8) & (9) & (10)
}
\startdata
\multicolumn{10}{c}{IC~1459 group} \\
\hline
ESO 406-G040 & 23h00m21s.99 & -37d12m04s.0 & IB & 44.3 & 1.5 & 1248 & 6.3$\pm$1.0 & 8.87 & 0.044$\pm$0.020 \\
ESO 406-G042 & 23h02m14s.24 & -37d05m01s.2 & SABm & 80.1 & 66.4 & 1375 & 19.2$\pm$1.6 & 9.42 & 0.169$\pm$0.025 \\
DUKST 406-083 & 23h02m00s.50 & -36d29m04s.5  & \nodata & 66.1 & 97.0 & 1624 & 2.8$\pm$1.0 & 7.50 & 0.021$\pm$0.003 \\
IC 5264 & 22h56m53s.02 & -36d33m15s.3 & Sab & 90.0 & 80.1 & 1940 & 9.7$\pm$1.2 & 10.15 & 0.144$\pm$0.004 \\
IC 5269B & 22h56m36s.72 & -36d14m59s.5 & SBc & 82.6 & 96.7 & 1638 & 30.0$\pm$2.0 & 9.89 & 0.178$\pm$0.026 \\
IC 5269C & 23h00m48s.17 & -35d22m13s.5 & Scd & 74.9 & 62.8 & 1796 & 14.7$\pm$1.6 & 9.46 & 0.121$\pm$0.022 \\
IC 5270 & 22h57m54s.88 & -35d51m29s.0 & SBc & 58.2 & 105.1 & 1929 & 66.6$\pm$3.7 & 9.94 & 0.740$\pm$0.102 \\
IC 5273 & 22h59m26s.72 & -37d42m10s.5 & SBc & 50.8 & 48.5 & 1286 & 38.1$\pm$2.2 & 10.34 & 1.758$\pm$0.529 \\
NGC 7418 & 22h56m36s.15 & -37d01m48s.2 & Sc & 40.0 & 137.7 & 1417 & 48.8$\pm$2.6 & 10.57 & 2.049$\pm$0.759 \\
NGC 7418A & 22h56m41s.23 & -36d46m21s.9 & Scd & 61.2 & 81.4 & 2102 & 40.1$\pm$2.3 & 9.30 & 0.456$\pm$0.033 \\
NGC 7421 & 22h56m54s.32 & -37d20m50s.7 & Sbc & 36.2 & 80.6 & 1801 & 7.5$\pm$1.2 & 10.32 & 0.274$\pm$0.041 \\
\hline
\multicolumn{10}{c}{NGC~4636 group} \\
\hline
EVCC 0854 & 12h32m58s.04 & +04d34m42s.5 & Sm  & 60.8 & \nodata & 1233 & 0.9$\pm$0.2 & 6.20 & 0.002$\pm$0.001 \\
EVCC 0962 & 12h37m29s.07 & +04d45m04s.3 & I & 62.4 & 12.4 & 1655 & 0.3$\pm$0.2 & 7.65 & 0.005$\pm$0.001 \\
IC 3474 & 12h32m36s.51 & +02d39m40s.9 & Scd & 87.1 & 42.5 & 1727 & 5.3$\pm$0.3 & 8.75 & \nodata\\
NGC 4496A & 12h31m39s.26 & +03d56m22s.8 & Scd & 38.5 & 67.5 & 1730 & 19.0$\pm$0.9 & 9.58 & 0.345$\pm$0.018 \\
NGC 4517 & 12h32m45s.48 & +00d06m52s.5 & Sc & 90.0 & 84.4 & 1131 & 38.1$\pm$1.4 & 10.27 & 0.633$\pm$0.079 \\
NGC 4517A & 12h32m27s.97 & +00d23m25s.7 & Sd & 75.2 & 30.1 & 1509 & 14.9$\pm$0.7 & 9.33 & 0.104$\pm$0.015 \\
NGC 4527 & 12h34m08s.47 & +02d39m13s.9 & SABb & 81.2 & 69.5 & 1736 & 41.9$\pm$1.4 & 10.30 & 1.895$\pm$0.943 \\
NGC 4536 & 12h34m27s.09 & +02d11m16s.8 & SABb & 73.1 & 120.7 & 1808 & 31.9$\pm$1.1 & 10.13 & 2.171$\pm$0.959 \\
NGC 4592 & 12h39m18s.74 & -00d31m54s.6 & Sd & 90.0 & 94.2 & 1069 & 75.0$\pm$3.3 & 9.38 & 0.305$\pm$0.016 \\
NGC 4632 & 12h42m32s.34 & -00d04m54s.2 & Sc & 70.5 & 58.9 & 1723 & 20.2$\pm$0.8 & 9.42 & 0.258$\pm$0.028 \\
NGC 4666 & 12h45m08s.64 & -00d27m42s.5 & SABc & 69.6 & 40.6 & 1533 & 36.9$\pm$1.4 & 10.18 & 2.193$\pm$1.025 \\
NGC 4688 & 12h47m46s.61 & +04d20m12s.8 & Sc & 23.7 & \nodata & 986 & 12.6$\pm$1.0 & 8.69 & 0.182$\pm$0.028 \\
NGC 4772 & 12h53m29s.17 & +02d10m06s.3 & SABa & 67.3 & 155.1 & 1040 & 1.5$\pm$0.2 & 10.28 & 0.029$\pm$0.006 \\
UGC 07715 & 12h33m55s.68 & +03d32m45s.6 & I & 20.9 & \nodata & 1138 & 0.7$\pm$0.1 & 8.65 & 0.012$\pm$0.002 \\
UGC 07780 & 12h36m42s.09 & +03d06m30s.7 & Sm & 90.0 & 48.2 & 1441 & 2.9$\pm$0.2 & 7.97 & 0.008$\pm$0.007 \\
UGC 07824 & 12h39m50s.42 & +01d40m20s.7 & SBm & 73.0 & 80.9 & 1227 & 2.1$\pm$0.2 & 8.70 & 0.009$\pm$0.003 \\
UGC 07841 & 12h41m11s.59 & +01d24m37s.1 & Sb & 62.3 & 144.4 & 1737 & 3.8$\pm$0.3 & 8.92 & \nodata \\
UGC 07911 & 12h44m28s.78 & +00d28m05s.4 & Sm & 66.8 & 15.0 & 1183 & 5.5$\pm$0.3 & 9.58 & 0.077$\pm$0.012 \\
UGC 07982 & 12h49m50s.25 & +02d51m06s.9 & Sb & 87.1 & 0.8 & 1155 & 3.3$\pm$0.2 & 9.39 & 0.009$\pm$0.001 \\
UGC 08041 & 12h55m12s.64 & +00d06m60s.0 & SBcd & 54.0 & 168.3 & 1359 & 9.2$\pm$0.7 & 9.53 & 0.094$\pm$0.014 \\
\enddata
\tablecomments{(1) Galaxy name; (2) R.A. (J2000); (3) decl. (J2000); (4) morphological type; (5) inclination angle; (6) position angle; Columns (2) - (6) from HyperLeda (\url{http://leda.univ-lyon1.fr/}; \citealt{makarov2014}) (7) optical velocity \citep{kilborn2009}; (8) {\HI} gas mass, {\HI} gas masses from Parkes observations \citep{kilborn2009}; (9) stellar mass, stellar masses estimated from the WISE data; (10) star formation rate (SFR), star formation rates derived using the GALEX FUV/NUV data and the WISE 22 $\mu$m data.}

\end{deluxetable*}

\begin{deluxetable*}{lcccc}
\tabletypesize{\footnotesize}
\tablecaption{CO Observational Parameters of Sample Galaxies in Two Groups \label{tab:table2}}
\tablehead{
\multicolumn{1}{l}{Name} & \multicolumn{1}{c}{$b_{maj}$$\times$$b_{min}$, $b_{\rm PA}$} & \multicolumn{1}{c}{$\sigma_{rms}$} & \multicolumn{1}{c}{$N_{\rm mosaic}$} & \multicolumn{1}{c}{{\HI} Imaging Data} \\
& ($\arcsec\times\arcsec$, deg) & (mJy~beam$^{-1}$) &  &  \\
(1) & (2) & (3) & \multicolumn{1}{c}{(4)} & \multicolumn{1}{c}{(5)}
}
\startdata
\multicolumn{5}{c}{IC~1459 Group} \\
\hline
ESO 406-G040 & 14.2$\times$8.9, 102.8 & 14.6 & 1 & ASKAP BETA \\
ESO 406-G042 & 14.1$\times$8.9, 102.3 & 13.9 & 1 & ASKAP BETA \\
DUKST 406-083 & 14.2$\times$8.9, 103.5 & 13.9 & 1 & \nodata \\
IC 5264 & 14.2$\times$8.9, 103.2 & 13.1 & 1 & ASKAP BETA \\
IC 5269B & 14.2$\times$8.9, 101.8 & 10.7 & 3 & ASKAP BETA \\
IC 5269C & 14.2$\times$8.8, 103.0 & 13.9 & 1 & ASKAP BETA \\
IC 5270 & 14.2$\times$8.9, 102.3 & 10.1 & 3 & ASKAP BETA \\
IC 5273 & 14.1$\times$9.0, 100.2 & 11.5 & 2 & ASKAP BETA \\
NGC 7418 & 14.1$\times$9.0, 101.3 & 9.2 & 5 & ASKAP BETA \\
NGC 7418A & 14.2$\times$8.9, 103.5 & 13.1 & 1 & ASKAP BETA \\
NGC 7421 & 14.2$\times$8.9, 103.2 & 13.8 & 1 & ASKAP BETA \\
\hline
\multicolumn{5}{c}{NGC~4636 Group} \\
\hline
EVCC 0854 & 14.1$\times$10.1, 97.0 & 13.9 & 1 & WALLABY pilot \\
EVCC 0962 & 14.2$\times$9.9, 101.4 & 13.3 & 1 & WALLABY pilot \\
IC 3474 & 14.2$\times$9.8, 102.0 & 13.7 & 1 & WALLABY pilot \\
NGC 4496A & 14.1$\times$9.9, 101.8 & 8.8 & 6 &  WALLABY pilot \\
NGC 4517 & 14.1$\times$9.7, 94.9 & 11.4 & 8 & WALLABY pilot \\
NGC 4517A & 14.1$\times$9.7, 97.0 & 14.3 & 1 & WALLABY pilot \\
NGC 4527 & 14.1$\times$9.9, 96.8 & 11.7 & 5 &  WALLABY pilot \\
NGC 4536 & 14.1$\times$9.8, 99.8 & 8.7 & 13 & VIVA \\
NGC 4592 & 14.0$\times$9.5, 100.5 & 11.4 & 3 &  WALLABY pilot \\
NGC 4632 & 14.0$\times$9.7, 95.6 & 10.9 & 3 &  WALLABY pilot \\
NGC 4666 & 13.9$\times$9.6, 99.4 & 11.1 & 3 &  WALLABY pilot \\
NGC 4688 & 14.1$\times$10.0, 91.1 & 10.1 & 5 &  WALLABY pilot \\
NGC 4772 & 14.1$\times$9.8, 92.7 & 12.1 & 3 &  VIVA \\
UGC 07715 & 14.1$\times$10.0, 96.1 & 13.9 & 1 & WALLABY pilot \\
UGC 07780 & 14.1$\times$10.0, 96.0 & 14.7 & 1 & WALLABY pilot \\
UGC 07824 & 14.1$\times$9.9, 94.8 & 14.5 & 1 & \nodata \\
UGC 07841 & 14.0$\times$9.8, 95.7 & 14.2 & 1 & WALLABY pilot \\
UGC 07911 & 14.1$\times$9.8, 93.4 & 14.5 & 1 & WALLABY pilot \\
UGC 07982 & 14.0$\times$9.9, 92.6 & 11.7 & 3 & WALLABY pilot \\
UGC 08041 & 13.9$\times$9.6, 98.3 & 11.2 & 3 & \nodata \\
\enddata
\tablecomments{
(1) Galaxy name; (2) beam size (major and minor axes), beam position angle; (3) rms noise level per channel of the CO data; (4) the number of mosaic fields; (5) the {\HI} imaging data that we use in Figure~\ref{fig:hico}. See the detailed descriptions of {\HI} imaging data in Section~\ref{subsec:add_data}.
}
\end{deluxetable*}

\begin{deluxetable*}{lccccrrrr}
\tabletypesize{\footnotesize}
\tablecaption{CO Data Parameters of Sample Galaxies in Two Groups \label{tab:table3}}

\tablehead{
\noalign{\vskip 2mm}
\multicolumn{1}{c}{Name} & \multicolumn{1}{c}{$W_{\rm 50, HI}$} & \multicolumn{1}{c}{$W_{\rm 20, CO}$} & \multicolumn{1}{c}{$W_{\rm 50, CO}$} & \multicolumn{1}{c}{CO Peak Flux} & \multicolumn{1}{c}{CO Flux} & \multicolumn{1}{c}{log $L_{\rm CO}^{'}$} & \multicolumn{1}{c}{log $M_{\rm H2, MW}$} & \multicolumn{1}{c}{log $M_{\rm H2, var}$} \\
& (km~s$^{-1}$) & (km~s$^{-1}$) & (km~s$^{-1}$) & (Jy~km~s$^{-1}$~beam$^{-1}$) & (Jy~km~s$^{-1}$) & \multicolumn{1}{c}{(K~km~s$^{-1}$~pc$^{2}$)} & \multicolumn{1}{c}{(\Msun)} & \multicolumn{1}{c}{(\Msun)} \\
\multicolumn{1}{c}{(1)} & (2) & (3) & (4) & \multicolumn{1}{c}{(5)} & \multicolumn{1}{c}{(6)} & \multicolumn{1}{c}{(7)} & \multicolumn{1}{c}{(8)} & \multicolumn{1}{c}{(9)}
}
\startdata
\multicolumn{9}{c}{IC~1459 Group} \\
\hline
ESO 406-G040 & 42 & \nodata & \nodata & \nodata & $<$1.3 & $<$6.36 & $<$7.00 & $<$7.19 \\
ESO 406-G042 & 121 & \nodata & \nodata & \nodata & $<$2.1 & $<$6.57 & $<$7.21 & $<$7.24 \\
DUKST 406-083 & 83 & \nodata & \nodata & \nodata & $<$1.7 & $<$6.49 & $<$7.12 & $<$8.42 \\
IC 5264 & 94 & 356 & 331 & 17.6 & 54.6$\pm$3.6 & 7.96$\pm$0.03 & 8.60$\pm$0.03 & 8.54$\pm$0.03 \\
IC 5269B & 228 & 156 & 149 & 1.8 & 4.3$\pm$0.7 & 6.89$\pm$0.74 & 7.53$\pm$0.07 & 7.49$\pm$0.07 \\
IC 5269C & 172 & \nodata & \nodata & \nodata & $<$2.45 &  $<$6.64 &  $<$7.28 & $<$7.31 \\
IC 5270 & 101 & 230 & 198 & 26.1 & 100.4$\pm$5.5 & 8.26$\pm$0.02 & 8.90$\pm$0.02 & 8.85$\pm$0.02 \\
IC 5273 & 196 & 152 & 110 & 13.5 & 60.3$\pm$3.7 & 8.04$\pm$0.03 & 8.67$\pm$0.03 & 8.60$\pm$0.03 \\
NGC 7418 & 207 & 206 & 175 & 34.6 & 302.2$\pm$15.4 & 8.74$\pm$0.02 & 9.37$\pm$0.02 & 9.29$\pm$0.02 \\
NGC 7418A & 180 & \nodata & \nodata & \nodata &$<$2.4 & $<$6.63 & $<$7.27 & $<$7.32 \\
NGC 7421 & 129 & 151 & 70 & 4.3 & 37.6$\pm$2.7 & 7.83$\pm$0.03 & 8.47$\pm$0.03 & 8.40$\pm$0.03 \\
\hline
\multicolumn{9}{c}{NGC~4636 Group} \\
\hline
EVCC 0854 & 62 & \nodata & \nodata & \nodata & $<$1.5 & $<$5.82 & $<$6.45 & $<$7.75 \\
EVCC 0962 & 27 & \nodata & \nodata & \nodata & $<$0.9 & $<$5.62 & $<$6.26 & $<$7.55 \\
IC 3474 & 147 & \nodata & \nodata & \nodata & $<$2.2 & $<$6.00 & $<$6.64 & $<$6.88 \\
NGC 4496A & 154 & 127 & 102 & 4.6 & 36.3$\pm$2.4 & 7.21$\pm$0.03 & 7.85$\pm$0.03 & 7.86$\pm$0.03 \\
NGC 4517 & 301 & 298 & 275 & 28.6 & 449.4$\pm$22.9 & 8.31$\pm$0.02 & 8.95$\pm$0.02 & 8.88$\pm$0.02 \\
NGC 4517A & 154 & \nodata & \nodata & \nodata & $<$2.4 & $<$6.03 & $<$6.66 & $<$6.72 \\
NGC 4527 & 356 & 411 & 370 & 372.6 & 1638.0$\pm$82.1 & 8.87$\pm$0.02 & 9.51$\pm$0.02 & 9.44$\pm$0.02 \\
NGC 4536 & 324 & 354 & 325 & 214.5 & 608.9$\pm$30.6 & 8.44$\pm$0.02 & 9.08$\pm$0.02 & 9.02$\pm$0.02 \\
NGC 4592 & 198 & 109 & 40 & 1.9 & 13.3$\pm$1.7 & 6.78$\pm$0.06 & 7.42$\pm$0.06 & 7.46$\pm$0.06 \\
NGC 4632 & 223 & 236 & 205 & 13.4 & 115.8$\pm$6.4 & 7.72$\pm$0.02 & 8.36$\pm$0.02 & 8.39$\pm$0.02 \\
NGC 4666 & 324 & 409 & 387 & 154.3 & 1471.7$\pm$73.7 & 8.82$\pm$0.02 & 9.46$\pm$0.02 & 9.40$\pm$0.02 \\
NGC 4688 & 44 & \nodata & \nodata & \nodata & $<$0.9 & $<$5.61 & $<$6.25 & $<$6.52 \\
NGC 4772 & 38 & 156 & 52 & 4.5 & 27.7$\pm$2.2 & 7.10$\pm$0.04 & 7.74$\pm$0.04 & 7.66$\pm$0.04 \\
UGC 07715 & 29 & \nodata & \nodata & \nodata & $<$1.0 & $<$5.66 & $<$6.29 & $<$6.59 \\
UGC 07780 & 117 & \nodata & \nodata & \nodata & $<$2.1 & $<$5.98 & $<$6.22 & $<$7.49 \\
UGC 07824 & 101 & \nodata & \nodata & \nodata & $<$2.0 & $<$5.95 & $<$6.58 & $<$6.85 \\
UGC 07841 & 104 & \nodata & \nodata & \nodata & $<$1.9 & $<$5.94 & $<$6.58 & $<$6.75 \\
UGC 07911 & 107 & \nodata & \nodata & \nodata & $<$2.0 & $<$5.96 & $<$6.60 & $<$6.60 \\
UGC 07982 & 214 & 202 & 176 & 6.4 & 41.3$\pm$2.8 & 7.27$\pm$0.03 & 7.91$\pm$0.03 & 7.95$\pm$0.03 \\
UGC 08041 & 175 & 129 & 102 & 2.3 & 8.00$\pm$1.1 & 6.56$\pm$0.06 & 7.20$\pm$0.06 & 7.21$\pm$0.06 \\
\enddata
\tablecomments{(1) Galaxy name; (2) the {\HI} linewidths measured at 50\% of the peak flux \citep{kilborn2009}; (3) the CO linewidths measured at 20\% and 50\% of the peak flux using SoFiA; (4) the peak value of CO intensity map; (5) the total CO flux; (6) the total CO luminosity; (7) \& (8) the total molecular gas mass derived from the CO luminosity using the constant CO-to-H$_{2}$ conversion factor ($\alpha_{\rm CO}$~=~4.35 {\Msun}~pc$^{-2}$~(K~km~s$^{-1}$)$^{-1}$) and the metallicity-dependent CO-to-H$_{2}$ conversion factor.}
\end{deluxetable*}

\section{Observations and data} \label{sec:obs}
\subsection{Observations}
Our CO imaging observations (project ID: 2019.1.01804.S; PI: B. Lee) of group galaxies were carried out using the ALMA/ACA in Cycle 7 (2019 October to 2019 December). While the ACA consists of 12 antennas of 7 m diameter (the 7m array) and 4 antennas of 12 m diameter (the total power array), we only used the 7m array, with 9$-$11 antennas in our observations. The mean precipitable water vapor value was 3.6 mm during the observations. 

Four spectral windows (SPW 1, 2, 3 ,4) were set up to observe $^{12}$CO($J$=1--0) ($\nu_{\rm rest}$ = 115.271~GHz) line and CN($N$=1--0; $J$=3/2--1/2) ($\nu_{\rm rest}$ = 113.491~GHz) line in the upper sideband (SPW 1 and 2) and 3 mm continuum emission in the lower sideband (SPW 3 and 4). Each SPW has a total bandwidth of 1875 MHz ($\sim$5000~km~s$^{-1}$) and a channel width of $\sim$1.13 MHz ($\sim$3~km~s$^{-1}$). 

The size of the primary beam is $\sim$87$\arcsec$. The most typical extent of CO gas in late-type galaxies is 50\% $-$ 70\% of the inner part of the optical disk \citep[e.g.,][]{schruba2011}. In order to cover the entire CO disk, 15 out of 31 galaxies were observed in mosaic mode. The number of mosaic fields varies from two to 13 pointings, which depends on the apparent size of each target galaxy. The rest (16 targets) of the sample were observed by a single pointing. The largest recoverable angular scale (i.e., the maximum recoverable scale (MRS)) in our ACA observations is 59$\arcsec$, corresponding to $\sim$7.8~kpc at 27.2~Mpc (I1459G) and $\sim$3.9~kpc at 13.6~Mpc (N4636G), respectively. With this MRS, we expected that our proposed observations using only the 7m array could recover most of the total flux. Thus, we did not include the total power array. However, we may possibly miss some CO flux. The missing flux problem is discussed in Section \ref{subsec:coprop} and \ref{subsec:scaling}.

Based on the prediction for the molecular gas surface density at the edge of the star-forming disk (e.g., $\sim$4~\Msun~pc$^{-2}$; \citealt{schaye2004}), we aim to reach a gas surface density of 4~\Msun~pc$^{-2}$ in 5$\sigma$ in a beam with a velocity resolution of 20 km~s$^{-1}$. Adopting the Milky Way CO-to-H$_{2}$ conversion factor ($\alpha_{\rm CO}$~=~4.35 {\Msun}~pc$^{-2}$~(K~km~s$^{-1}$)$^{-1}$; \citealt{strong1996,bolatto2013}), the requested sensitivity is 14.6 mJy~beam$^{-1}$ over a velocity resolution of 20 km~s$^{-1}$.

\subsection{Data Reduction and Imaging} \label{subsec:data_red}
The ACA data were calibrated using the standard ALMA pipeline in the Common Astronomy Software Applications package (CASA, version: 5.6.1-8; \citealt{mcmullin2007}). After calibration, the continuum is subtracted by fitting the line-free channels with the {\tt uvcontsub} task. Cleaned CO data cubes of individual samples are created using the {\tt tclean} task in the CASA package. The clean regions are carefully selected by visual inspection for every channel of the cubes. In particular, the mosaic images of 15 galaxies observed with multiple pointings were produced by setting the sub-parameter {\tt gridder}={\tt ‘mosaic’} in the {\tt tclean} task. To increase the signal-to-noise ratio (S/N) of the data and to recover faint CO emission, natural weighting was applied and the channel width of the cleaned cubes was binned into a velocity resolution of 20 km~s~$^{-1}$. The final data cubes have a synthesized beam of $\sim$14$\arcsec$$\times$$\sim$9$\arcsec$ with a pixel size of 2$\arcsec$. The noise level is measured using line-free channels. The typical root mean square (rms) noise level is $\sim$12.4 mJy beam$^{-1}$ over a channel width of 20~km~s$^{-1}$, which is measured prior to the primary beam correction. The synthesized beam sizes and rms noise levels for individual galaxies are listed in Table~\ref{tab:table2}. The final data cubes are corrected for primary beam attenuation with the {\tt impbcor} task. 

For each galaxy, we obtained a detection mask from the cleaned data cube, using the SoFiA (Source Finding Application) software \citep{serra2015a,westmeier2021} with a 3$\sigma$ threshold for reliable detection. In particular, the data cube was convolved with several smoothing kernels, which are Gaussian filters with FWHM equivalent to 0, 3, and 7 pixels on the sky plane, and a threshold-based detection mask is produced for each smoothed version of the cube. The final detection mask is a combination of all these masks. Such a smooth + clip (S + C) procedure is good at detecting low density and diffuse structures. Then, a reliability parameter of the detection should be calculated by SoFiA by comparing the statistics of distribution of negative and positive pixels, based on the assumption that the noise has a rough symmetric distribution around zero, and thus the noise pattern of positive pixels should be similar to that of negative pixels. We refer the readers to \cite{serra2015a} for more details of the algorithm. Because the field of view is relatively small for some of our observations, there is not sufficient statistics to robustly derive the reliability parameter for every galaxy. Therefore, we further manually inspect each detection mask to remove false detections. By applying the final detection masks for the sample galaxies, we generated integrated CO intensity maps (0th moment), velocity field maps (1st moment), and velocity dispersion maps (2nd moment). 

The CN line data were reduced into final products (i.e., data cubes, moment maps) in the same manner as the CO data. In addition to the line data, we created 3 mm continuum images by averaging the line-free channels in four SPWs. The results of the CN and continuum data are in Appendices~\ref{app:cn} and \ref{app:conti}.

\subsection{Ancillary Data} \label{subsec:add_data}
To estimate the stellar masses ($M_\star$), SFRs, and {\HI} masses of galaxies in our sample and to investigate {\HI} distribution of the sample galaxies, we made use of the complementary data including the far-ultraviolet (FUV), near-ultraviolet (NUV), infrared, and {\HI} images. For calculating the stellar masses and SFRs, we assume a Kroupa initial mass function (IMF; \citealt{kroupa2001}). First, to obtain an estimate of the stellar masses, we utilize the photometric pipeline of \cite{wang2017} with the 3.4~$\mu$m data (W1 band) and 4.6~$\mu$m data (W2 band) of the Wide-field Infrared Survey Explorer (WISE; \citealt{wright2010}). We derive the stellar mass of the sample galaxies using the W1 luminosity with a W1-W2 color dependent mass-to-light ratio \citep{jarrett2017} (for details, see Section 3 of \citealt{wang2017}). The typical error of the stellar mass is 0.15 dex. With a combination of the Galaxy Evolution Explorer (GALEX) FUV data \citep{martin2005} and the 22 $\mu$m data (W4 band) from WISE \citep{wright2010}, we also derive the dust attenuation corrected SFR by using the equations of \cite{calzetti2013}. The FUV and W4 fluxes trace the dust free and dust attenuated part of the total SFR in a galaxy, respectively. When the FUV fluxes have too low S/N ($<$1), we use GALEX NUV fluxes to estimate the dust free part of the SFR.

For some of our sample galaxies, there are recent high-resolution {\HI} imaging data. For the I1459G sample, 10 out of 11 galaxies were detected in the ASKAP BETA (the Boolardy Engineering Test Array) survey \citep{serra2015}. In the case of the N4636G sample, 18 out of 20 galaxies were detected in the WALLABY pilot survey \citep{koribalski2020}. For NGC~4536 and NGC~4772, we use the {\HI} imaging data of the VIVA (the VLA Imaging survey of Virgo galaxies in Atomic gas, \citealt{chung2009}). In total, 28 out of 31 sample galaxies ($\sim$90\%) have resolved {\HI} images. The spatial resolutions for the ASKAP BETA, WALLABY pilot survey, and VIVA are 60$\arcsec$ (7.92~kpc), 30$\arcsec$ (1.98~kpc), and 18$\arcsec$ (1.19~kpc), respectively. These {\HI} imaging data are useful for a comparison of the CO distribution of group galaxies. We adopted the {\HI} gas masses from previous Parkes observations \citep{kilborn2009} for homogeneity and flux completeness. The stellar masses, SFRs, and {\HI} gas masses of our sample are summarized in Table~\ref{tab:table1}.

\section{Results} \label{sec:res}
\subsection{CO Detections}
\begin{figure*}[!htbp]
\begin{center}
\includegraphics[width=1.00\textwidth]{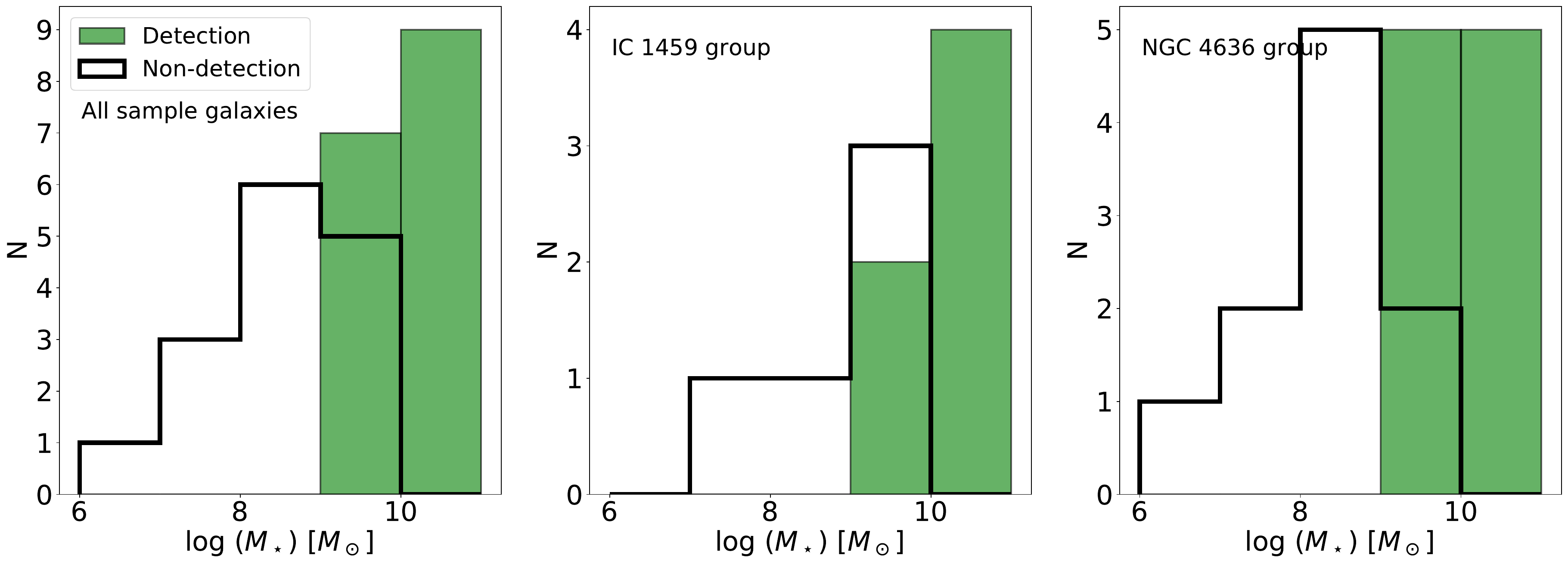}
\caption{The histograms show the stellar mass distributions of CO-detected (green) and non-CO-detected (black solid line) group samples. Left panel: all sample galaxies, middle panel: the I1459G, right panel: the N4636G. The stellar mass range of galaxies with CO detections is $\sim$$10^{9}-10^{10}${\Msun}. 
\label{fig:fig3}}
\end{center}
\end{figure*}

\begin{figure*}[!htbp]
\begin{center}
\includegraphics[width=0.77\textwidth]{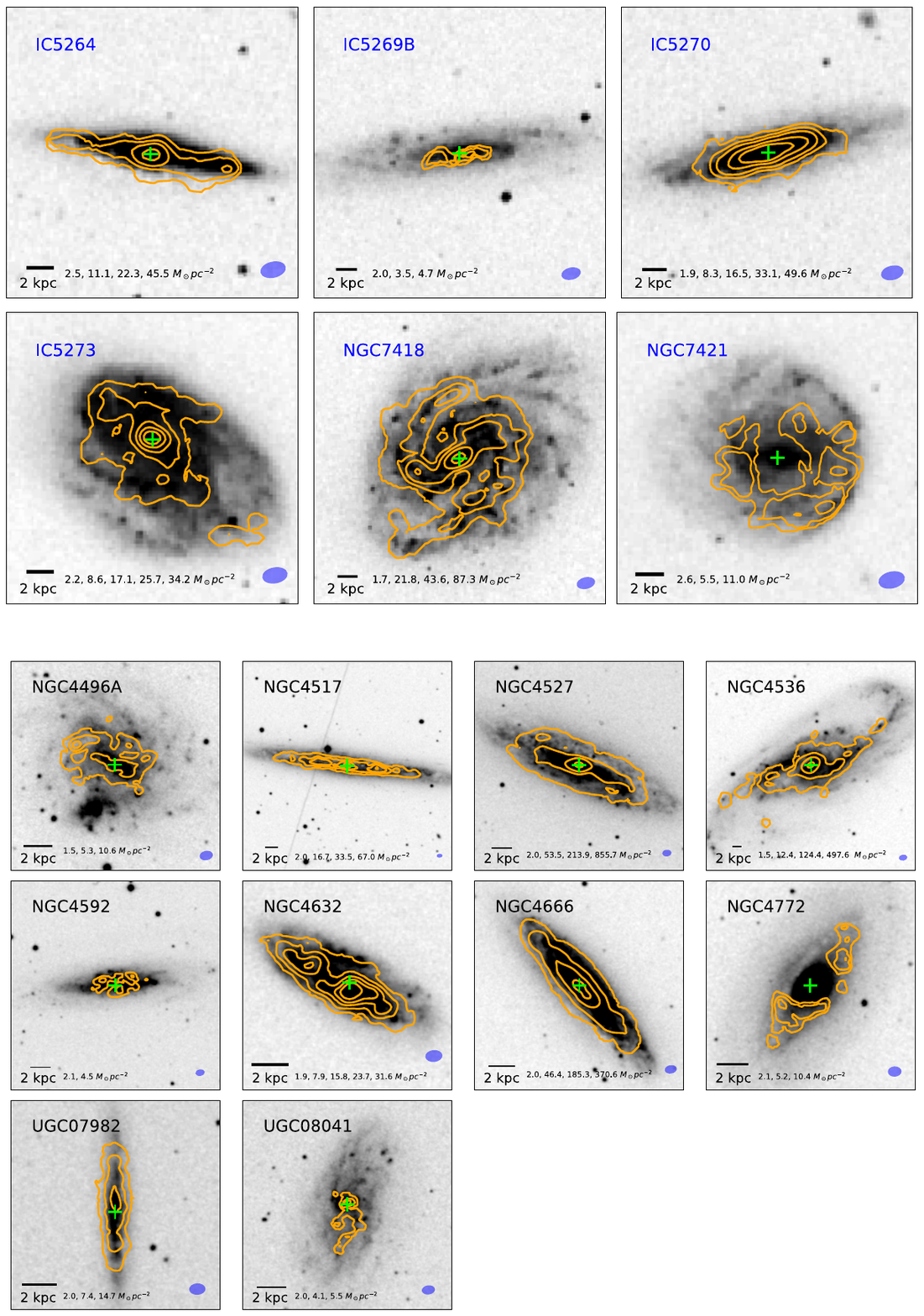}
\caption{The CO distribution (contours) of 16 group galaxies is overlaid on the Digitized Sky Survey 2 (DSS2, \url{https://archive.stsci.edu/dss/index.html}) blue images. The contour levels from outer to inner, indicating the molecular gas surface density, are shown at the bottom of each panel. The surface density of the molecular gas is calculated by adopting $\alpha_{\rm CO}$~=~4.35 {\Msun}~pc$^{-2}$~(K~km~s$^{-1}$)$^{-1}$ \citep{strong1996,bolatto2013}. In this calculation, we did not apply an inclination correction. Galaxies in the I1459G are shown in the first two rows (their names in blue), and galaxies in the N4636G are shown in the second three rows (their names in black). The green cross indicates the stellar disk center, which is obtained from {\it Spitzer} 3.6 $\mu$m data \citep{salo2015}. The bar in the bottom-left corner represents the physical scale of 2~kpc. The synthesized beam is shown in the bottom-right corner.  \label{fig:fig4}}
\end{center}
\end{figure*}

We detected CO emission in 16 out of 31 galaxies from both the I1459G (6/11) and the N4636G (10/20). The stellar mass range of galaxies with CO detection is $\sim$10$^{9}$$-$10$^{10}${\Msun} (Figure~\ref{fig:fig3}); all of these are spiral galaxies. Figure~\ref{fig:fig4} shows the CO distribution overlaid on DSS2 blue optical images. On the other hand, galaxies without CO detection tend to have lower stellar masses ($<$9.4$\times$10$^{9}${\Msun}; see Figure~\ref{fig:fig3}), and their morphological types are dwarfs and spirals (Figure~\ref{fig:nonc0} in Appendix~\ref{app:nonco}). Both Figures~\ref{fig:fig1} and \ref{fig:fig2} show the distributions of all sample galaxies on the projected sky plane and on the projected PSD. Red and black open circles indicate CO detections and nondetections of CO, respectively. Interestingly, most of the CO-detected galaxies of the N4636G tend to be at large group-centric radii, while those in the I1459G tend to be at smaller group-centric radii. We discuss this in detail in Section~\ref{subsec:pre}.

\subsection{CO Properties} \label{subsec:coprop}
The integrated CO flux ($S_{\rm CO}$) are measured in Jy~km~s$^{-1}$ using

\begin{equation}
S_{\rm CO} = (\Sigma~F_{\rm CO})~\Delta v, 
\label{eqn:coflux}
\end{equation}

\noindent where $F_{CO}$ is the total flux of CO in each channel, $\Delta v$ is the velocity resolution (20 km~s$^{-1}$) of the CO data cube. For the integrated CO flux, all pixels within the detection mask are summed over channels. The uncertainty of the integrated CO flux is calculated by

\begin{equation}
\sigma (S_{\rm CO}) = \sqrt{(\Sigma~{\sigma_{\rm rms}}^{2}N_{\rm Beam})~{\Delta v}^{2} + ({S_{\rm CO}}^{2}/400)},
\label{eqn:coerr}
\end{equation}

\noindent where $\sigma_{\rm rms}$ is the rms noise level of data cube, and $N_{\rm Beam}$ is the number of beams in the emission region of each channel. The first term of equation~\ref{eqn:coerr} is the measurement error of the integrated CO flux, and the second term is a typical absolute flux accuracy (5\% in Band 3, ALMA cycle 7 proposer's guide\footnote{\url{https://almascience.nao.ac.jp/documents-and-tools/cycle7/alma-proposers-guide}}). The integrated CO flux and its uncertainty are summarized in Table~\ref{tab:table3}. 

For galaxies with nondetections, the $3\sigma$ upper limits of CO flux are estimated using the rms noise level. To calculate the upper limits, we assume that the size of the CO-emitting area is the same as the beam size and the line width of the CO disk corresponds to an FWHM of {\HI} gas. We adopt values of the FWHM of {\HI} gas (Table~\ref{tab:table3}) from the GEMS-{\HI} observations of \cite{kilborn2009}. 

In our sample, CO emission of five galaxies (NGC~4517, NGC~4527, NGC~4536, NGC~4632, NGC~4666) in the N4636G were also detected in previous single-dish observations \citep{boselli2014a,sorai2019}. Except for NGC~4517, four galaxies have resolved CO maps from the CO Multi-line Imaging of Nearby Galaxies (COMING) project \citep{sorai2019}. To estimate how much of the total CO flux we recover in our ACA observations, we compared the CO flux of our ACA observations with that of previous single-dish observations from the literature. The flux ratio between our ACA data and the single-dish data ranges from 0.5 to 0.9. The averaged flux ratio of three galaxies (NGC~4536, NGC~4632, NGC~4666) is $\sim$0.9, but two galaxies (NGC~4517, NGC~4527) show relatively low flux ratios (NGC~4517: 0.5 and NGC 4527: 0.7). We may miss some CO flux of large-scale structures that are larger than the maximum recoverable scale (59$\arcsec$) of our ACA observations, especially for galaxies with a large angular size of the optical disk (see also Appendix~D of \cite{leroy2021} for a detailed description of the missing flux as a function of CO disk size).

The CO luminosity ($L^{'}_{\rm CO}$) is calculated using the following equation \citep{solomon2005}:

\begin{equation}
L^{'}_{\rm CO}=3.25\times10^7 S_{\rm CO}~\nu_{\rm obs}^{-2}~D_{\rm L}^2~(1+\textit{z})^{-3} 
\label{eqn:colum}
\end{equation}

\noindent in K~km~s$^{-1}$~pc$^{2}$, where $S_{\rm CO}$ is the integrated CO flux in Jy~km~s$^{-1}$, $D_{\rm L}$ is the luminosity distance in Mpc, $\nu_{\rm obs}$ is the observing frequency in GHz, and \textit{z} is the redshift. Using $L^{'}_{\rm CO}$, the H$_{2}$ masses for our sample galaxies are determined as

\begin{equation}
M_{\rm H_{2}} = \alpha_{\rm CO} L^{'}_{\rm CO},
\label{eqn:h2mass}
\end{equation}

\noindent where $\alpha_{\rm CO}$ is a CO-to-H$_{2}$ conversion factor in {\Msun}~pc$^{-2}$~(K~km~s$^{-1}$)$^{-1}$.

We apply two different CO-to-H$_{2}$ conversion factors. First, we use the metallicity-dependent CO-to-H$_{2}$ conversion factor calculated using the Equation (25) of \cite{accurso2017}. Following the approach of \cite{zabel2019}, we do not consider the distance from the main sequence ($\Delta$MS). The $\Delta$MS parameter does not significantly influence the results of calculation of the metallicity-dependent CO-to-H$_{2}$ conversion factor \citep{zabel2019}. We also follow the approach of \cite{zabel2019} to calculate metallicities of our samples because we do not have metallicities of individual galaxies. To derive metallicities of individual galaxies, we use the mass-metallicity relation of \cite{sanchez2017}. As a result, the metallicity-dependent CO-to-H$_{2}$ conversion factor here mainly depends on the stellar masses of galaxies. Therefore, the low-mass galaxies have higher conversion factors. Secondly, we also adopt $\alpha_{\rm CO}$~=~4.35 {\Msun}~pc$^{-2}$~(K~km~s$^{-1}$)$^{-1}$ ($X_{\rm CO} = 2 \times 10^{20}$~cm$^{-2}$ (K~km~s$^{-1})^{-1}$; \citealt{strong1996,bolatto2013}). This CO-to-H$_{2}$ conversion factor includes a factor of 1.36 for helium abundance correction.

The CO linewidth is measured at 20\% ($W_{\rm 20, CO}$) and 50\% ($W_{\rm 50, CO}$) of the peak flux using SoFiA. The CO linewidths, CO luminosities, and H$_{2}$ masses for our sample of group galaxies are summarized in Table~\ref{tab:table3}. The details of the CO data for individual group galaxies with CO detections are described in Appendix.

\subsection{Peculiar CO Structures} \label{subsec:pecco}
We find that some of our group samples show peculiar CO structures in their CO intensity maps, CO velocity maps, and position-velocity diagrams (PVDs), as seen in Figure~\ref{fig:fig4} and the CO atlas of our group members in Appendix~\ref{app:codata}. These peculiarities are such as (1) a highly asymmetric CO distribution, (2) a hole in the central region, (3) a large offset ($>$1 kpc) between the CO peak and the optical center, (4) a bar- or ring-like structure, and (5) a high molecular gas surface density ($>$100~{\Msun}~pc$^{-2}$) in the central region.
\paragraph{\textbf{$\bullet$ Asymmetric CO distribution}} IC~5264, IC~5273, and NGC~7418 in the I1459G show highly asymmetric CO distributions. For IC~5264, the CO disk in the west side is shrunken, compared to the extent of the CO disk in the east side (49$\arcsec$ ($\sim$6.5 kpc) in the west side versus 57$\arcsec$ ($\sim$7.5 kpc) in the east side, Figure~\ref{fig:fig4}). The CO gas also appears to be extended toward the southwest. The 60\% of the total CO flux is measured in the southern part below the major axis of IC~5264, compared to the CO flux in the northern part. In IC~5273, a CO clump is located at the southwest edge of the stellar disk (Figure~\ref{fig:fig4}); its distance from the optical center is 8.8~kpc. The CO flux of this clump is $\sim$3 Jy~km~s$^{-1}$, which corresponds to $\sim$5\% of the total CO flux. While the CO distribution of NGC~7418 follows well the spiral arms of the stellar disk, a long CO structure in the southeast part of the CO disk is extended up to $\sim$11.4 kpc from the center of the stellar disk. On the opposite side, however, the extent of the CO disk is about 7.8 kpc.
Two galaxies (NGC~4632 and UGC~08041) in the N4636G also show asymmetric CO distributions. The CO disk of NGC~4632 is more extended toward the northeast side (80$\arcsec$ ($\sim$5.3 kpc) in the northeast versus 61$\arcsec$ ($\sim$4.0 kpc) in the southwest). UGC~08041 has a very extended CO structure in the southern part (53$\arcsec$ ($\sim$3.5 kpc) in the south versus 13$\arcsec$ ($\sim$0.9 kpc) in the north). 

\paragraph{\textbf{$\bullet$ No CO emission in the central region and large offset between the CO peak and the optical center}} As seen in Figure~\ref{fig:fig4}, our ACA CO data do not show any CO emission in the central region of two galaxies (NGC~7421 in the I1459G and NGC~4772 in the N4636G), contrary to our other samples. Instead, NGC~7421 has a relatively strong CO emission in the southwest region of the optical image, with $\sim$50\% of the total CO flux. In the case of NGC~4772, there are two discrete CO regions. In particular, the southeast region has $\sim$65\% of the total CO flux. 
In addition to no CO emission in the central region, these two galaxies show a large offset between the CO peak position and the optical center (see Figure~\ref{fig:app_8} and \ref{fig:app_17}). The offset distances are 4.8 kpc in NGC~7421 and 2.1 kpc (NGC~4772), respectively. Interestingly, the CO peak of NGC~4496A is also found at the northeast edge of the CO disk (see Figure~\ref{fig:app_10}), and the distance between the peak position and the center is $\sim$3 kpc. 

\paragraph{\textbf{$\bullet$ Bar- or ring-like structure and high surface density in the central region}} The CO PVDs of NGC~4527 and NGC~4536 in the N4636G clearly show a steep velocity gradient in the inner region (see Figure~\ref{fig:app_12} and \ref{fig:app_13}). This indicates a presence of a bar- or ring-like structure in the central region \citep[e.g.,][]{alatalo2013}. The rapidly increasing velocity structures of these two galaxies are also seen in the central region of their velocity field maps (Figure~\ref{fig:app_12} and \ref{fig:app_13}). In addition to the presence of distinct CO structure in the central region, these two samples and NGC~4666 show relatively high molecular surface density ($>$100~{\Msun}~pc$^{-2}$), compared to other member galaxies (Figure~\ref{fig:fig4}).

\section{Discussion} \label{sec:dis}
Using the {\HI} and CO imaging data, we discuss how the group environments affect the distributions of cold ISM components of galaxies in Section~\ref{subsec:hico}. In Section~\ref{subsec:scaling}, we present scaling relations of the global properties of our group sample, and compare them with the global properties in the xCOLD GASS sample. Finally, we discuss preprocessing in the group environment in Section~\ref{subsec:pre}. Note that although some of our targets are thought to be affected by various group environmental processes, we briefly discuss external mechanisms for individual group members in this study. Instead, distinguishing between various environmental processes for individual members is studied in more detail in a following work \citep[e.g.,][]{lin2023}.

\subsection{{\HI} and CO Distributions of Group Galaxies} \label{subsec:hico}
\begin{figure*}[!htbp]
\begin{center}
\includegraphics[width=0.77\textwidth]{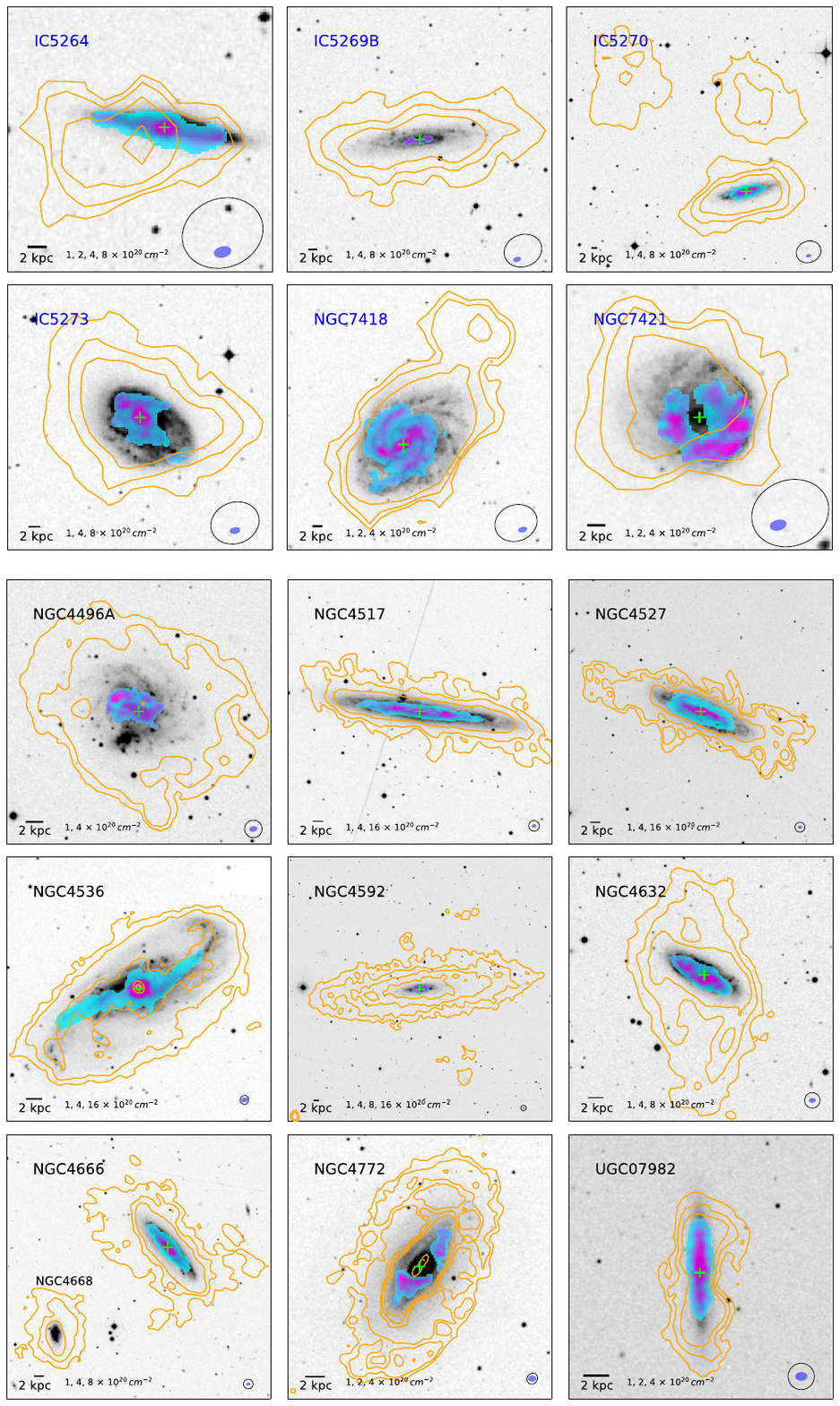}
\caption{{\HI} (contours) and CO distributions (color scale) of 15 group galaxies (the first two rows: 6 galaxies in the I1459G, and the second three rows: 9 galaxies in the N4636G) is overlaid on their optical images (DSS2 blue). In each CO map, while a cyan color indicates a low-intensity value, a magenta color shows a relatively high-intensity value. Contour levels of the {\HI} gas are shown at the bottom of each panel. These {\HI} images are from the ASKAP BETA data \citep{serra2015} for the I1459G, the WALLABY pilot data \citep{koribalski2020} for the N4636G, and the VIVA data \citep{chung2009} for NGC~4536 and NGC~4772 in the N4636G. The green cross indicates the stellar disk center. The bar in the bottom-left corner represents the physical scale of 2~kpc. The synthesized beams of CO (blue ellipse) and {\HI} (black open ellipse) observations are shown at the bottom-right corner. \label{fig:hico}}
\end{center}
\end{figure*}


15 out of 16 group galaxies with CO detection in our ACA observations also have {\HI} imaging data \citep{chung2009,serra2015,koribalski2020}, and Figure~\ref{fig:hico} shows their {\HI} and CO distributions overlaid on their optical images (DSS2 blue).

Based on their {\HI} and CO distributions, we have classified the 15 group galaxies into three categories: (i) peculiar distributions in both {\HI} and CO (IC~5264, IC~5273, NGC~7418, NGC~7421, NGC~4632, NGC~4772), (ii) peculiar distribution in {\HI} (IC~5270, NGC~4666, UGC~07982), (iii) relatively symmetric distributions in both {\HI} and CO (IC~5269B, NGC~4527, NGC~4536, NGC~4592). Interesting features of group members are briefly summarized in Table~\ref{tab:table_new}.

\subsubsection{Peculiar Distributions in Both CO and {\HI}} \label{asymcohi}
\paragraph{\textbf{$\bullet$ Asymmetric structure}}
Four galaxies (IC~5264, IC~5273, NGC~7418, NGC~7421) show asymmetric morphologies in both CO and {\HI}. In particular, their {\HI} distributions are extended toward one side. On the opposite side of the long {\HI} extensions, the {HI} disks are truncated near or within the stellar disk.

For IC~5264, as shown in Figure~\ref{fig:hico}, the {\HI} gas is more extended toward the southeast. On the opposite side, the {\HI} disk is truncated within the stellar disk. This asymmetric {\HI} distribution is analogous to the CO distribution of IC~5264 as described in Section~\ref{subsec:pecco}. Interestingly, a locally strong CO emission is found at the edge of the truncated {\HI} disk. In NGC~7421, the {\HI} disk is pushed away to the northeast. On the opposite side, there is a relatively strong CO emission with the CO peak (Figure~\ref{fig:hico}). In the case of NGC~7418, while a long {\HI} tail is seen in the northwest, the {\HI} disk is compressed in the southeast. In particular, an extended CO emission is found at the site of {\HI} compression in NGC~7418 (Figure~\ref{fig:hico}).

In these three galaxies, {\HI} compression at the truncated side of the {\HI} disk may lead to an increase of {\HI} gas density, which likely triggers an efficient transition from {\HI} gas to H$_{2}$ gas \citep[][]{chung2014,lizee2021}. Consequently, this conversion process is likely to result in the local enhancement of CO (IC~5264 and NGC~7421) and the extended CO structure (NGC~7418). In addition, external perturbations can directly compress CO gas. These phenomena are already known from previous studies of cluster galaxies \citep[][]{chung2014,lee2017,lizee2021}, but similar results are now found for our group galaxies. The morphological correlations and connections between two different ISM components suggest that group environmental processes can significantly affect both diffuse {\HI} gas and dense CO gas. 

IC~5273 shows highly asymmetric {\HI} and CO morphologies, but there seems to be no similarity or no connection between the {\HI} distribution and the CO distribution, in contrast with the above three galaxies (IC~5264, NGC~7418, and NGC~7421). This suggests that this galaxy may be affected by the environment in a different way. In addition to asymmetric distributions of both {\HI} and CO, IC~5273, NGC~7418, NGC~7421 also show clearly a lopsided feature in their optical images, which suggests that these three galaxies are likely to be affected by tidal interactions. 

\paragraph{\textbf{$\bullet$ Ring-like structure}} 
NGC~4632 and NGC~4772 have large {\HI} outer ring structures, as shown in Figure~\ref{fig:hico}. In particular, the position angle (PA) of {\HI} outer ring of NGC~4632 largely deviates from that of the inner {\HI} disk and the stellar disk. The PA of {\HI} outer ring of NGC~4772 is also slightly different from the inner {\HI} disk \citep[][]{chung2009}. The origin of {\HI} ring structures could be accretions, minor mergers, and tidal interactions with other galaxies \citep[e.g., see][and references therein]{buta1996,barnes1999,bettoni2010}. Moreover, both NGC~4632 and NGC~4772 show irregular CO distributions in the inner {\HI} disks, as described in Section~\ref{subsec:pecco}. The external perturbations seem to cause not only large {\HI} outer ring structures but also irregular CO distributions in these two galaxies.

\subsubsection{Peculiar Distribution in {\HI}} \label{hi}
The {\HI} morphologies of three galaxies (IC~5270, NGC~4666, UGC~07982) appear to be asymmetric, but their CO distributions are smooth and undisturbed. Near the north of IC~5270, there are two {\HI} clouds, as seen in Figure~\ref{fig:hico}. Although the {\HI} disk does not look asymmetric, these two {\HI} clouds are possibly stripped from IC~5270 \citep{serra2015}, which is suggestive of evidence for tidal interactions \citep{serra2015}. In NGC~4666, the outer part of {\HI} disk extends farther out toward its neighbor galaxy (NGC~4668), which indicates that the {\HI} morphology is strongly disturbed by the interaction with the neighboring galaxy (Figure~\ref{fig:hico}). The direction of the {\HI} tail of NGC~4668 is also toward NGC~4666. The previous {\HI} observations with the VLA also show clear signs of interaction between NGC~4666 and NGC~4668 (see more details in Figure~5 of \citealt{walter2004}). In UGC~07982, the northern part of the {\HI} disk is truncated within the stellar disk (Figure~\ref{fig:hico}). In addition, the {\HI} gas is slightly extended toward the west side. However, the stellar disk of UGC~07982 looks undisturbed. This suggests that UGC~07982 is likely to be undergoing ram pressure stripping (RPS) in the N4636G. Indeed, this galaxy is identified as the RPS galaxy, based on an analysis of the ram pressure level against the restoring force in the disk for UGC~07982. Further details of the RPS for UGC~07982 and other galaxies in the N4636G are presented in \cite{lin2023}.

\subsubsection{Symmetric/Smooth Distributions in Both {\HI} and CO} \label{smdisk}
Four galaxies (IC~5269B, NGC~4527, NGC~4536, NGC~4592) show relatively symmetric/smooth distributions in both {\HI} and CO, compared to other group samples mentioned above. Although the CO distributions of some galaxies (e.g., IC~5269B and NGC~4592) appear to be somewhat clumpy due to the low S/N of the CO data, the {\HI} disks of four group samples are symmetric and smooth, indicating that there may be no external perturbations.

\subsubsection{Asymmetry Parameter for the CO Intensity Map}
In addition to probing peculiar gas distributions qualitatively, we also calculate the asymmetry parameter ($A_{\rm map}$) using the CO intensity maps to estimate the degree of asymmetry quantitatively. The asymmetry of the CO image is calculated using the following equation \citep{conselice2000, holwerda2011a, giese2016}:

\begin{equation}
A_{\rm map} = \frac{\Sigma_{i ,j}~| I (i, j) - I_{180} (i, j) |}{2~\Sigma_{i ,j}~| I (i, j) |},
\end{equation}

\noindent where $I (i, j)$ is the CO intensity map, and $I_{180} (i, j)$ is the same CO intensity map rotated by 180$^{\degr}$ with respect to the center of the stellar disk. A high value indicates a high asymmetry of CO distribution. The CO asymmetry values for 16 galaxies are summarized in Table~\ref{tab:table_new}. Group members (IC~5264, IC~5273, NGC~7418, NGC~7421, NGC 4632, NGC~4772) showing asymmetric morphologies from both CO and {\HI} images tend to have higher asymmetry values, compared to other members. On average, the asymmetry value for these six galaxies is 0.46. On the other hand, the mean value for samples (NGC~4517, NGC~4527, NGC~4536) with no signs of interactions from both CO/{\HI} morphologies is 0.18, and galaxies (IC~5270, NGC~4666, UGC~07982) showing asymmetric morphology in only {\HI} have the mean value of 0.17. As previous studies with {\HI} images found higher asymmetry values from galaxies in a dense environment \citep{holwerda2011b, reynolds2020}, it is also expected that CO asymmetry values tend to be high in galaxies undergoing environmental processes (e.g., tidal interaction and RPS). Indeed, our results support the notion that high CO asymmetry values can be found in the group environment.

However, some (e.g., IC~5269B, NGC~4592) of our group members also tend to have high values of the CO asymmetry parameter although these galaxies are not likely to be affected by environmental processes, based on their symmetric {\HI} and optical morphologies. Instead, in these galaxies, clumpy CO structures due to low S/N of the CO data could result in high values in the CO asymmetry. 

In this analysis for the CO asymmetry parameter, our results could be biased due to the small sample size (16 group members for the asymmetry analysis). Therefore, more group galaxies are required to obtain a statistically robust result.

\begin{deluxetable*}{lccc}
\tabletypesize{\footnotesize}
\tablecaption{Information for CO and/or {\HI} Distributions \label{tab:table_new}}

\tablehead{
\multicolumn{1}{l}{Name} & \multicolumn{1}{c}{CO Asymmetry Value} & \multicolumn{1}{c}{Peculiar Distribution} & \multicolumn{1}{c}{Notes}
}
\startdata
\multicolumn{4}{c}{IC~1459 Group} \\
\hline
IC 5264 & 0.30 & CO, {\HI} & asymmetric CO and {\HI} distributions \\
IC 5269B & 0.50 & \nodata & low S/N in CO data \\
IC 5270 & 0.10 & {\HI} & two {\HI} clouds near IC~5270 \\
IC 5273 & 0.38 & CO, {\HI} & CO clump (southwest) \\
NGC 7418 & 0.34 & CO, {\HI} & extended CO structure (southeast), {\HI} tail (northwest) \\
NGC 7421 & 0.68 & CO, {\HI} & strong local CO emission (southwest) \\
\hline
\multicolumn{4}{c}{NGC~4636 Group} \\
\hline
NGC 4496A & 0.49 & CO & off-center CO peak (northeast)  \\
NGC 4517 & 0.20 & \nodata & symmetric CO and {\HI} distributions \\
NGC 4527 & 0.17 & \nodata & symmetric CO and {\HI} distributions \\
NGC 4536 & 0.16 & \nodata & symmetric CO and {\HI} distributions \\
NGC 4592 & 0.50 & \nodata & clumpy CO distribution, low S/N in CO data \\
NGC 4632 & 0.51 & CO, {\HI} &  asymmetric CO distribution, {\HI} polar ring structure \\
NGC 4666 & 0.15 & {\HI} & asymmetric {\HI} distribution, close neighbor galaxy \\
NGC 4772 & 0.57 & CO, {\HI} & asymmetric CO distribution, {\HI} outer ring structure \\
UGC 07982 & 0.25 & {\HI} & asymmetric {\HI} distribution \\
UGC 08041 & 0.70 & CO & asymmetric CO distribution, low S/N in CO data, no {\HI} image \\
\enddata
\tablecomments{
(1) Galaxy name; (2) CO asymmetry value; (3) peculiar distributions of CO and/or {\HI}; (4) notes for CO and/or {\HI} distributions of group members.
}
\end{deluxetable*}

\subsection{Comparisons of Global Properties between Group Galaxies and xCOLD GASS Galaxies} \label{subsec:scaling}

\begin{figure*}[!htbp]
\begin{center}
\includegraphics[width=1.00\textwidth]{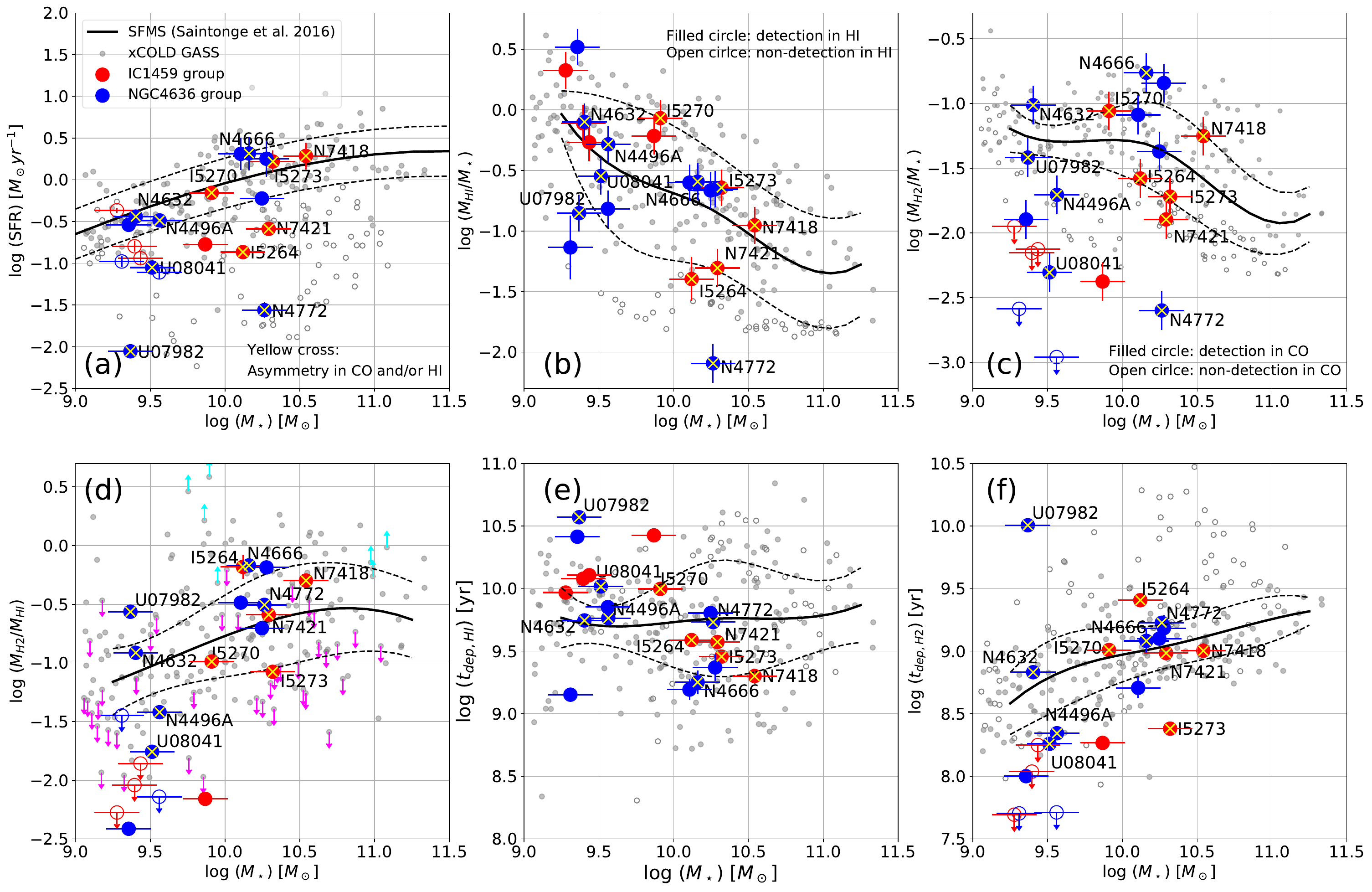}
\caption{Comparisons of global properties between our group galaxies and the xCOLD GASS galaxies. (a) SFR, (b) {\HI} gas fraction ($M_{\rm\tiny {\HI}}/M_\star$), (c) H$_{2}$ gas fraction ($M_{\rm H2}/M_\star$), (d) the ratio of molecular to atomic gas ($M_{\rm H2}/M_{\rm\tiny {\HI}}$), (e) {\HI} gas depletion time ($\tau_{\rm dep, \tiny {HI}}$), (f) H$_{2}$ gas depletion time ($\tau_{\rm dep, H2}$) as a function of the stellar mass ($M_\star$). The red and blue filled circles indicate our sample galaxies with CO detection in the I1495G and the N4636G, respectively. The upper limit for nondetection of CO is shown as the open circles. Group samples with their names, which show asymmetric {\HI} and/or CO distributions, are marked by yellow crosses. The xCOLD GASS sample are shown as gray filled circles ({\HI} or CO detections) and open gray circles (non-detection in {\HI} or nondetection in CO, respectively). In panel (a), the star-forming main sequence, which is determined by \cite{saintonge2016}, is indicated by a solid line, and the 0.3 dex scatter is shown in dashed lines. Filled circles indicate CO detections while nondetections in CO are shown in open circles. In panels (b)$-$(f), the median trends (solid lines) and the 1$\sigma$ standard deviations (dashed lines) are derived from the xCOLD GASS sample. In panel (d), the upper limit (i.e., nondetection in CO) and the lower limit (i.e., nondetection in {\HI}) are represented by small magenta and cyan arrows, respectively. We limited our group sample to relatively massive galaxies (log~($M_\star$/{\Msun}) $>$ 9) because all group samples with CO detection have their stellar mass of log~($M_\star$/{\Msun}) $>$ 9. \label{fig:fig6}}
\end{center}
\end{figure*}

In this section, we present scaling relations of global measurements (e.g., SFR, gas fraction, gas depletion time) in our group galaxies, and compare global properties of group galaxies with those of galaxies in a low-density environment. From the comparison, we can investigate how the group environment affects global properties of group galaxies. For this comparison, we use 240 isolated galaxies, which are selected based on the Sloan Digital Sky Survey (SDSS) DR7 group catalog of \cite{yang2007}, from the extended CO Legacy Database for GASS (xCOLD GASS; \citealt{saintonge2017}) with {\HI} information from the extended GALEX Arecibo SDSS Survey (xGASS; \citealt{catinella2018}). The sample of 240 local isolated galaxies (0.01 $< z <$ 0.05 ) uniformly covering the stellar mass range (9 $<$ log~($M_\star$/{\Msun}) $<$ 11.5) were observed deeply in both {\HI} and CO, which can provide a reference for various scaling relations in local isolated galaxies. Hereafter, we refer to the sample of 240 galaxies as the xCOLD GASS sample. Note that we rescaled the measurements of stellar masses and SFRs of our group sample by dividing by 1.06 \citep[e.g.,][]{salim2007,zahid2012}, to compare our group sample with the xCOLD GASS sample under the same Chabrier IMF condition \citep{chabrier2003}.

Figure~\ref{fig:fig6} shows (a) SFR, (b) {\HI} gas fraction ($M_{\rm\tiny {\HI}}/M_\star$), (c) H$_{2}$ gas fraction ($M_{\rm H2}/M_\star$), (d) the ratio of H$_{2}$ gas to {\HI} gas ($M_{\rm H2}/M_{\rm\tiny {\HI}}$), (e) {\HI} gas depletion time ($\tau_{\rm dep, \tiny {HI}}$), (f) H$_{2}$ gas depletion time ($\tau_{\rm dep, H2}$) as a function of the stellar mass ($M_\star$) of galaxies. In Figure~\ref{fig:fig6}, we use the metallicity-dependent CO conversion factor to calculate the H$_{2}$ gas masses of the group sample and the xCOLD GASS sample. The red and blue filled circles indicate our group galaxies that have CO detection, in the I1495G and the N4636G, respectively. The group samples with non-detection in CO are shown as open circles. Note that in the following analysis of scaling relations, we limited our group sample objects with a stellar mass of log~($M_\star$/{\Msun}) $>$ 9 because all samples with CO detections have stellar masses of log~($M_\star$/{\Msun}) $>$ 9.  The xCOLD GASS galaxies are shown in the gray filled circles ({\HI} or CO detections) and gray open circles (nondetections in CO (Figure~\ref{fig:fig6} (a), (c), (f)) or nondetection in {\HI} (Figure~\ref{fig:fig6} (b), (e))), respectively. Note that fractions that we calculated in the following scaling relations also include nondetection cases.

\paragraph{\textbf{$\bullet$ Star formation activity}}
First, Figure~\ref{fig:fig6} (a) shows the distribution of SFRs of the group sample and the xCOLD GASS sample as a function of the stellar mass of the galaxies. A solid line indicates the star-forming main sequence (SFMS) determined by \cite{saintonge2016}, and dashed lines are the $\pm$0.3 dex scatter \citep[][]{speagle2014}. Galaxies lying below the solid line are less star-forming (i.e., relatively suppressed in star formation) than the galaxies lying above and on the line. Galaxies lying below the lower dashed line tend to leave the star-forming main sequence and get quenched in star formation. In Figure~\ref{fig:fig6} (a), we can see that our group samples cover both star-forming galaxies (48\%) and quenched galaxies (48\%). The ratio of star-forming galaxies in our group sample is consistent with that (50\%) in the xCOLD GASS sample. On the other hand, the fraction of quenched galaxies in our group sample is slightly higher than that (39\%) in the xCOLD GASS sample. Interestingly, eight out of 12 low-mass (log~($M_\star$/{\Msun}) $<$ 10) group members have suppressed SFR with respect to the main sequence. 

In the following, using scaling relations (median values (solid lines) and standard deviations (dashed lines), see Figure~\ref{fig:fig6} (b)$-$(f)), we compare global physical properties (e.g., gas fraction, gas depletion time) of our group galaxies to the isolated galaxies from the xCOLD GASS sample, in order to obtain clues why and how the galaxies get affected in the group environment. The different behavior of the low-mass galaxies also motivates us to investigate the low- and high-mass galaxies separately.

\paragraph{\textbf{$\bullet$ {\hbox{{H}\kern 0.2em{\sc i}}} gas and H$_{2}$ gas fractions}}
Figure~\ref{fig:fig6} (b) and (c) show the {\HI} gas fraction and the H$_{2}$ gas fraction, respectively. By using the median trends (solid lines) of the {\HI} gas and the H$_{2}$ gas fractions for isolated galaxies, with their 1$\sigma$ standard deviations (dashed lines), this figure shows that {\HI} (H$_{2}$)-deficient galaxies are below the lower dashed line, while normal galaxies are above that line. {\HI} (H$_{2}$)-rich galaxies are above the upper dashed line. We also compare the distribution of {\HI} and H$_{2}$ fractions of group samples with respect to the median trend as a function of stellar mass to identify potential systematic shifts.

In Figure~\ref{fig:fig6} (b), the group galaxies follow a similar distribution as the whole xCOLD GASS sample above and below the median relation of {\HI} mass fraction as a function of stellar mass. There are 17 group galaxies with normal or high {\HI} gas fraction (81\%), while 4 group galaxies (19\%) are below the -1$\sigma$ scatter, indicating {\HI}-deficient galaxies. In the xCOLD GASS sample (9.25 $<$ log~($M_\star$/{\Msun}) $<$ 11.25), the fractions of {\HI}-rich or normal, and {\HI}-deficient galaxies are 77\% and 23\%, respectively. We do not find that overall, the galaxies in our group sample are deficient in {\HI} gas, compared to the xCOLD GASS sample, even though previous {\HI} studies for group galaxies \citep[e.g.,][]{hess2013,brown2017} found evidence that group galaxies tend to be deficient in the {\HI} gas. This result is likely at least partly due to our sample selection of GEMS-{\HI} detected galaxies from the very beginning. Thus our results may be slightly biased toward the relatively {\HI}-rich systems. Nevertheless, {\HI}-deficient group members (e.g., IC~5264, NGC~4772, UGC~07982) tend to show highly asymmetric {\HI} morphology, implying that the {\HI} gas removal occurs due to the group environmental processes.

In Figure~\ref{fig:fig6} (c), we find a high fraction (57\%) of H$_{2}$-deficient galaxies in our group sample in comparison with the fraction (26\%) of H$_{2}$-deficient galaxies in the xCOLD GASS sample (9.25 $<$ log~($M_\star$/{\Msun}) $<$ 11.25). This result suggests that group members can be deficient in the H$_{2}$ gas content. On the other hand, \cite{martinez-badenes2012} showed that there is an excess of the H$_{2}$ gas content in compact group galaxies, compared to isolated galaxies. Interestingly, most of the low-mass members (log~($M_\star$/{\Msun}) $<$ 10), including the nondetections) in our group sample show a significant deficiency in H$_{2}$ gas (below the dashed line). Given that our sample is already biased toward the {\HI}-rich galaxies, such a deficiency in H$_{2}$ gas is striking.

\paragraph{\textbf{$\bullet$ The ratio of H$_{2}$ gas to {\hbox{{H}\kern 0.2em{\sc i}}} gas and {\hbox{{H}\kern 0.2em{\sc i}}}/H$_{2}$ depletion times}}
For the ratio of H$_{2}$ gas to {\HI} gas ($M_{\rm H2}/M_{\rm\tiny {\HI}}$), as shown in Figure~\ref{fig:fig6} (d), the high-mass galaxies (log~($M_\star$/{\Msun}) $>$ 10) of our group sample are in line with the trend of the xCOLD GASS sample. However, the ratios of H$_{2}$ gas to {\HI} gas in the low-mass galaxies (log~($M_\star$/{\Msun}) $<$ 10), including the nondetections of our group sample tend to be lower than the median trend of star-forming galaxies. A similar trend is found in the H$_{2}$ gas depletion time that shows, on average, a shorter depletion time of H$_{2}$ gas for the low-mass galaxies when compared to the star-forming galaxies of the same stellar mass (Figure~\ref{fig:fig6} (f)). As a result, this subset of the low-mass galaxies shows systematically longer (except for one outlier) depletion time for the {\HI} gas than the median value of the same stellar mass (Figure~\ref{fig:fig6} (e)).

\begin{figure*}[!htbp]
\begin{center}
\includegraphics[width=1.00\textwidth]{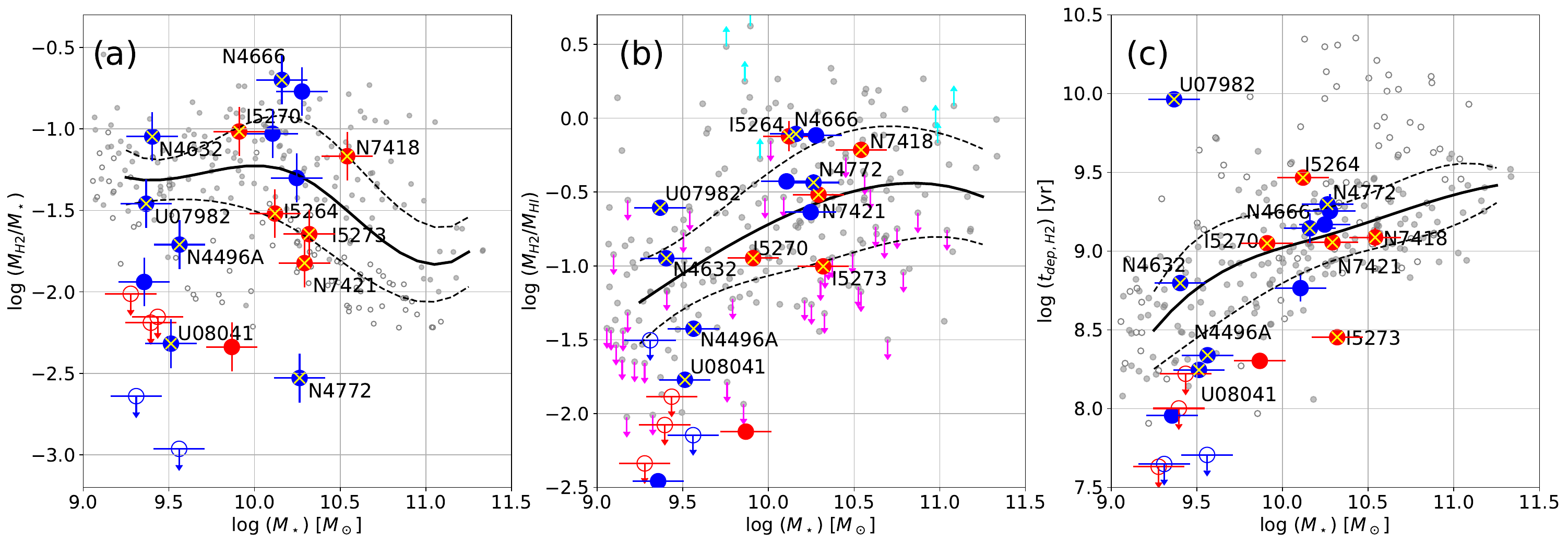}
\caption{Comparisons of global properties between our group galaxies and the xCOLD GASS galaxies. The H$_{2}$ gas mass is calculated with the constant CO-to-H$_{2}$ conversion factor ($\alpha_{\rm CO}$~=~4.35 {\Msun}~pc$^{-2}$~(K~km~s$^{-1}$)$^{-1}$ \citep{strong1996,bolatto2013}). (a) H$_{2}$ gas fraction ($M_{\rm H2}/M_\star$), (b) the ratio of molecular to atomic gas ($M_{\rm H2}/M_{\rm\tiny {\HI}}$), (c) H$_{2}$ gas depletion time ($\tau_{\rm dep, H2}$) as a function of the stellar mass ($M_\star$). All symbols are the same as the symbols in Figure~\ref{fig:fig6}.  \label{fig:fig7}}
\end{center}
\end{figure*}

\paragraph{\textbf{$\bullet$ A constant CO-to-H$_{2}$ conversion factor}}
We also calculate the H$_{2}$ gas fraction, the ratio of H$_{2}$ to {\HI}, and H$_{2}$ gas depletion time, using the constant CO-to-H$_{2}$ conversion factor ($\alpha_{\rm CO}$~=~4.35 {\Msun}~pc$^{-2}$~(K~km~s$^{-1}$)$^{-1}$), to investigate whether these quantities significantly changes with the constant CO-to-H$_{2}$ conversion factor, compared to the results with the metallicity-dependent CO-to-H$_{2}$ conversion factor. As seen in Figure~\ref{fig:fig7}, there are no significant differences in the scaling relations between these two different conversion factors. This indicates our conclusions are insensitive to the CO-to-H$_{2}$ conversion factor in a given stellar mass range. 

\paragraph{\textbf{$\bullet$ The missing flux problem}} 
For two galaxies (NGC~4517 and NGC~4527) whose total fluxes are significantly missed in our ACA observations (the flux ratio between our ACA data and the single-dish data: 0.5 (NGC~4517) and 0.7 (NGC~4527), see also Section~\ref{subsec:coprop}), we examine again the scaling relations using their total fluxes from the single-dish data, in order to test whether the missing flux issue in our ACA observations severely affects our results. Based on the results with the single-dish data, the missing flux issue of these two members, which belong to the high-mass subsample, does not change our conclusion in the analysis of scaling relations. 

Considering (1) the most typical extent of the CO disk (i.e., 50\% $-$ 70\% of the optical disk) in late-type galaxies \citep{schruba2011} and (2) the fact that the low-mass galaxies tend to have a smaller size of the stellar disk \citep{lorenzo2013}, the low-mass group members are expected to have smaller CO disks than the high-mass group galaxies. Consequently, we may miss relatively small portions of total CO fluxes of the low-mass group sample in our ACA observations (see also Appendix~D of \cite{leroy2021} for a detailed description of the missing flux as a function of CO disk size), compared to the high-mass group sample. Therefore, the deficiency of H$_{2}$ gas in the low-mass group members is unlikely due to the missing flux problem. As a result, although we may miss some CO fluxes from our group targets, we expect that this missing flux problem is not likely to change our conclusions. 

However, we cannot completely rule out the possibility of the missing flux problem can affecting our conclusions because we do not have accurate measurements of total CO fluxes in most group samples.

\subsubsection{Group Environmental Effects on Low- and High-mass Group Members}
For the high-mass galaxies, their distribution of all the quantities discussed above (if ignoring the marginally lower H$_{2}$ mass fraction) tend to be close to that of the xCOLD GASS sample. For the low-mass galaxies the majority of which have a suppressed SFR, a short H$_{2}$ depletion time, and a long {\HI} gas depletion time, the bottleneck in the baryonic flow from the {\HI} gas to the stars seems to be mostly at the conversion from {\HI} to H$_{2}$. We note that the missing flux problem from the observations with the interferometry is unlikely to explain the systematic differences that we observe here, as the average fraction of the missing flux is only 10\% and the problem should not be so serious for these small low-mass galaxies. It is known that the transition of {\HI} to H$_{2}$ is not efficient in low-mass galaxies as the pressure and metallicities are both lower \citep{leroy2005,krumholz2009,bolatto2011}. However, this effect does not explain the systematic difference that we observe either, as we are comparing with the median behavior of galaxies with the same stellar mass. 

One of the possible scenarios is that for these low-mass galaxies, which are more sensitive to the environmental effects due to their relatively shallow gravitational potential, the {\HI} gas is highly perturbed and becomes stable against gravitational collapse (i.e., a high $\sigma_{v}$ value in the Toomre Q parameter). The systematically low ratio of H$_{2}$ to {\HI} in these low-mass galaxies was not fully expected, as environmental effects could have enhanced such a conversion, when the ram pressure assists the gas to reach high densities \citep[e.g.,][]{mok2017}, or when the tidal effects induce shocks and condense the cold gas \citep[e.g.,][]{lizee2021}. It is therefore meaningful to find that in these low-mass galaxies, the actual situation is that the conversion of {\HI} to H$_{2}$ seems to be suppressed. It is also interesting to point out the tentatively shortened H$_{2}$ depletion time in the low-mass galaxies. It is possible that the molecular clouds once formed can be more massive and denser as a result of the higher Jeans mass in the perturbed interstellar medium \citep[][]{bournaud2010}, leading to more vigorous star-formation than in an unperturbed circumstance. This scenario is supported by studies based on observations of the dense molecular gas in merging systems \citep[e.g.,][]{juneau2009}. As a result, we observe a systematically shorter depletion time of the molecular gas for these low-mass group galaxies than for field galaxies. Although the suppressed {\HI}-to-molecular conversion and the shortened molecular depletion are opposite effects in the flow of baryons from the {\HI} to the stars, the former process seems to dominate over the latter, resulting in a systematically elongated {\HI} depletion time, and lower SFR than the SFMS for these low-mass galaxies, as found in this work.

If we look closely at the three massive, quenched galaxies (NGC~7421, IC~5264, and NGC~4772), they have relatively low {\HI} gas and H$_{2}$ gas fractions, when compared to the median behavior of the xCOLD GASS sample. In addition, their SFRs are lower than the SFMS galaxies. Thus these three galaxies are mostly quenched because their neutral gas reservoir significantly shrinks, possibly due to environmental gas stripping, or starvation accelerated by stripping. The necessity of losing gas reservoirs before the quenching of these high-mass galaxies is consistent with recent findings for normal field galaxies \citep[e.g.,][]{cortese2020,wang2020,guo2021}.

Overall, environmental effects seem to work on the low- and high-mass galaxies in different ways. The consequence is that the low-mass galaxies systematically and significantly drop in SFR because the conversion of {\HI} to H$_{2}$ is severely suppressed, while the high-mass galaxies tend to remain on the SFMS before the {\HI} reservoir significantly drops. However, our results in the analysis of the scaling relations suffer from a small sample size. Therefore, in order to verify this result and have a more robust conclusion, more CO observations for group galaxies are required.

In fact, individual galaxies in the two groups seem to be subject to different environmental effects. The locations of individual galaxies in each group are different. While galaxies in the I1459G are located in various regions from the group center to the outskirt of the group, many of the sample galaxies in the N4636G are located near the outskirts ($\sim$$R_{200}$) of the N4636G. In addition, the properties (e.g., group mass) of these two groups are different although both groups are loose galaxy groups. These different environmental effects make individual group galaxies have different global properties. We describe individual galaxies in each group in more detail in Appendix~\ref{app:comments}.

\subsection{Group Preprocessing} \label{subsec:pre}
We find that some galaxies in both the I1459G and the N4636G present highly asymmetric distributions in {\HI} and/or CO data. (e.g., IC~5264, IC~5273, NGC~7421, NGC~4632, NGC~4772, UGC~07982). Furthermore, some of them have low SFRs and low cold gas fractions. Our results suggest that group galaxies can be preprocessed by external mechanisms (tidal interactions and/or ram pressure stripping). In particular, strong local peaks in the CO disk and one-sided extended CO structures due to the external perturbations are the interesting findings in our study, which are supportive evidence for which the molecular gas can be also affected by the group environment.

 The N4636G seems to be falling into the Virgo cluster. Together with the asymmetric {\HI} and/or CO distributions and changes of global properties of galaxies in the N4636G, the fact that the N4636G has many early-type galaxies around the group center (see the right panel of Figure~\ref{fig:fig1}) and the higher {\HI} deficiency \citep{kilborn2009} indicates that galaxies can have been already processed by the group environment before they enter the cluster. Recent studies have also found that galaxies in the substructures (e.g., W cloud) and the filaments near the Virgo cluster tend to have a decrease of SFR and low gas fraction \citep{yoon2017,morokuma-matsui2021,mun2021,castignani2021}. As in other previous works, our findings support the group preprocessing scenario, which could be one of the important mechanisms in galaxy evolution with the environment. However, our study is limited to only two groups. Therefore, to have a robust conclusion about the group preprocessing, follow-up studies with more groups associated with the cluster are required.



As already mentioned in previous sections, our sample is already biased toward the {\HI}-rich systems in these two groups. This results in missing severely {\HI}-deficient members, especially in the N4636G. These {\HI}-deficient galaxies may still have the molecular gas within their stellar disks. To have a complete picture of the environmental effects on the group galaxies, these {\HI}-deficient members should also be studied together in future work.

\section{Summary and conclusions} \label{sec:sum}
We have presented the results of the CO imaging survey for 31 galaxies in the I1459G and the N4636G, using the ALMA/ACA. This is the first CO imaging survey for loose galaxy groups. The main scientific goal of this CO survey is to obtain an understanding of the group's environmental effects on molecular gas, star formation, and galaxy evolution. 

We obtained well-resolved CO imaging data ($\sim$0.7 $-$ 1.5 kpc; Figure~\ref{fig:fig4} and Appendix~\ref{app:codata}) for 16 out of 31 galaxies in both I1459G and N4636G from our ACA observations in ALMA Cycle 7. In the I1459G, six out of 11 galaxies have CO detection, and in the N4636G, 10 out of 20 galaxies have CO detection. Their stellar masses range from 10$^{9}$\Msun to 10$^{10}$\Msun. 

We find that {\HI} and/or CO distributions (Figure~\ref{fig:hico}) are asymmetric in our group galaxies. In particular, IC~5264, NGC~7421, and NGC~7418 have {\HI} tails and {\HI} compression. Their CO distributions are also highly asymmetric, which show extended CO structure and a local peak of CO emission. Our CO imaging data reveal the peculiar CO distributions of group galaxies, which are motivations for further study on the molecular gas of group galaxies with CO imaging observations.

In the comparison of scaling relations of global properties (e.g., SFR and gas fraction) between our group sample and the xCOLD GASS sample, overall, environmental effects seem to work on the low- and high-mass galaxies in different ways. The consequence is that the low-mass galaxies systematically and significantly drop in SFR because the conversion of {\HI} to H$_{2}$ seems to be severely suppressed, while the high-mass galaxies tend to remain along the SFMS before the {\HI} reservoir significantly drops (Figure~\ref{fig:fig6}). For some interesting group members (e.g., IC~5264, NGC~7421, NGC~4772, and UGC~07982) showing highly asymmetric morphologies in {\HI} and/or CO images, a significant decrease of SFRs and gas fractions is found. These results indicate that environmental processes (e.g., tidal interactions and ram pressure stripping) in a group can change the distributions of both molecular gas and {\HI} gas. This likely results in changes of global properties of group galaxies, such as a decrease of SFR, {\HI} and H$_{2}$ gas fractions. 

Our results suggest that group galaxies can be significantly processed by the group environment. In particular, the results of the N4636G provide supporting evidence for which the group preprocessing, one of the important mechanisms for galaxy evolution, can occur before groups enter a cluster. 

However, our conclusions are solely based on a small sample. To have a robust result of the group environmental effects on physical properties, especially the molecular gas, of galaxies, more CO observations for group members are required for future studies. 

In the near future, we plan to make the products (e.g., cubes, moments maps, PVDs) from our ACA CO observations of group galaxies accessible online.

\acknowledgments
We thank the anonymous referee for helpful suggestions and comments. BL acknowledges support from the National Science Foundation of China (12073002, 11721303, 11991052) and the National Key R\&D Program of China (2016YFA0400702). BL is supported by the Boya Fellowship at Peking University. BL gratefully thanks Hyein Yoon for useful discussions. This work was partly supported by the Korea Astronomy and Space Science Institute grant funded by the Korea government(MSIT) (Project No. 2022-1-840-05). Support for this work was also provided by the National Research Foundation of Korea to the grant No. 2018R1D1A1B07048314.

Y.K. was supported by the National Research Foundation of Korea (NRF) grant funded by the Korean government (MSIT) (No. 2021R1C1C2091550), and acknowledges the support from China Postdoc Science General (2020M670022) and Special (2020T130018) Grants funded by the China Postdoctoral Science Foundation.

Parts of this research were supported by the Australian Research Council Centre of Excellence for All Sky Astrophysics in 3 Dimensions (ASTRO 3D), through project number CE170100013.

LC is the recipient of an Australian Research Council Future Fellowship (FT180100066) funded by the Australian Government.

This project has received funding from the European Research Council (ERC) under the European Union’s Horizon 2020 research and innovation programme (grant agreement no. 679627; project name FORNAX).

KS acknowledges support from the Natural Sciences and Engineering Research Council of Canada (NSERC).

AB acknowledges support from the Centre National d'Etudes Spatiales (CNES), France.

LVM acknowledges financial support from the State Agency for Research of the Spanish Ministry of Science, Innovation and Universities through the ``Center of Excellence Severo Ochoa'' awarded to the Instituto de Astrofisica de Andalucia (SEV-2017-0709), from grant RTI2018-096228-B-C31 (Ministry of Science, Innovation and Universities/State Agency for Research/European Regional Development Funds, European Union), and grant IAA4SKA (Ref. P18-RT-3082) from the Consejeria de Transformacion Economica, Industria, Conocimiento y Universidades de la Junta de Andalucia and the European Regional Development Fund from the European Union.

FB acknowledges funding from the European Research Council (ERC) under the European Union’s Horizon 2020 research and innovation programme (grant agreement No.726384/Empire).

This paper makes use of the following ALMA data: ADS/JAO.ALMA\#2019.1.01804.S. ALMA is a partnership of ESO (representing its member states), NSF (USA) and NINS (Japan), together with NRC (Canada), MOST and ASIAA (Taiwan), and KASI (Republic of Korea), in cooperation with the Republic of Chile. The Joint ALMA Observatory is operated by ESO, AUI/NRAO and NAOJ. The National Radio Astronomy Observatory is a facility of the National Science Foundation operated under cooperative agreement by Associated Universities, Inc.

The Australian SKA Pathfinder is part of the Australia Telescope National Facility (https://ror.org/05qajvd42) which is managed by CSIRO. Operation of ASKAP is funded by the Australian Government with support from the National Collaborative Research Infrastructure Strategy. ASKAP uses the resources of the Pawsey Supercomputing Centre. Establishment of ASKAP, the Murchison Radio-astronomy Observatory and the Pawsey Supercomputing Centre are initiatives of the Australian Government, with support from the Government of Western Australia and the Science and Industry Endowment Fund. We acknowledge the Wajarri Yamatji people as the traditional owners of the Observatory site.

We acknowledge the usage of the HyperLeda database (http://leda.univ-lyon1.fr).

%

\vspace{5mm}
\facilities{ALMA/ACA}


\software{
Astropy \citep{astropy:2013, astropy:2018},  
CASA \citep{mcmullin2007}, 
Matplotlib \citep{Hunter:2007}, 
NumPy \citep{harris2020array}, 
SciPy \citep{2020SciPy-NMeth},
SoFiA \citep{serra2015a,westmeier2021}
}




\appendix
\section{Locations of CO-detected group members on the projected sky}
\begin{figure*}
\begin{center}
\includegraphics[width=1.00\textwidth]{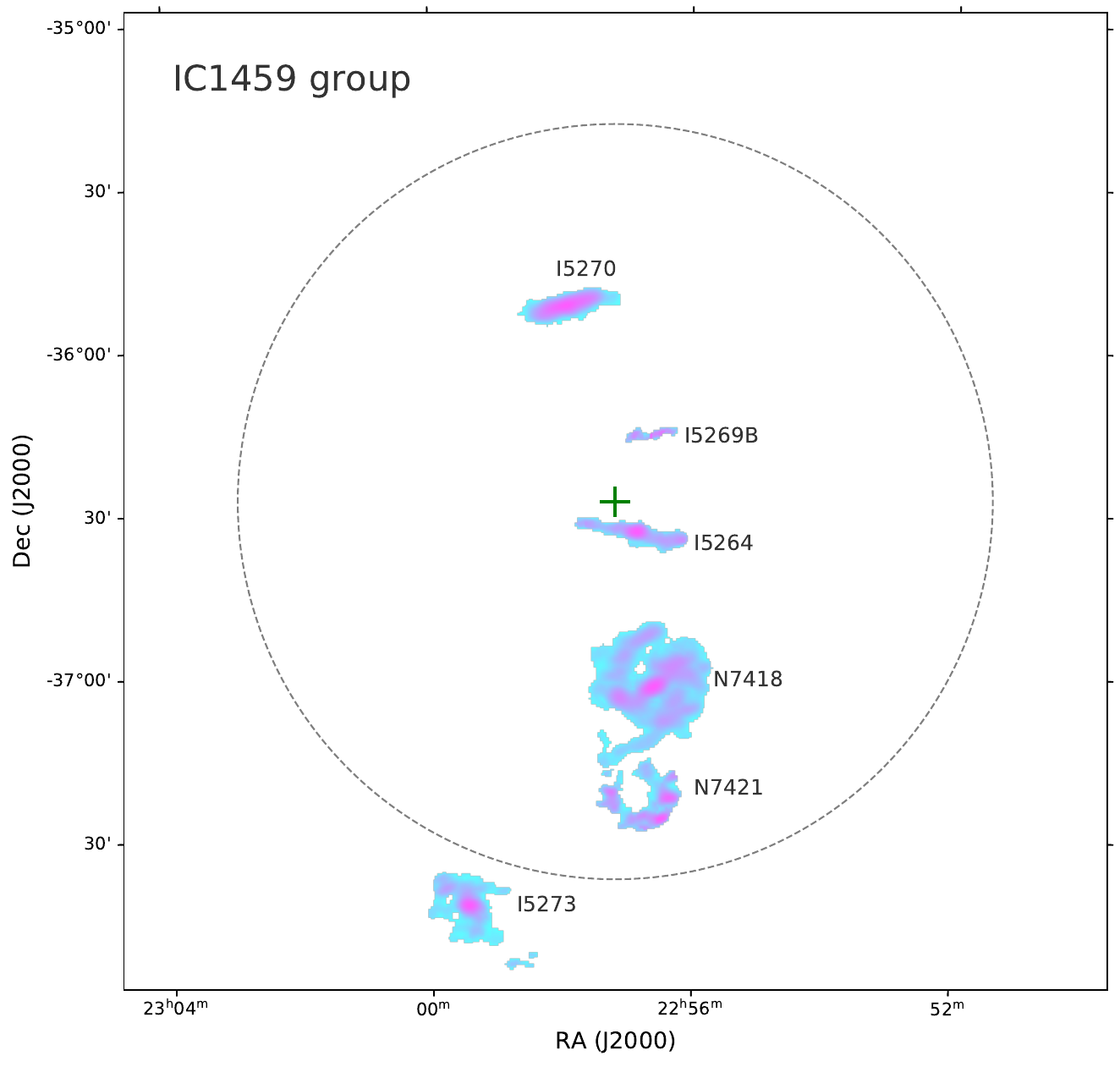}
\caption{CO-detected members in the I1459G}
\label{fig:app_i1459}
\end{center}
\end{figure*}

\begin{figure*}
\begin{center}
\includegraphics[width=1.00\textwidth]{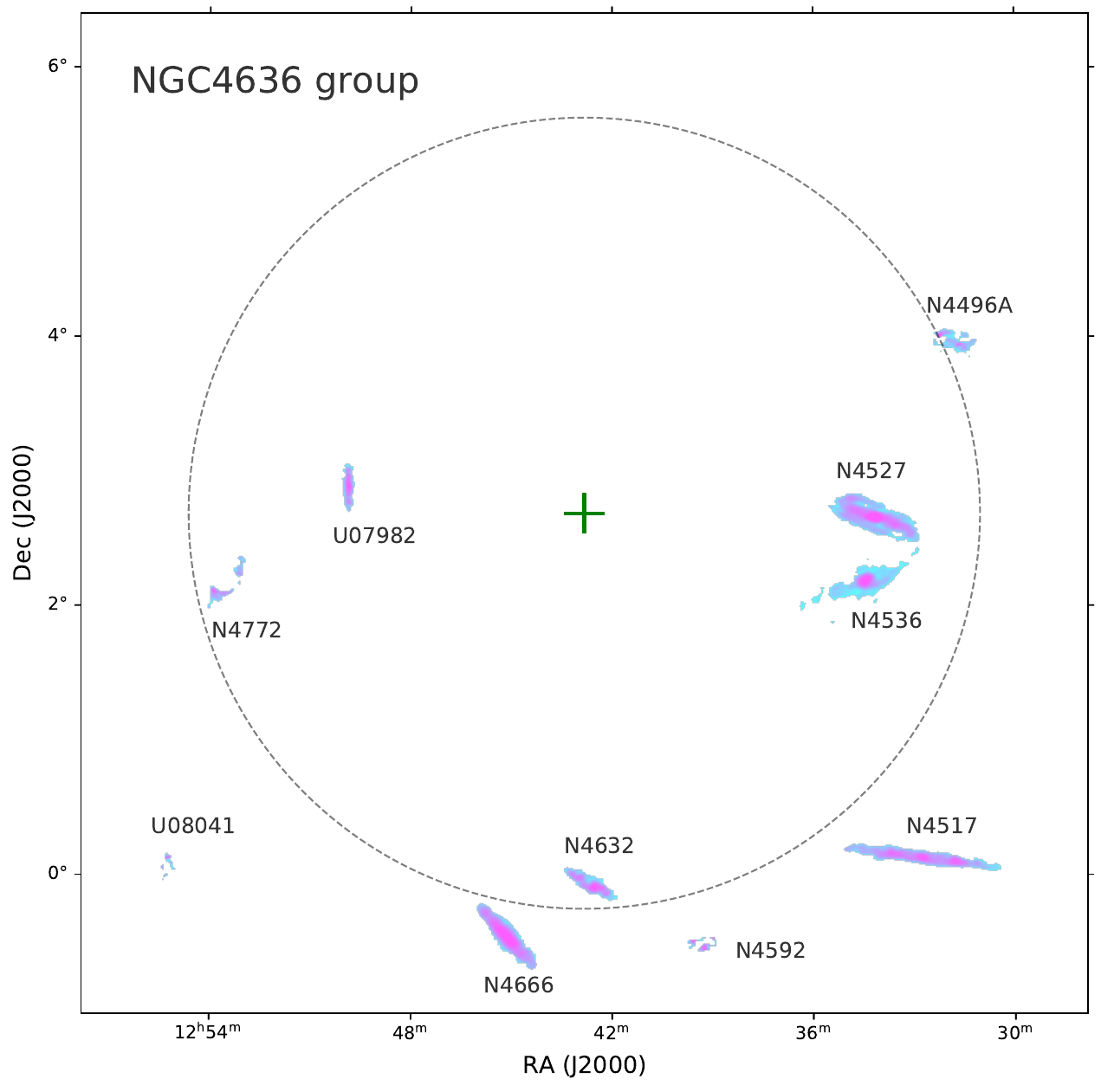}
\caption{CO-detected members in the N4636G}
\label{fig:app_n4636}
\end{center}
\end{figure*}

Figures~\ref{fig:app_i1459} and \ref{fig:app_n4636} show locations of CO-detected group galaxies on the projected sky plane and their CO distributions (color scale). The CO intensity maps are enlarged by a factor of 12. A dashed circle indicates $R_{200}$. A green cross shows location of the BGG. While a cyan color indicates a low-intensity value, a magenta color shows a relatively high-intensity value. 

\section{CO atlas of galaxies in the IC~1459 group and the NGC~4636 group} \label{app:codata}
We present the ACA CO data for each galaxy of the group sample (Figures~10$-$25). Each figure shows the CO intensity map, velocity field map, CO PVD, CO line profile together with the optical image. The details of each panel in the figure are described as follows. The large dashed circle(s) are the primary beam(s) of the ACA observations. Top left panel: the CO intensity map (orange contours) is overlaid on the DSS2 blue (gray scale). Top middle panel: the CO intensity map in color scale. The contour levels indicating the molecular gas surface density are shown at the top of each panel. The surface density of the molecular gas is calculated by adopting $\alpha_{\rm CO}$~=~4.35 {\Msun}~pc$^{-2}$~(K~km~s$^{-1}$)$^{-1}$ \citep{strong1996,bolatto2013}. In this calculation, we did not apply an inclination correction. Top right panel: CO velocity field map (color scale) with contours (20 km~s$^{-1}$ interval). Bottom middle panel: CO PVD. The PVD is obtained by cutting through the major axis of the galaxy with 16$\arcsec$ width in the CO data cube. Bottom right panel: CO line profile (blue color). For NGC~4527, NGC~4536, and NGC~4666, their CN line profiles (red color) are also overlaid on their CO line profiles (blue color). The CN profiles are magnified by a factor of 40. The cross indicates the stellar disk center. The bar at the bottom-left corner represents the physical scale of 2 kpc. The synthesized beam (blue ellipse) of CO data is shown at the bottom-right corner.

\begin{figure*}
\begin{center}
\includegraphics[width=0.85\textwidth]{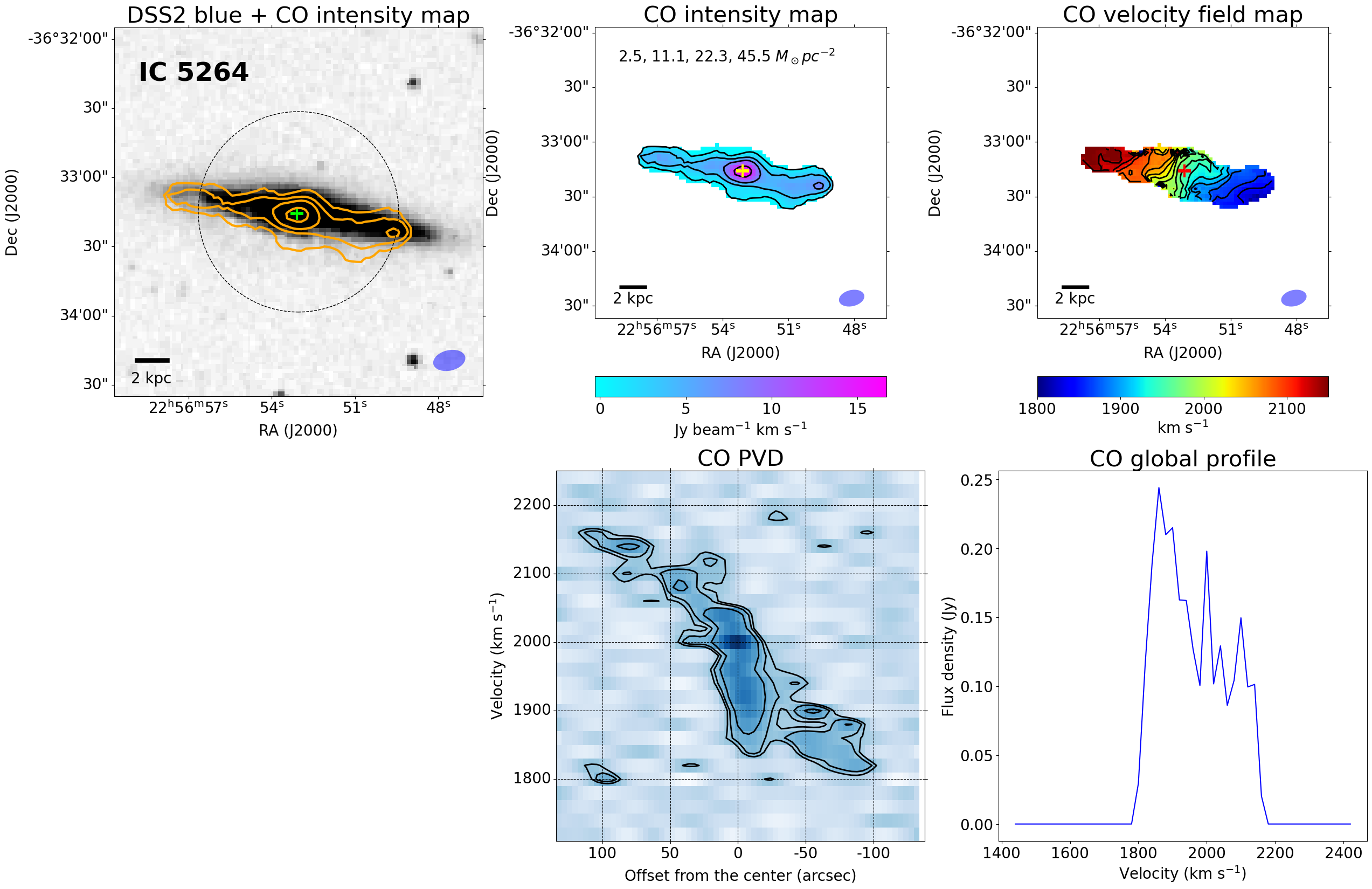}
\caption{IC~5264}
\label{fig:app_1}
\end{center}
\end{figure*}

\begin{figure*}
\begin{center}
\includegraphics[width=0.85\textwidth]{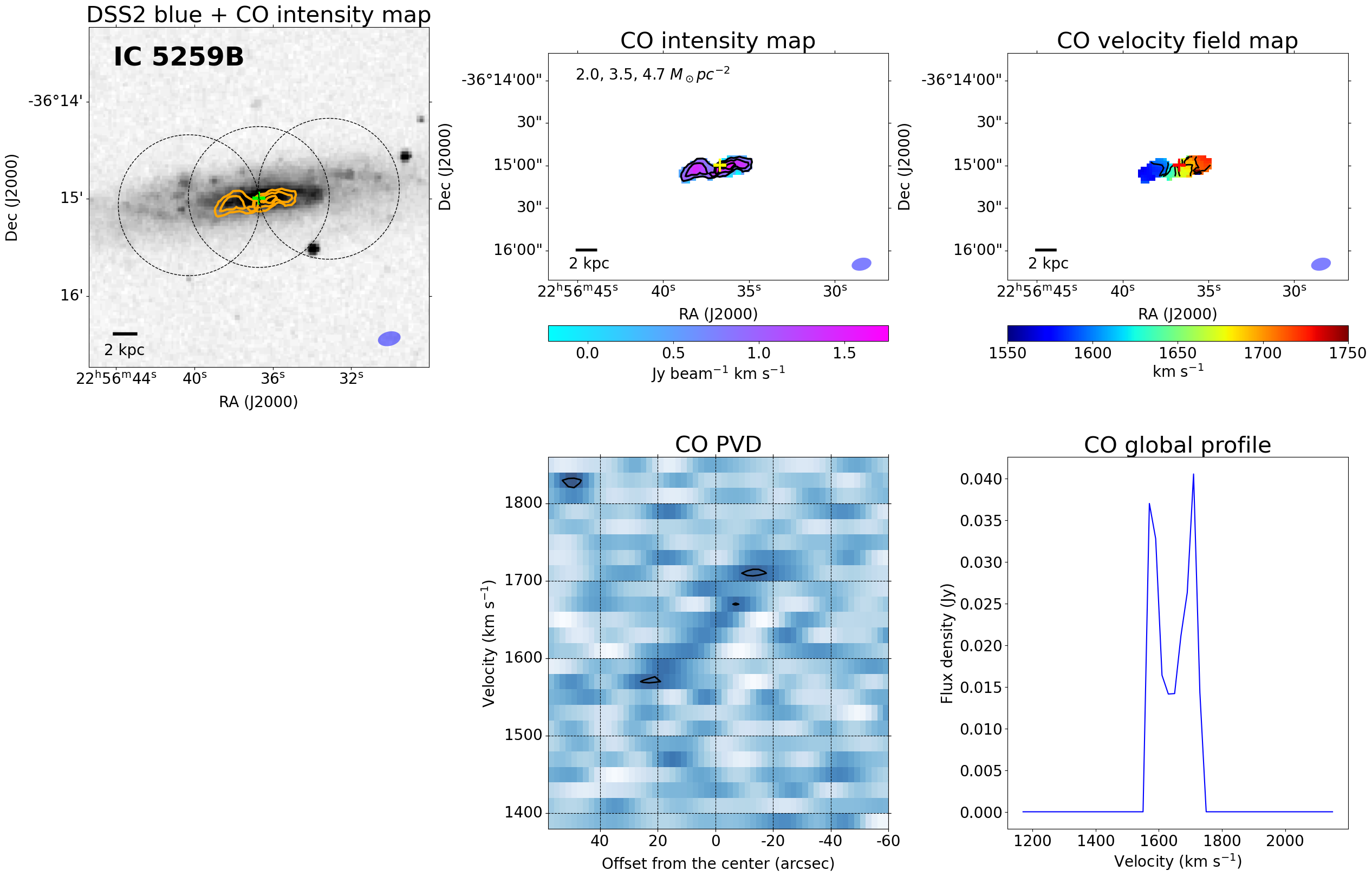}
\caption{IC~5269B}
\label{fig:app_2}
\end{center}
\end{figure*}

\begin{figure*}
\begin{center}
\includegraphics[width=0.85\textwidth]{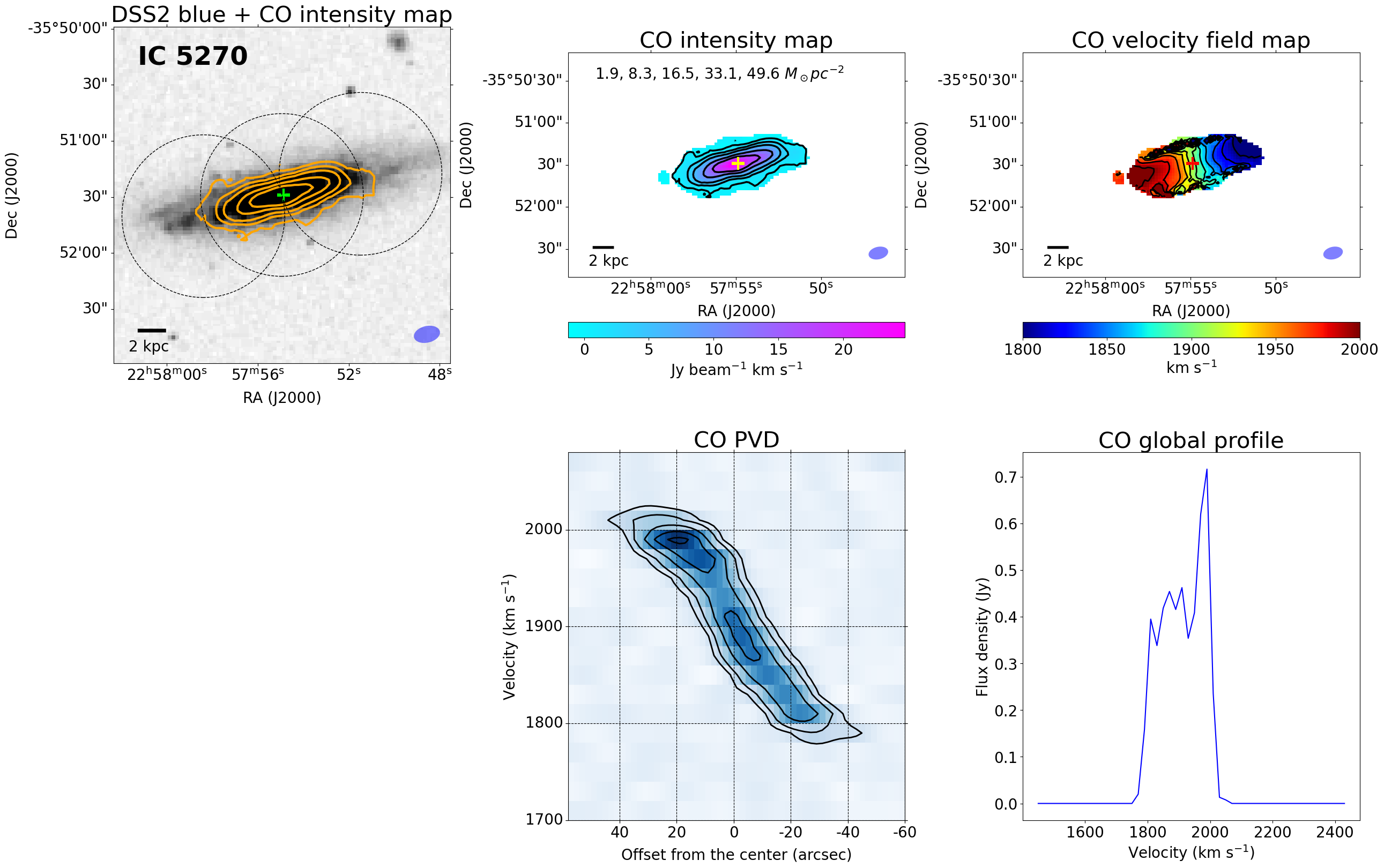}
\caption{IC~5270}
\label{fig:app_3}
\end{center}
\end{figure*}

\begin{figure*}
\begin{center}
\includegraphics[width=0.85\textwidth]{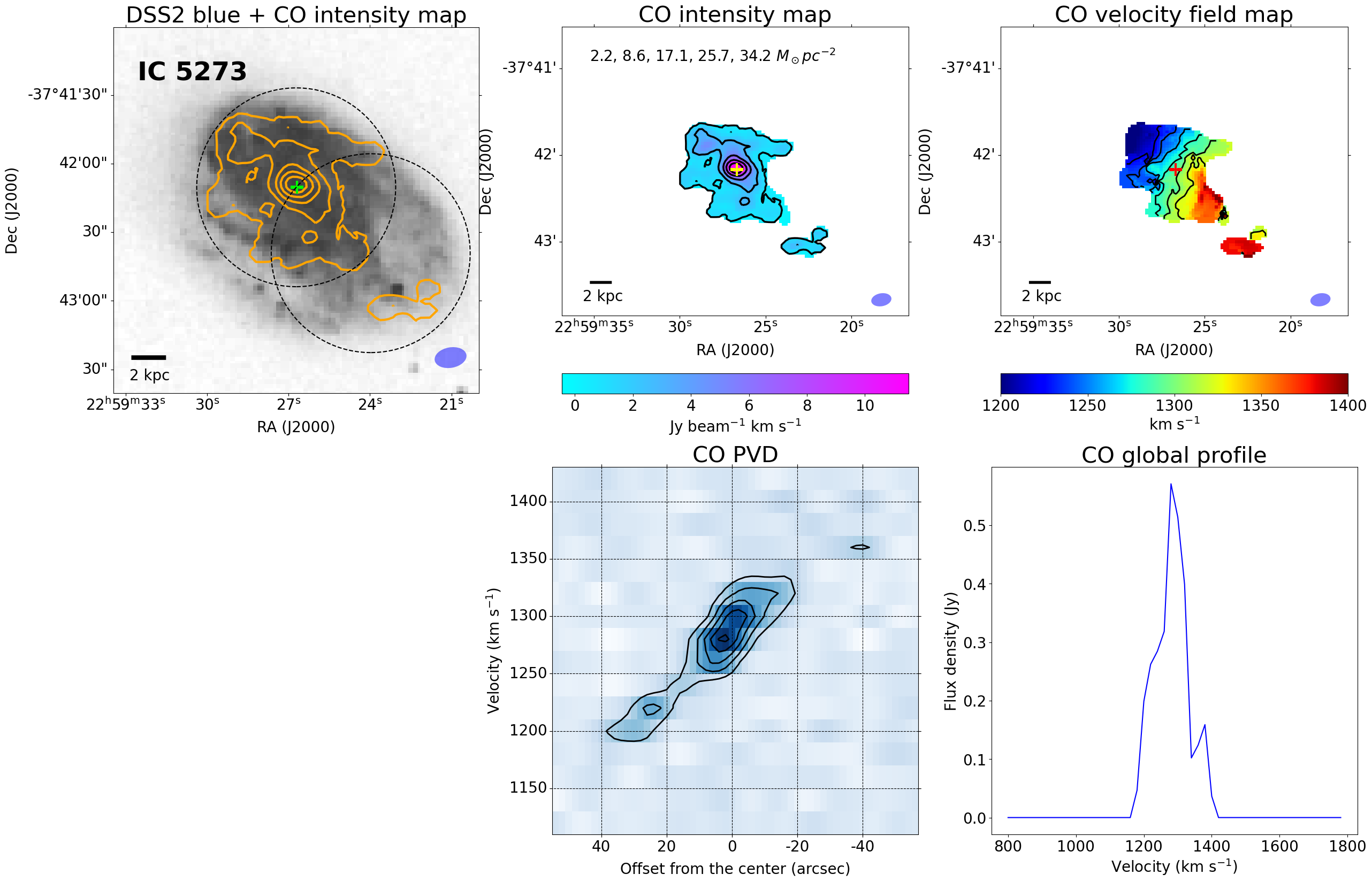}
\caption{IC~5273}
\label{fig:app_5}
\end{center}
\end{figure*}

\begin{figure*}
\begin{center}
\includegraphics[width=0.85\textwidth]{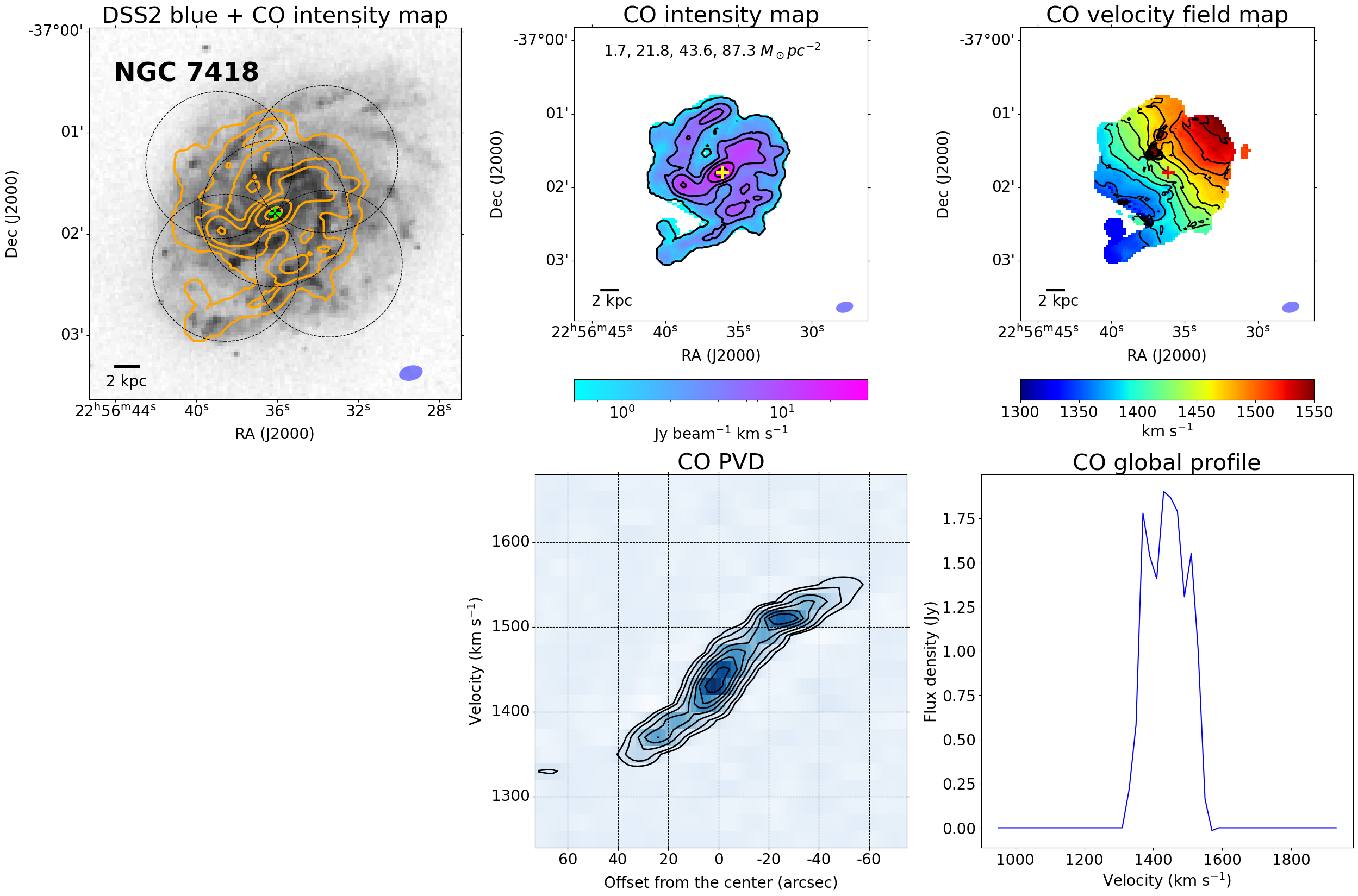}
\caption{NGC~7418}
\label{fig:app_7}
\end{center}
\end{figure*}

\begin{figure*}
\begin{center}
\includegraphics[width=0.85\textwidth]{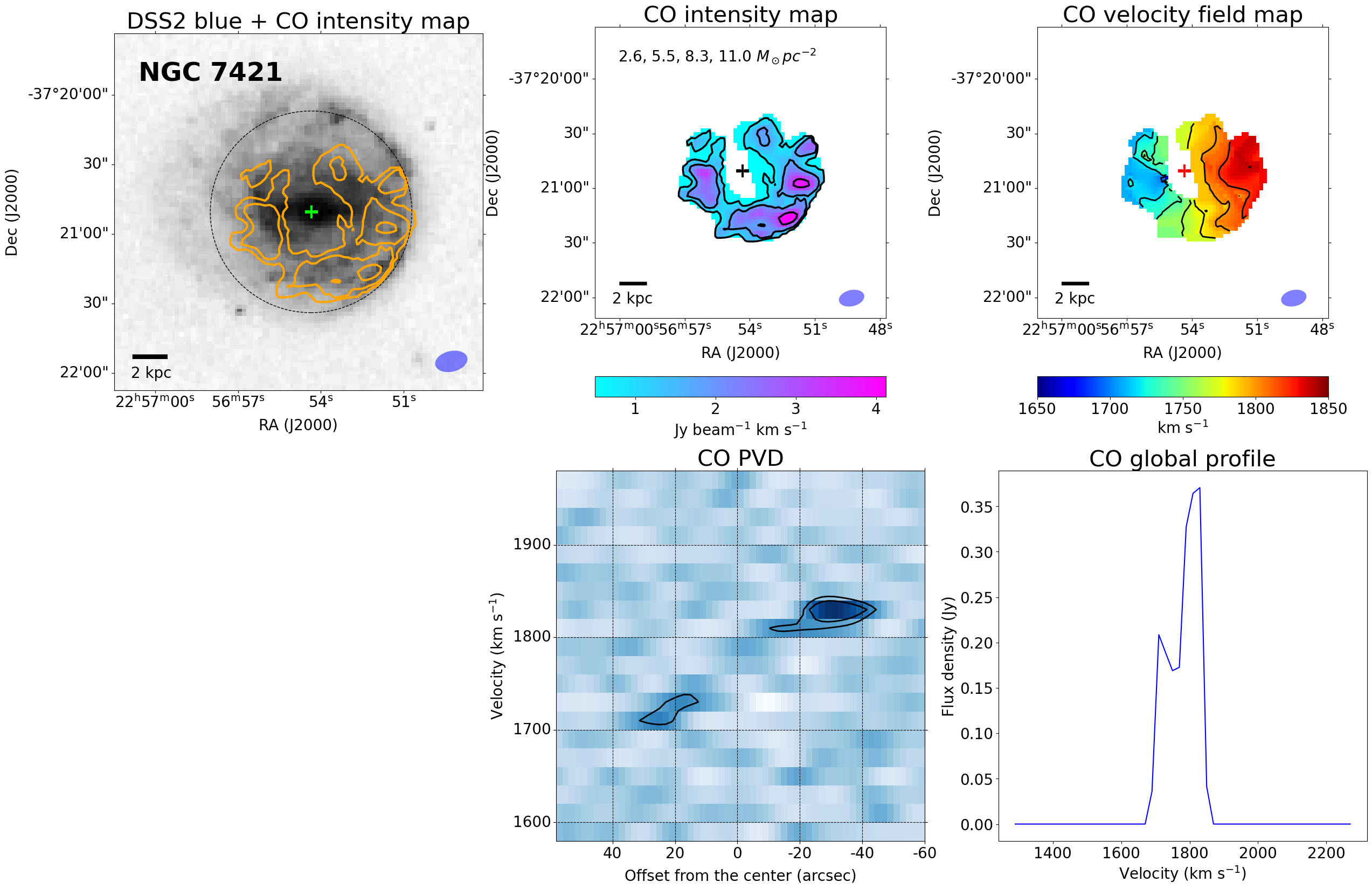}
\caption{NGC~7421}
\label{fig:app_8}
\end{center}
\end{figure*}

\begin{figure*}
\begin{center}
\includegraphics[width=0.85\textwidth]{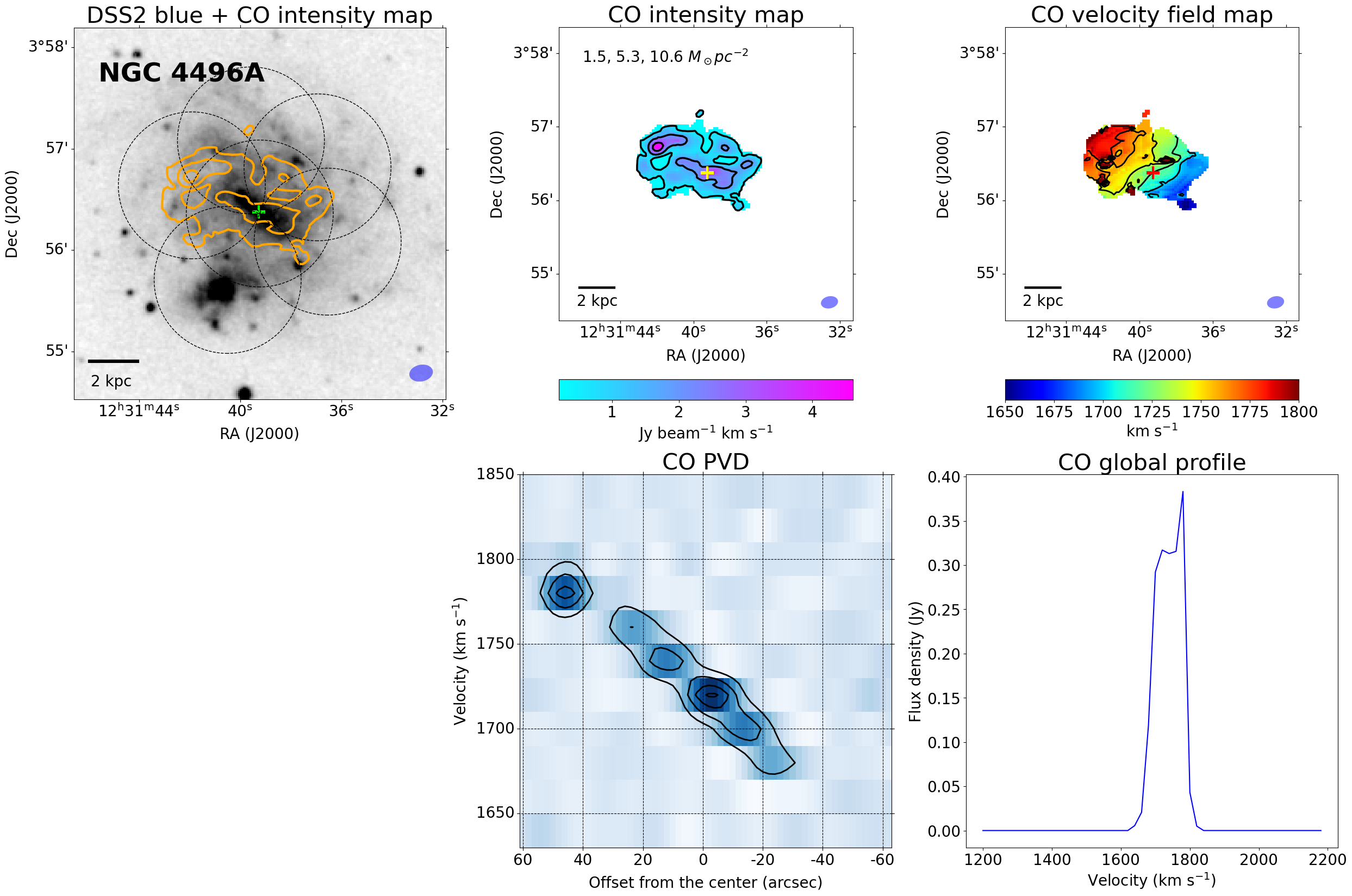}
\caption{NGC~4496}
\label{fig:app_10}
\end{center}
\end{figure*}

\begin{figure*}
\begin{center}
\includegraphics[width=0.85\textwidth]{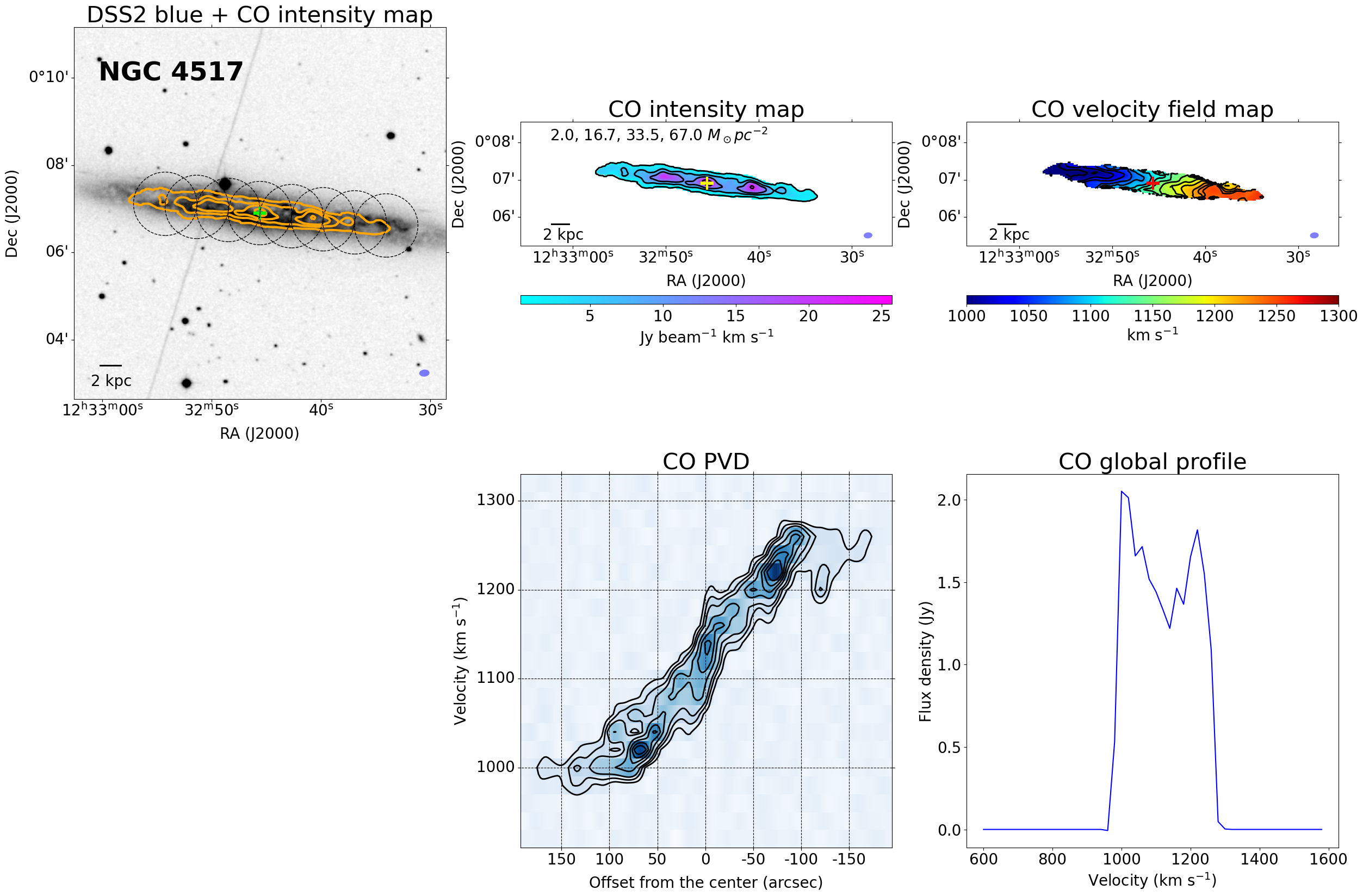}
\caption{NGC~4517}
\label{fig:app_11}
\end{center}
\end{figure*}

\begin{figure*}
\begin{center}
\includegraphics[width=0.85\textwidth]{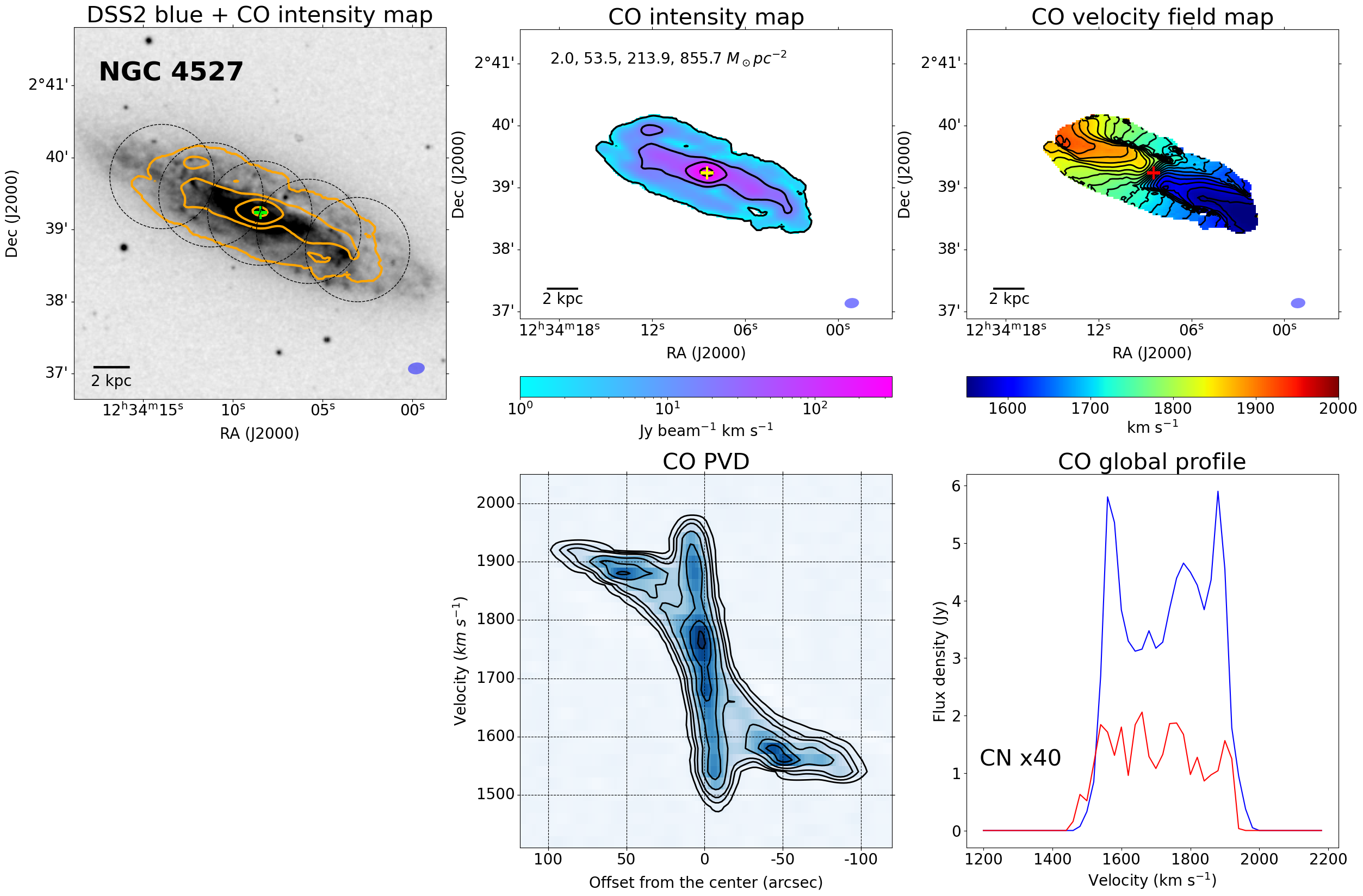}
\caption{NGC~4527}
\label{fig:app_12}
\end{center}
\end{figure*}

\begin{figure*}
\begin{center}
\includegraphics[width=0.85\textwidth]{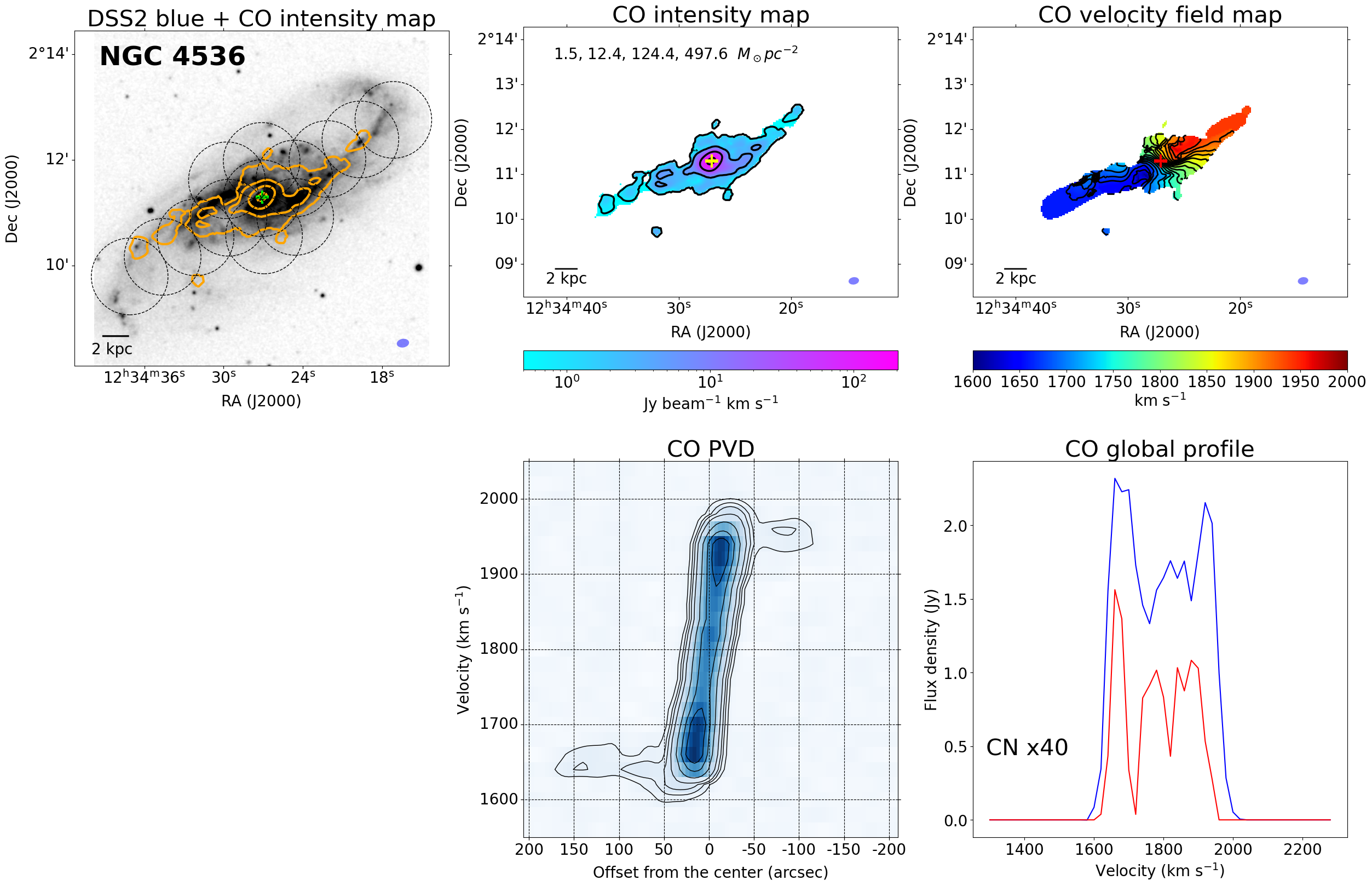}
\caption{NGC~4536}
\label{fig:app_13}
\end{center}
\end{figure*}

\begin{figure*}
\begin{center}
\includegraphics[width=0.85\textwidth]{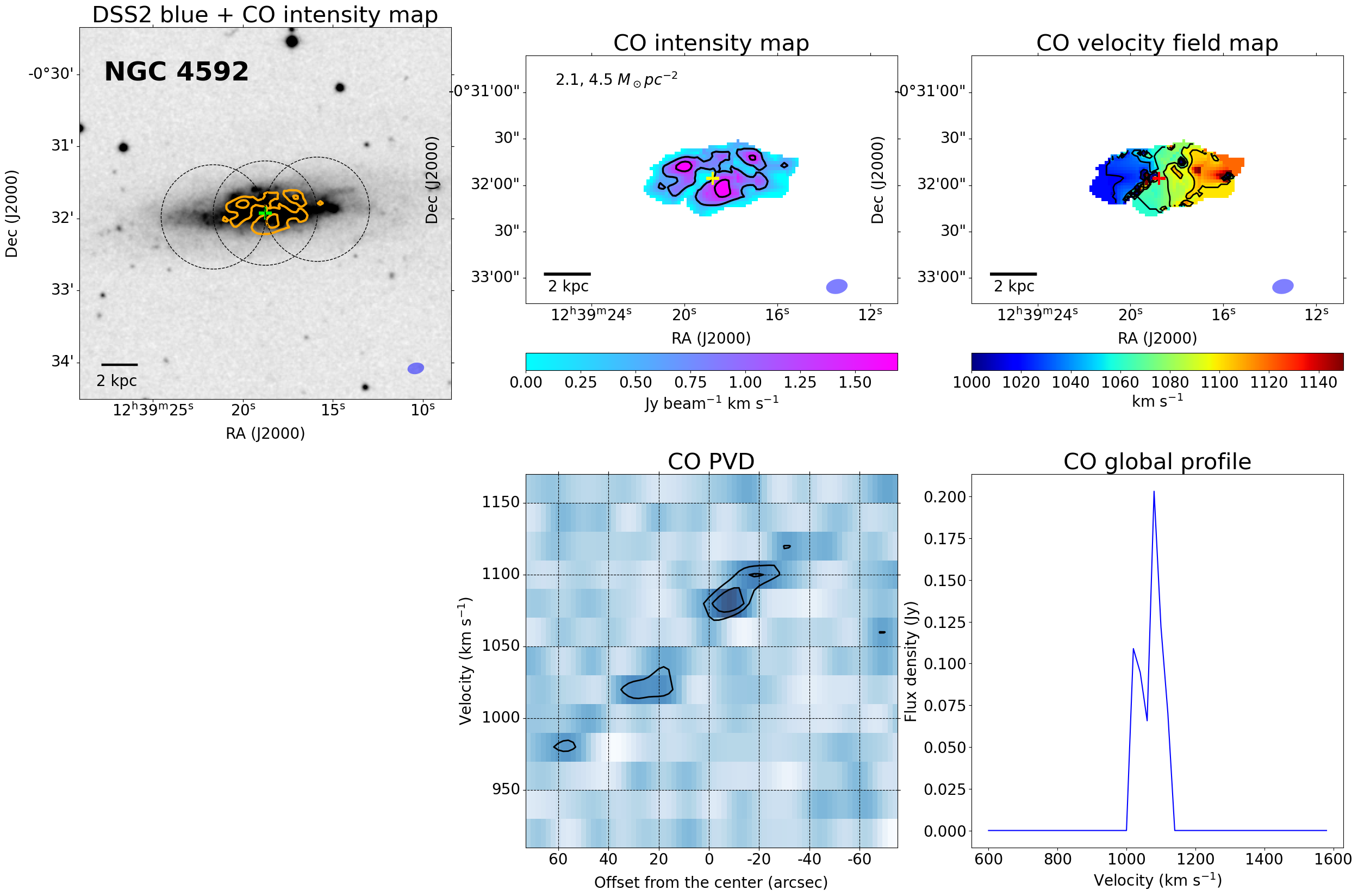}
\caption{NGC~4592}
\label{fig:app_14}
\end{center}
\end{figure*}

\begin{figure*}
\begin{center}
\includegraphics[width=0.85\textwidth]{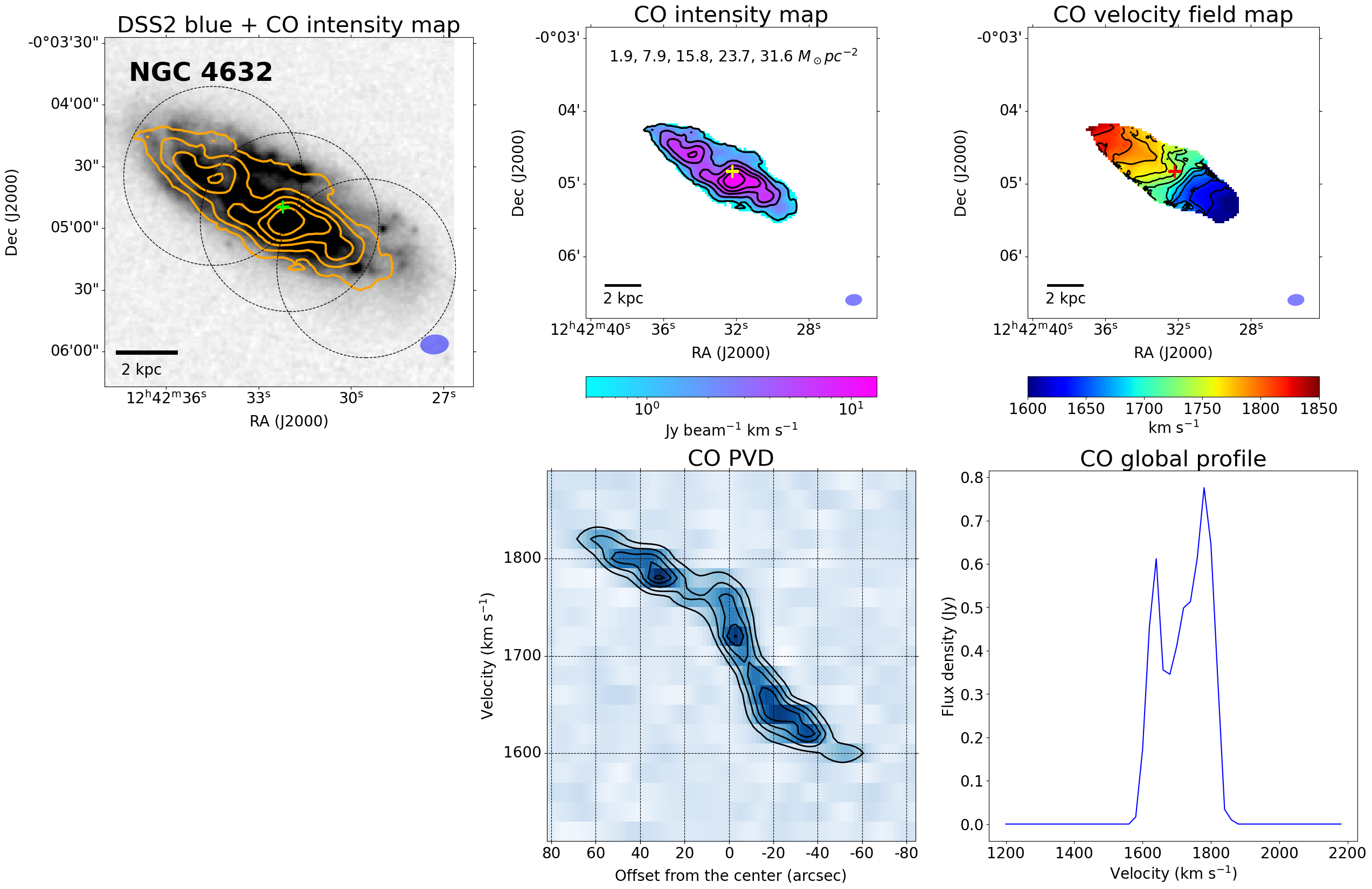}
\caption{NGC~4632}
\label{fig:app_15}
\end{center}
\end{figure*}

\begin{figure*}
\begin{center}
\includegraphics[width=0.85\textwidth]{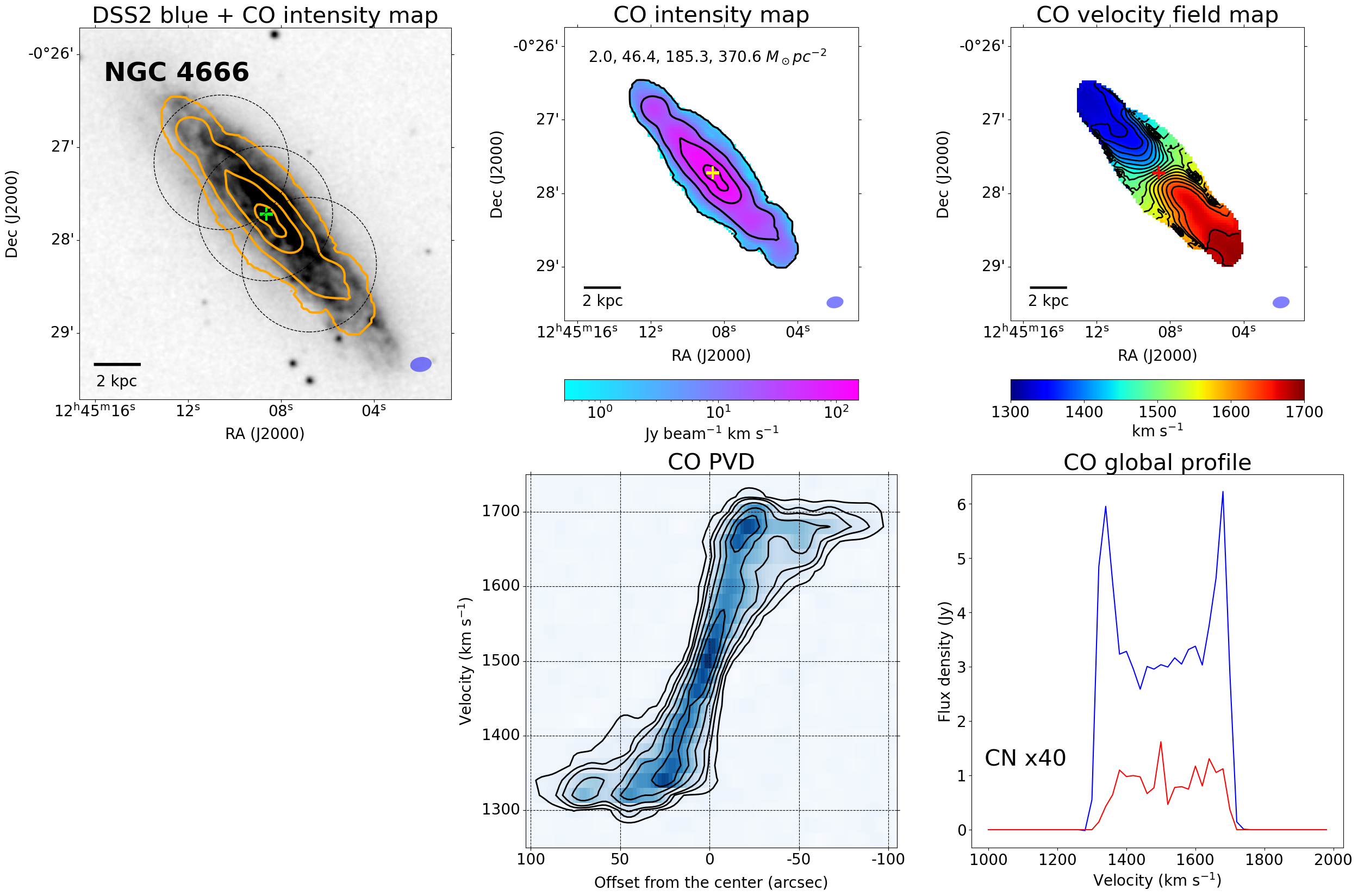}
\caption{NGC~4666}
\label{fig:app_16}
\end{center}
\end{figure*}

\begin{figure*}
\begin{center}
\includegraphics[width=0.85\textwidth]{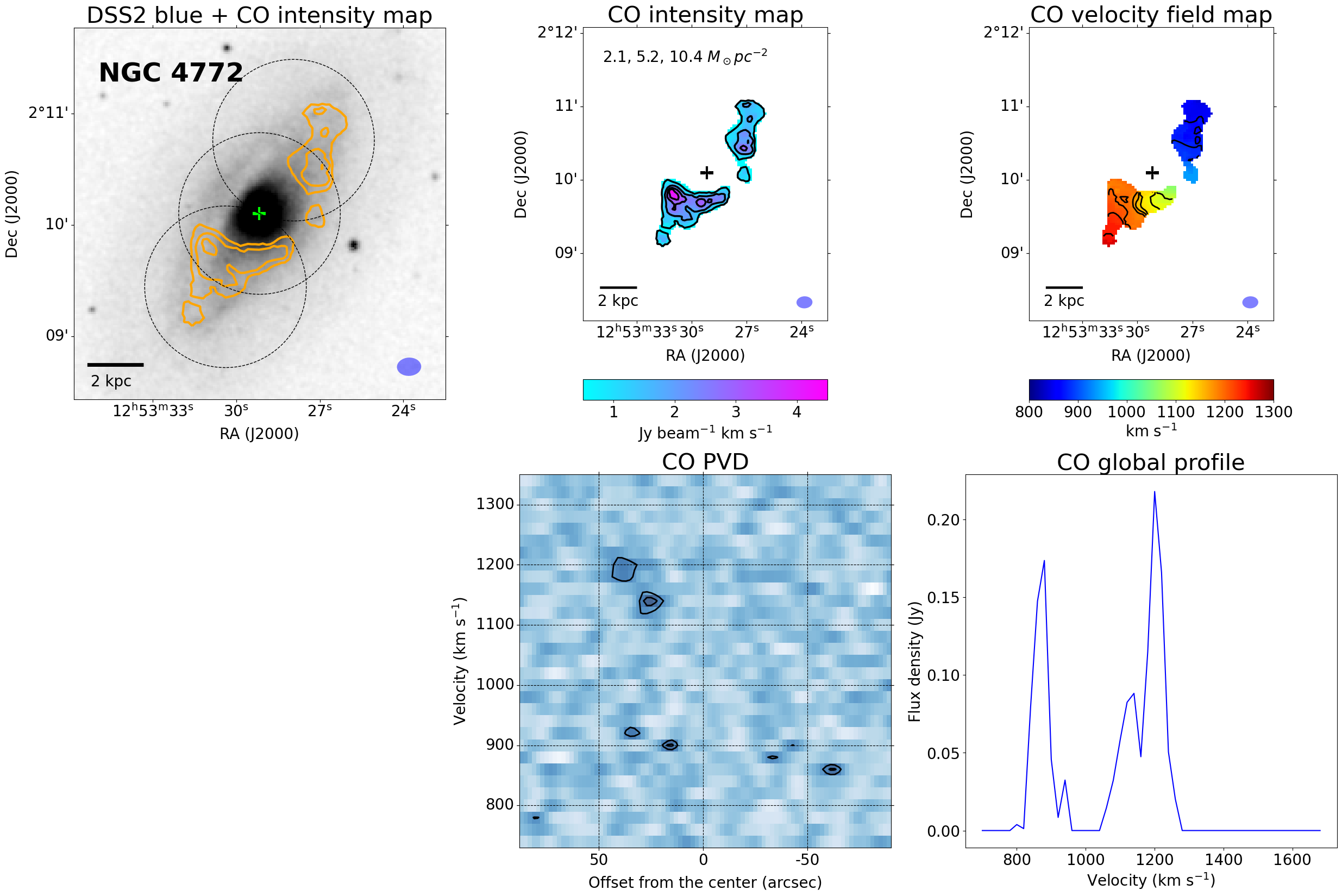}
\caption{NGC~4772}
\label{fig:app_17}
\end{center}
\end{figure*}

\begin{figure*}
\begin{center}
\includegraphics[width=0.85\textwidth]{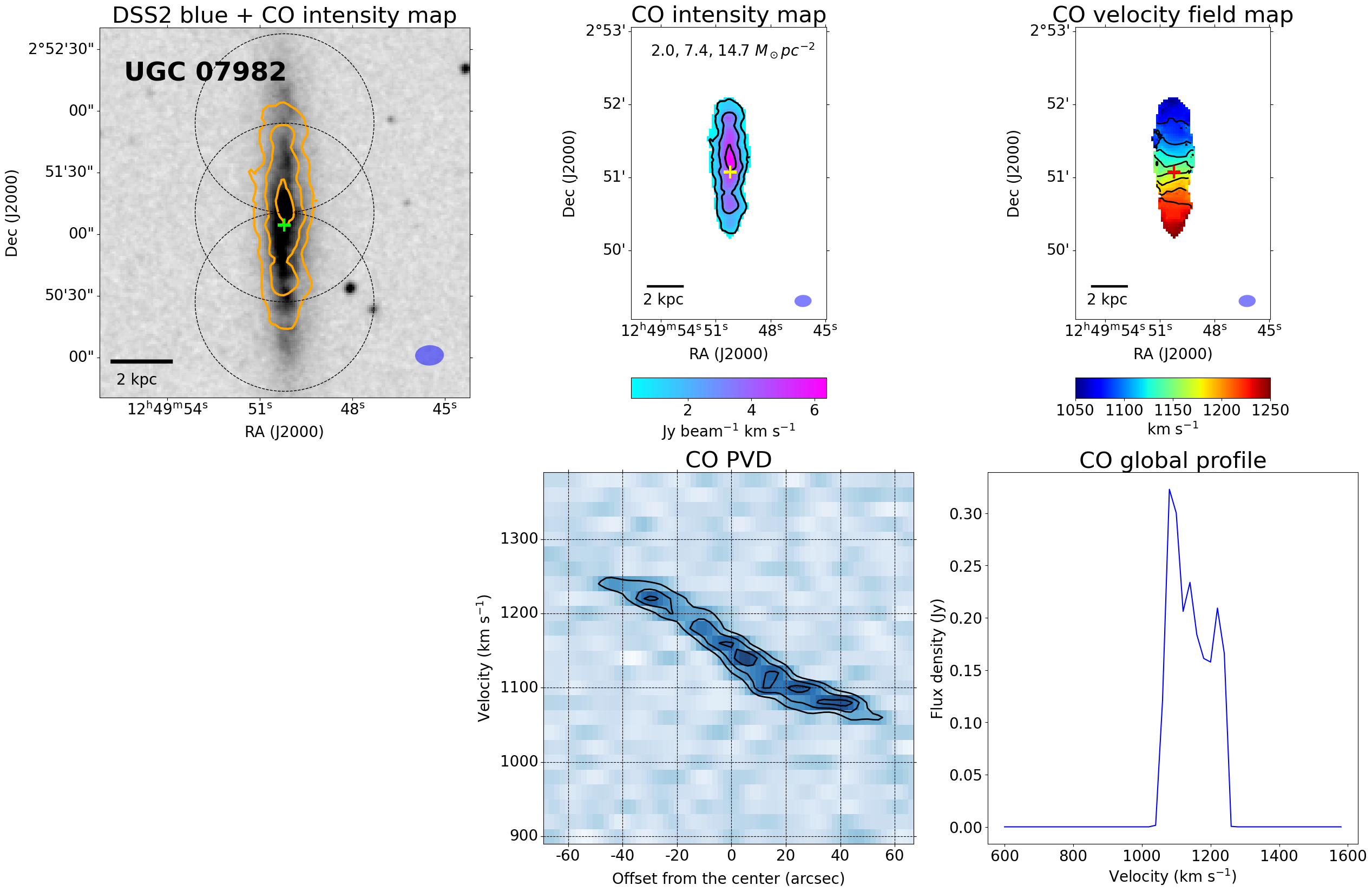}
\caption{UGC~07982}
\label{fig:app_18}
\end{center}
\end{figure*}

\begin{figure*}
\begin{center}
\includegraphics[width=0.85\textwidth]{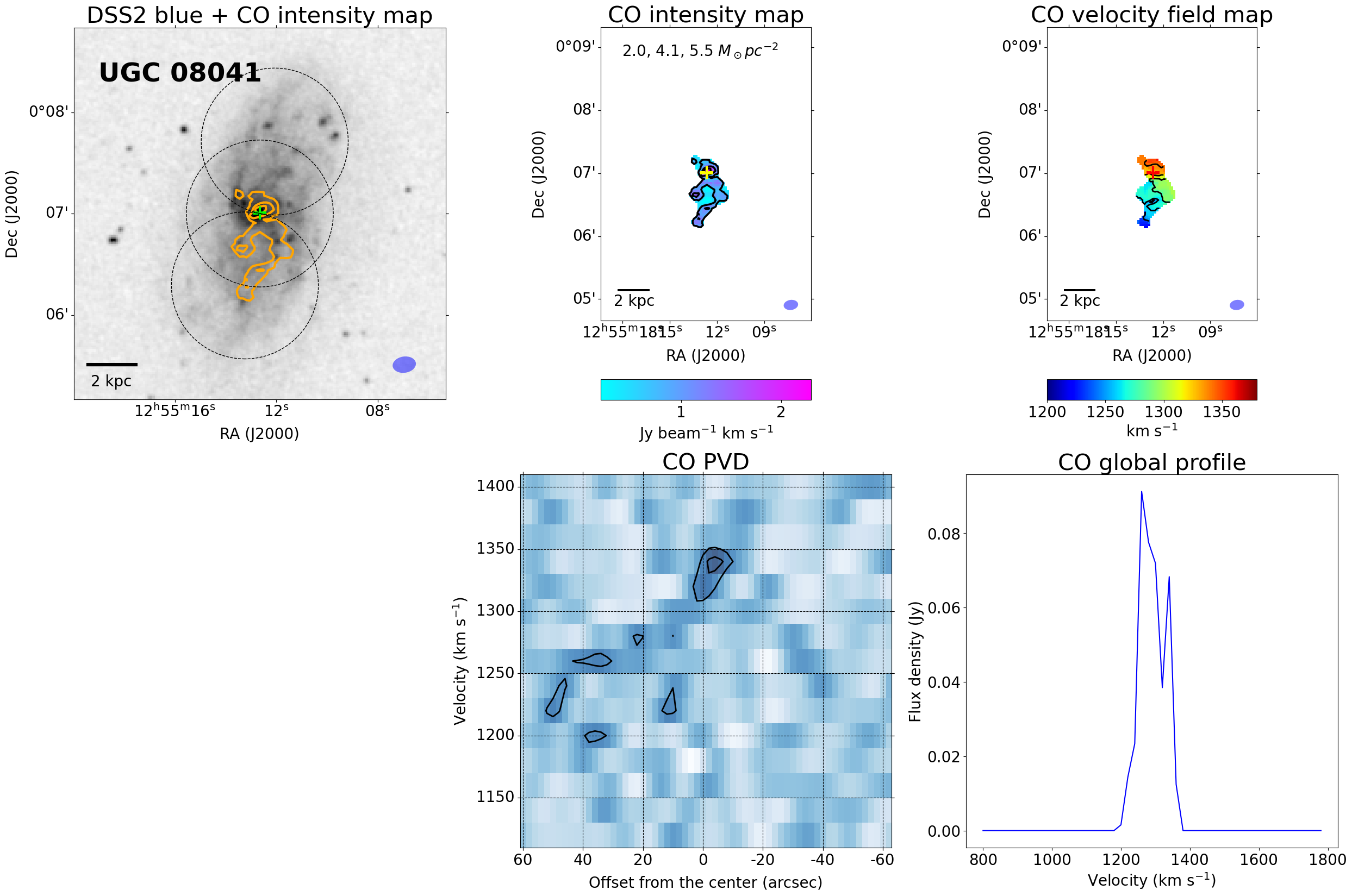}
\caption{UGC~08041}
\label{fig:app_19}
\end{center}
\end{figure*}

\section{Comments on individual galaxies with CO detections in the IC~1459 group and the NGC~4636 group} \label{app:comments}

\noindent\textbf{IC~5264} As seen in Figure~\ref{fig:app_1}, the CO morphology is very asymmetric. As described in Section~\ref{subsec:pecco}, the CO disk is more extended in the east side than in the west side. Meanwhile, a local peak of CO is found at the edge of the CO disk in the west side. In addition, the CO gas appears to be pushed down to the south from the stellar disk. The CO kinematics are disturbed, as seen in both the velocity map and the PVD in Figure~\ref{fig:app_1}. The asymmetry is also found in the CO global profile of IC~5264. \\
IC~5264 is very close to the BGG (IC~1459) of the group (a projected distance of $\sim$27.8 kpc from the group center), which seems to be located within the stellar halo of IC~1459 \citep{iodice2020}. This implies that IC~5264 is interacting with the BGG (IC~1459) tidally and with the hot halo of the BGG hydrodynamically (e.g., ram pressure stripping). Indeed, \cite{iodice2020} found the warped structure in the stellar component from VST (VLT Survey Telescope) observations. \\
As seen in Figure~\ref{fig:fig6}, IC~5264 has a lower {\HI} gas fraction than the median trend of the xCOLD GASS sample. In addition, the H$_{2}$ gas fraction of IC~5264 is close to the lower dashed line, which indicates a relatively low H$_{2}$ gas fraction. This decrease of the cold gas ({\HI} and H$_{2}$) contents due to external perturbations likely causes its lower SFR, and this suppressed star formation naturally leads to a long H$_{2}$ gas depletion time. IC~5264 is likely to have already been significantly processed in the group environment.  \\

\noindent\textbf{IC~5269B} In our ACA observations, the CO emission is found in the central region of the stellar disk, although IC~5269B has an extended {\HI} disk (see Figures~\ref{fig:hico} and \ref{fig:app_2}). \\

\noindent\textbf{IC~5270} The CO distribution is smooth, and the CO velocity field map seems to be regular (Figure~\ref{fig:app_3}). However, some asymmetries are seen in the PVD and the CO global profile (Figure~\ref{fig:app_3}). In the global profile, more CO flux is found around $\sim$2000 km~s$^{-1}$. IC~5270 has slightly high {\HI}/H$_{2}$ gas fractions (Figure~\ref{fig:fig6} (b) and (c)) and normal SFR (Figure~\ref{fig:fig6} (a)). \\

\noindent\textbf{IC~5273} Even though IC~5273 is located at the outskirt of the I1459G, the CO morphology is highly asymmetric (Figure~\ref{fig:app_5}). A locally strong CO emission is found at the northwest side of the CO disk (see also Figure~\ref{fig:hico}). On the other hand, in the southwestern edge of the stellar disk, a CO clump (or patch) is found. IC~5273 shows both asymmetric stellar and CO distributions, strongly suggesting that IC~5273 is affected by external perturbations. However, the {\HI} fraction (Figure~\ref{fig:fig6} (b)) and SFR (Figure~\ref{fig:fig6} (a)) of this galaxy lie on the median trend and the SFMS, respectively. \\

\noindent\textbf{NGC~7418} Relatively strong CO emission is seen along the spiral arm (Figures~\ref{fig:hico} and \ref{fig:app_7}). There is an extended CO emission at the southeast where the {\HI} compression is seen (see Figure~\ref{fig:hico}). NGC~7418 also shows an asymmetric stellar disk from the optical image, together with asymmetric CO and {\HI} distributions, indicating that this galaxy has experienced environmental processes in the I1459G. The CO kinematics seems to be not much disturbed, based on the CO velocity field map showing a regular rotating CO disk. Although NGC~7418 shows sings of interactions from {\HI}, CO, and optical images, this galaxy has typical {\HI} gas fraction (Figure~\ref{fig:fig6} (b)) and normal SFR (Figure~\ref{fig:fig6}~(a)). However, its H$_{2}$ gas fraction is slightly higher (Figure~\ref{fig:fig6} (c)).  \\

\noindent\textbf{NGC~7421} This galaxy shows a hole in the central region of the CO intensity map (Figure~\ref{fig:app_8}). Interestingly, the CO distribution is very asymmetric, with locally strong CO emission in the west side of the CO disk (Figures~\ref{fig:hico} and \ref{fig:app_8}). This feature is often seen in ram pressure stripping galaxies \citep[][]{lee2017,george2019,lizee2021}. Ram pressure can compress the molecular gas \citep[e.g.][]{lee2017}. Thus, NGC~7421 is thought to be likely affected by ram pressure, based on its {\HI} image (see also Figure~\ref{fig:hico}) \citep[e.g.,][]{ryder1997,serra2015}. However, \cite{oosterloo2018} suggests that it is expected that the strength of ram pressure is much weaker than in the galaxy cluster (e.g., Virgo cluster), which implies that the ram pressure stripping may not be the main mechanism to affect NGC~7421. A stellar disk of NGC~7421 appears to be extended to the east. The asymmetric stellar disk and disturbed {\HI}/CO morphologies suggest that a tidal interaction may affect the cold gas components and stellar disk of NGC~7421, as suggested in other previous studies \citep[][]{oosterloo2018,iodice2020}. NGC~7421 has a slightly low {\HI} fraction (Figure~\ref{fig:fig6} (b)), and its H$_{2}$ gas fraction and SFR are much lower than the median trend of the xCOLD GASS sample. These results suggest that the group environment significantly affect not only the distributions of cold gas contents ({\HI} and CO) but also global properties (i.e., removal of cold gas and suppression of SFR) of this galaxy. \\

\noindent\textbf{NGC~4496A} NGC~4496A shows asymmetric CO distribution with a local CO peak in the northeast of the CO disk (Figure~\ref{fig:app_10}) although NGC~4496A seems to not be affected by external perturbations, based on its symmetric stellar and {\HI} disks (see its optical and {\HI} images in Figure~\ref{fig:hico}). The asymmetric CO distribution of this galaxy may be due to internal mechanisms. NGC~4496A has a typical SFR (Figure~\ref{fig:fig6} (a)) and {\HI} gas fraction (Figure~\ref{fig:fig6} (b)), but its H$_{2}$ gas fraction is much lower than the median trend of the xCOLD GASS sample (Figure~\ref{fig:fig6} (c)). \\
On the projected sky plane, NGC~4496A has a close companion (NGC~4496B) that is located at the southern part of the stellar disk of NGC~4496A. However, the system velocity ($\sim$4530 km~s$^{-1}$) of NGC~4496B significantly differs from the system velocity ($\sim$1730 km~s$^{-1}$) of NGC~4496A, which indicates that two galaxies are unlikely to interact with each other. Indeed, there are no signs of interaction in the {\HI} image of NGC~4496A (Figure~\ref{fig:hico}). \\

\noindent\textbf{NGC~4517} Overall, the CO morphology and the CO kinematics are regular, as seen in Figure~\ref{fig:app_11}. However, the kinematics of CO gas at the west side (from -100$\arcsec$ to -150$\arcsec$ in the PVD) of the CO disk deviates from the regularly rotating CO disk. This deviation is also seen in the velocity field map. \\

\noindent\textbf{NGC~4527} The CO morphology and CO kinematics appear to be undisturbed (Figure~\ref{fig:app_12}). The CO PVD clearly shows a steep velocity gradient in the inner part. This indicates a presence of a bar-like structure in the central region \citep[e.g.,][]{alatalo2013}. The rapidly increasing velocity component is also seen in the central region of the velocity field map. This distinct velocity structure was also reported in previous high-resolution CO observation \citep{shibatsuka2003}. Figure~\ref{fig:app_12} presents the line profiles of both CO (blue line) and CN lines (red line). The velocity range of the CN emission is comparable to that of the CO emission. The details of the CN structure of NGC~4527 are in Appendix~\ref{app:cn}. \\

\noindent\textbf{NGC~4536} The CO gas seems to be concentrated in the central region with a strong peak, and the relatively faint CO emission appears to extend along the stellar arms. Some CO patches are also found in the outer part (Figure~\ref{fig:app_13}). Previous high-resolution CO observations ($\sim$2$\arcsec$) using the NMA (the Nobeyama Millimeterwave Array) and the OVRO (the Owens Valley Radio Observatory) millimeterwave interferometer reported the gas concentration in the nuclear disk \citep{sofue2003, jogee2005}. Recently, this galaxy is also mapped in CO (2$-$1) transition in the Virgo Environment Traced in CO Survey (VERTICO; \citealt{brown2021}). As for NGC~4527, the CN emission is detected in the central region of NGC~4536 (see Figures~\ref{fig:app_13} and \ref{fig:cn}). \\

\noindent\textbf{NGC~4592} The CO distribution of NGC~4592 looks clumpy, as seen in Figure~\ref{fig:app_14}. As seen in Figure~\ref{fig:hico}, its {\HI} disk is symmetric/smooth and extends beyond its stellar disk. \\

\begin{figure*}[hbt!]
\begin{center}
\includegraphics[width=1.0\textwidth]{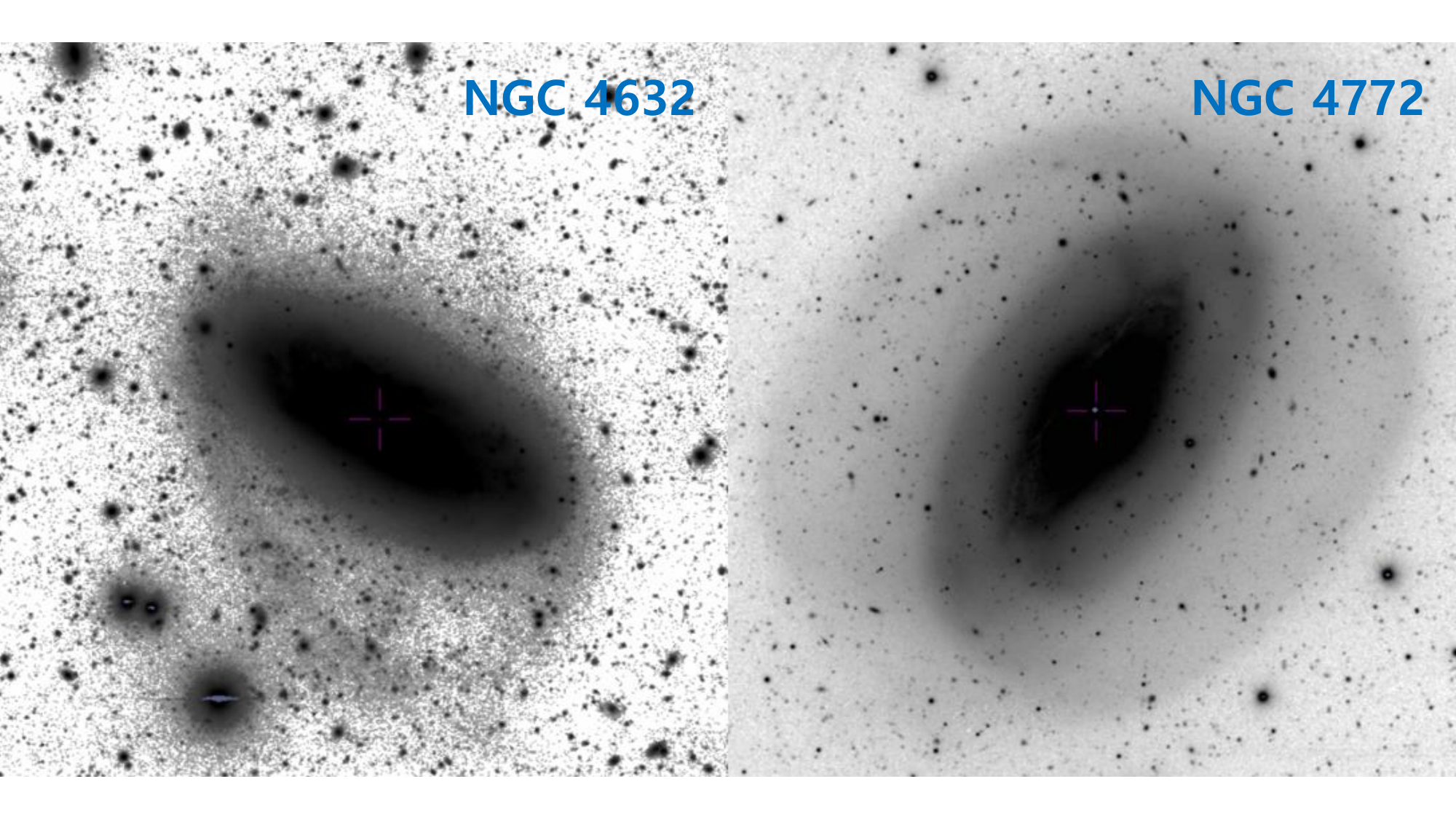}
\caption{Deep optical images (the Hyper Surprime-Cam (HSC) $r$-band wide DR2) of NGC~4632 (left) and NGC~4772 (right) from the HSC Strategic Program (HSC-SSP; \citealt{aihara2019}), which are extracted using Aladin Sky Atlas \citep{bonnarel2000}. }   \label{fig:hsc}
\end{center}
\end{figure*}

\noindent\textbf{NGC~4632} The CO distribution and the CO kinematics are asymmetric. In particular, the CO distribution seems to resemble the complex structure of the inner stellar disk (Figure~\ref{fig:app_15}). In the east side, the CO disk is more extended and presents a strong local peak. This asymmetric structure is also clearly seen in the PVD of NGC~4632. Interestingly, NGC~4632 has a polar ring-like structure in the {\HI} gas, and its deep optical image (Figure~\ref{fig:hsc}) also shows the polar ring structure in the south. External mechanisms (e.g., minor merger) may cause the formation of the ring structure of NGC~4632. Additional analysis on the deep imaging and kinematics of the {\HI} ring is underway (X. Chen et al. 2022, in preparation and N. Deg et al. 2022, in preparation). \\
Although NGC~4632 shows clear signs of external perturbations with the polar ring structure and asymmetric distributions of the CO gas and the stellar component, this galaxy has a typical SFR (Figure~\ref{fig:fig6} (a)), and a normal {\HI} gas fraction (Figure~\ref{fig:fig6} (b)). However, its H$_{2}$ gas fraction is above +1$\sigma$ scatter, indicating more abundant H$_{2}$ gas (Figure~\ref{fig:fig6}~(c)). \\

\noindent\textbf{NGC~4666} Overall, the CO morphology and the CO kinematics are regular, as seen in Figure~\ref{fig:app_16}. In addition, the CN emission is also detected in the central region of NGC~4666 (see Figures~\ref{fig:app_16} and \ref{fig:cn}). The details of the CN structure of NGC~4666 are in Appendix~\ref{app:cn}. \\
As discussed in Section~\ref{subsec:hico}, NGC~4666 is interacting with its neighbor galaxy (NGC~4668). Interestingly, while NGC~4666 has a typical {\HI} gas fraction (Figure~\ref{fig:fig6} (b)), its SFR is relatively high (Figure~\ref{fig:fig6} (a)), and the H$_{2}$ gas fraction is much higher than the median trend of the xCOLD GASS sample ((Figure~\ref{fig:fig6} (c)). In this case, tidal interactions may help to convert the {\HI} gas into the H$_{2}$ gas efficiently \citep{kaneko2017} and enhance the star formation activity, as reported in previous studies \citep[e.g.,][]{pan2018,renaud2019}. \\

\noindent\textbf{NGC~4772} The CO gas is found along the inner {\HI} disk, as discussed in Section~\ref{subsec:hico}. We did not detect CO emission in the bulge region of NGC~4772 (Figure~\ref{fig:app_17}). The VERTICO project \citep{brown2021} also observed CO(2--1) line of this galaxy, and its CO(2--1) distribution is similar to the CO(1--0) distribution in our ACA observation. As seen in the {\HI} image (Figure~\ref{fig:hico}), NGC~4772 has a large {\HI} outer ring. Furthermore, this galaxy shows a diffuse extended stellar rings from a deep optical image (Figure~\ref{fig:hsc}). Based on these complex structures, NGC~4772 is thought to experience a minor merger \citep{haynes2000,chung2009}. \\
NGC~4772 present significantly low SFR ((Figure~\ref{fig:fig6} (a)) and low {\HI}/H$_{2}$ gas fractions ((Figure~\ref{fig:fig6} (b) and (c)). The decrease of the cold gas contents likely results in the suppression of star formation activity in NGC~4772. \\

\noindent\textbf{UGC~07982} Overall, the CO morphology and the CO kinematics are regular, as seen in Figure~\ref{fig:app_18}. However, as discussed in Section~\ref{subsec:hico}, this galaxy is likely to be experiencing RPS in the N4636G, based on its asymmetric {\HI} morphology and the analysis of the ram pressure strength against the restoring force in the disk for UGC~07982 \citep{lin2023}. 
UGC~07982 has a very low SFR (Figure~\ref{fig:fig6} (a)) and a low {\HI} gas fraction (Figure~\ref{fig:fig6}~(b)), compared to the SFMS galaxies. In particular, this galaxy has quite a long H$_{2}$ gas depletion time (Figure~\ref{fig:fig6} (f)), indicating a low star formation efficiency. In this case, the external perturbation (i.e., RPS) may prevent the molecular gas from forming new stars by injecting additional turbulent motion into the molecular gas, as suggested by previous studies \citep{alatalo2015,mok2016,lee2017}. \\

\noindent\textbf{UGC~08041} As described in Section~\ref{subsec:pecco}, the CO distribution of UGC~08041 is irregular, showing that the CO emission is much extended toward the south (Figure~\ref{fig:app_19}). UGC~08041 shows typical SFR (Figure~\ref{fig:fig6} (a)) and normal {\HI} gas fraction (Figure~\ref{fig:fig6} (b)). However, its H$_{2}$ gas fraction is quite low (Figure~\ref{fig:fig6} (c)), compared to the median trend of the xCOLD GASS sample. \\

\section{CN line detection} \label{app:cn}
\begin{figure*}[!htbp]
\begin{center}
\includegraphics[width=1.0\textwidth]{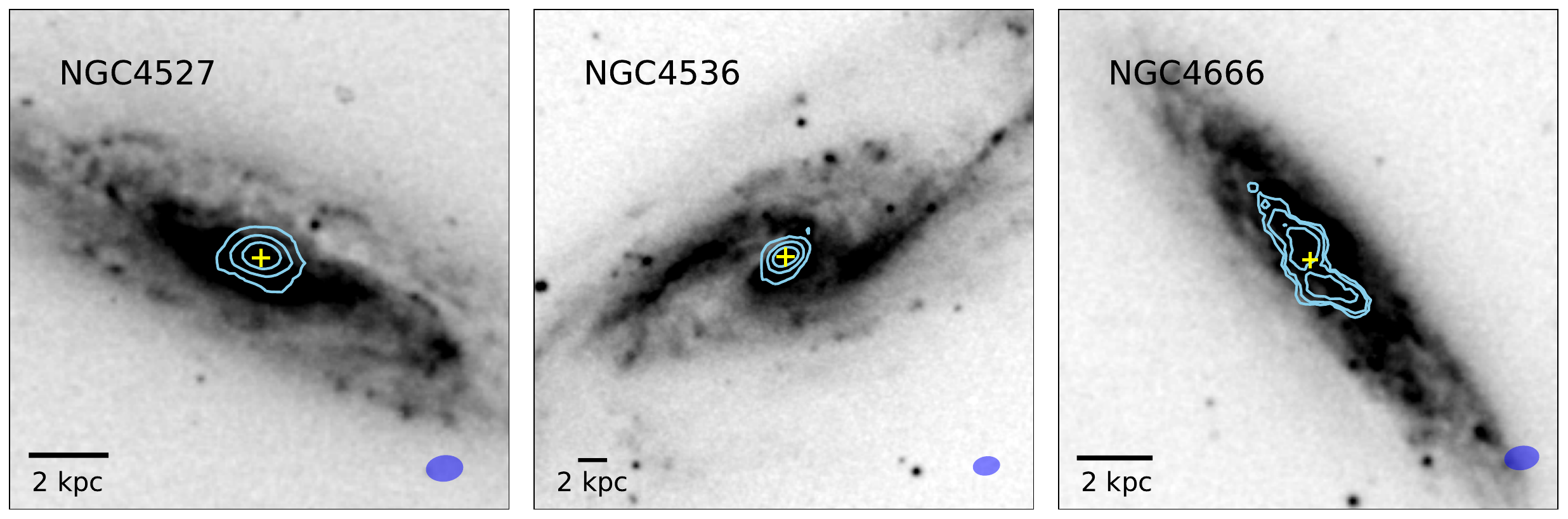}
\caption{CN distribution (contours) of three group galaxies (NGC~4527, NGC~4536, and NGC~4666) is overlaid on the DSS2 blue images. The contour levels are 30\% and 70\% of the peak emission (Table~\ref{tab:table4}) for three group galaxies, and the outermost contour corresponds to 3$\sigma$ value of each CN map. The yellow cross indicates the stellar disk center. The bar in the bottom-left corner represents the physical scale of 2~kpc. The synthesized beam of the CN data is shown in the bottom-right corner.  \label{fig:cn}}

\end{center}
\end{figure*}

The CN line is one of the molecular lines that trace dense molecular gas. This molecule is abundant in photon-dominated regions (PDRs) because the CN molecule can be created by a photo-dissociation process \citep[e.g.,][]{rodriguez-franco1998,boger2005}. In our ACA observations, three galaxies (NGC~4527, NGC~4536, and NGC~4666) in the N4636G have CN detection. These three galaxies show active star-formation activity (SFR $\sim$2~{\Msun}~yr$^{-1}$, see also Table~\ref{tab:table1}) \citep{ dahlem1997, shibatsuka2003, jogee2005} together with nuclear activity in the galactic center (e.g., AGN) \citep{ho1997, shibatsuka2003, persic2004, mcalpine2011}. Figure~\ref{fig:cn} reveals CN distribution of three galaxies. NGC~4527 and NGC~4536 show compact distribution of CN emission in the central region, while NGC~4666 show relatively extended CN distribution. For NGC~4527, the linewidth of CN is slightly larger than that of CO. Meanwhile, for NGC~4536 and NGC~4666, the linewidths of CN are narrow, compared to the linewidths of CO (see Tables~\ref{tab:table3} and \ref{tab:table4}, and Figures~\ref{fig:app_12}, \ref{fig:app_13}, and \ref{fig:app_16}). The integrated CN fluxes, peak intensities of the intensity maps, the CN linewidths, rms noise per channel of the CN data cubes, and information on the synthesized beam of the CN data are summarized in Table~\ref{tab:table4}.

\begin{deluxetable}{lcccccc}[!htbp] 
\tablecaption{CN Data Parameters of Three Galaxies \label{tab:table4}}
\tablehead{
\multicolumn{1}{l}{Name} & \multicolumn{1}{c}{$b_{maj}$$\times$$b_{min}$, $b_{PA}$} & \multicolumn{1}{c}{$\sigma_{\rm rms, CN}$} & \multicolumn{1}{c}{$W_{\rm 20, CN}$} & \multicolumn{1}{c}{$W_{\rm 50, CN}$} & \multicolumn{1}{c}{CN Peak Flux} & \multicolumn{1}{c}{CN Flux} \\
& ($\arcsec\times\arcsec$, deg) & (mJy~beam$^{-1}$) & (km~s$^{-1}$) & (km~s$^{-1}$) & (Jy~km~s$^{-1}$~beam$^{-1}$) & (Jy~km~s$^{-1}$) \\
(1) & (2) & (3) & (4) & (5) & (6) & (7)
}
\startdata
NGC 4527 & 14.3$\times$10.1, 96.1 & 8.0 & 463 & 408 & 12.5 & 15.6$\pm$1.3 \\
NGC 4536 & 14.3$\times$10.0, 99.2 & 6.0 & 303 & 264 & 5.8 & 6.4$\pm$0.7 \\
NGC 4666 & 14.1$\times$9.7, 99.1 & 8.0 & 370 & 321 & 2.7 & 8.5$\pm$1.0
\enddata
\tablecomments{(1) Galaxy name; (2) beam size (major and minor axes), beam position angle; (3) rms noise level per channel of the CN data; (4) and (5) the CN linewidths measured at 20\% and 50\% of the peak flux using the SoFiA; (6) the peak value of CN intensity map; (7) the total CN flux.}
\end{deluxetable}

\section{3 mm continuum detection} \label{app:conti}

\begin{figure*}[!htbp]
\begin{center}
\includegraphics[width=1.0\textwidth]{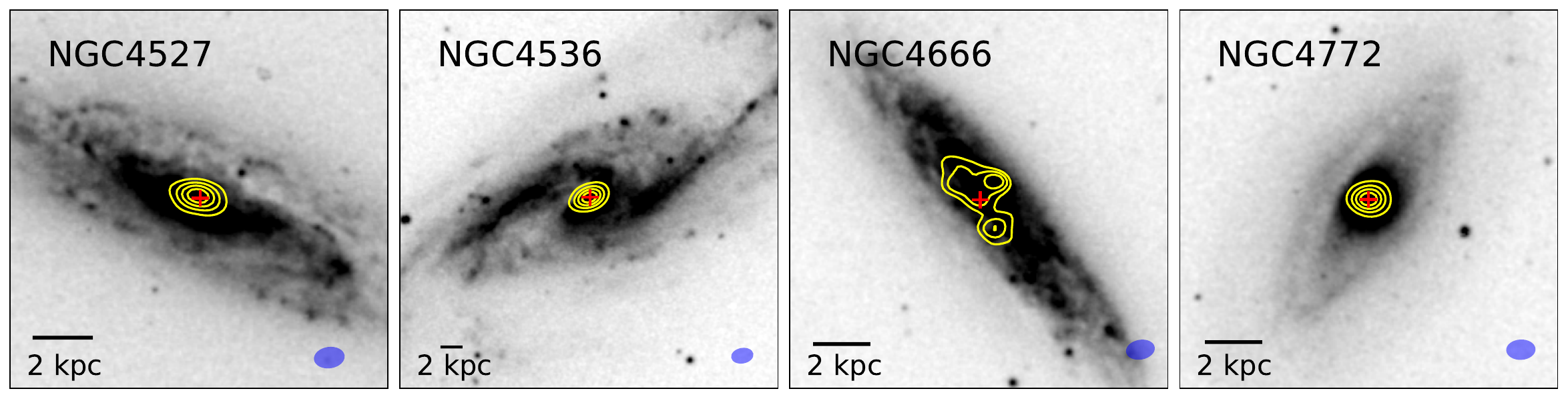}
\caption{3 mm continuum distribution (contours) of four group galaxies (NGC~4527, NGC~4536, NGC~4666, and NGC~4772) is overlaid on the DSS2 blue images. The contour levels are 20, 40, 60, and 80\% of the peak emission (Table~\ref{tab:table5}) for three galaxies. For NGC~4666, the contour levels are 40, 60, and 80\% of the peak emission (2.4 mJy~beam$^{-1}$). The red cross indicates the stellar disk center. The bar in the bottom-left corner represents the physical scale of 2~kpc. The synthesized beam of the 3 mm continuum data is shown in the bottom-right corner.  \label{fig:cont}}
\end{center}
\end{figure*}

Four galaxies (NGC~4527, NGC~4536, NGC~4666, and NGC~4772) of the N4636G have a 3~mm continuum detection. The 3~mm continuum emission of NGC~4772 was detected in the previous Nobeyama Millimeter Array (NMA) observations. For NGC~4772, its 3~mm flux density (6.0 $\pm$ 0.6 mJy) in our ACA observations is consistent with the 3~mm flux density (5.2 $\pm$ 2.9 mJy) in the NMA observations \citep{doi2011}. Figure~\ref{fig:cont} shows the 3~mm continuum distributions of the four galaxies. The 3~mm continuum distribution is compact in three galaxies (NGC~4527, NGC~4536, and NGC~4772). NGC~4666 shows the extended continuum distribution, similar to the CN distribution. All four galaxies have nuclear activity in the galactic center. NGC~4527 is the transition object between the starburst and the low-ionization nuclear emission-line region (LINER) \citep{ho1997}. NGC~4536 hosts the low luminosity AGN (LLAGN) together with active star formation in the central region \citep{satyapal2008, mcalpine2011}. In addition, the obscured AGN is found in NGC~4666 \citep{dahlem1997}. NGC~4772 has also LLAGN \citep{doi2011}. The 3~mm flux densities, peak flux densities of the 3~mm continuum, rms noise and synthesized beam of the 3~mm data are summarized in Table~\ref{tab:table5}.

\begin{deluxetable}{lcccccc}[!htbp]
\tablecaption{3~mm Continuum Data Parameters of Four Galaxies \label{tab:table5}}
\tablehead{
\multicolumn{1}{l}{Name} & \multicolumn{1}{c}{$b_{maj}$$\times$$b_{min}$, $b_{PA}$} & \multicolumn{1}{c}{$\sigma_{\rm rms, continuum}$} & \multicolumn{1}{c}{Peak Flux Density} & \multicolumn{1}{c}{Flux Density} & \multicolumn{1}{c}{Nuclear Activity} & \multicolumn{1}{c}{References}\\
& ($\arcsec\times\arcsec$, deg) & (mJy~beam$^{-1}$) & (mJy~beam$^{-1}$) & (mJy) &  & \\
(1) & (2) & (3) & (4) & (5) & (6) & (7)
}
\startdata
NGC 4527 & 15.5$\times$10.7, 97.6 & 0.32 & 5.1 & 6.1$\pm$0.9 & starburst/LINER & (a)\\
NGC 4536 & 15.5$\times$10.7, 100.3 & 0.20 & 9.1 & 11.9$\pm$0.9 & LLAGN & (b), (c)\\
NGC 4666 & 15.4$\times$10.5, 100.4 & 0.28 & 2.4 & 9.3$\pm$1.2 & obscured AGN &  (d)\\
NGC 4772 & 15.5$\times$10.8, 92.6 & 0.24 & 5.9 & 6.0$\pm$0.6 & LLAGN & (e) 
\enddata
\tablecomments{(1) Galaxy name; (2) beam size (major and minor axes), beam position angle; (3) rms noise level of the 3~mm continuum data; (4) the peak value of 3~mm continuum map; (5) the total 3~mm continuum flux; (6) Type of the nuclear activity; (7) References for types of the nuclear activity: (a) \cite{ho1997}, (b) \cite{satyapal2008}, (c) \cite{mcalpine2011}, (d) \cite{dahlem1997}, (e) \cite{doi2011}} 
\end{deluxetable}

\section{Optical Images of Our Group Sample with Nondetections of CO} \label{app:nonco}
\begin{figure*}[!htbp]
\begin{center}
\includegraphics[width=0.79\textwidth]{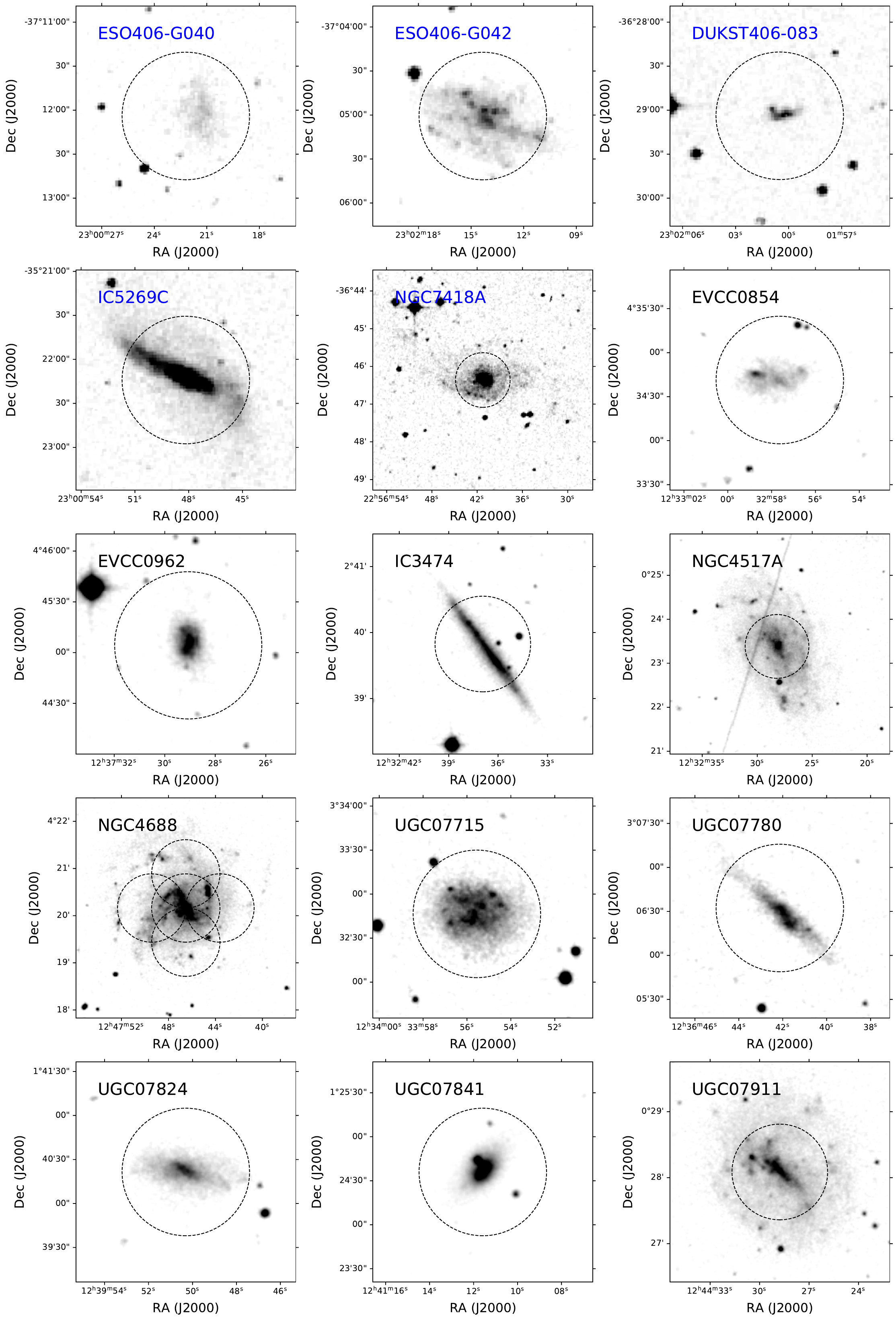}
\caption{Optical images (DSS2 blue) of the galaxies with non-detection of CO in the I1459G and the N4636G. The dashed circle(s) are the primary beam(s) of the ACA observations. The names of the group members in the I1459G are shown in blue color.   \label{fig:nonc0}}

\end{center}
\end{figure*}

Figure~\ref{fig:nonc0} shows optical images of group members with nondetection of CO in the I1459G and the N4636G.

\clearpage
\bibliographystyle{aasjournal}
\bibliography{group_co_paper_refs}

\begin{thebibliography}{}
\expandafter\ifx\csname natexlab\endcsname\relax\def\natexlab#1{#1}\fi
\providecommand{\url}[1]{\href{#1}{#1}}
\providecommand{\dodoi}[1]{doi:~\href{http://doi.org/#1}{\nolinkurl{#1}}}
\providecommand{\doeprint}[1]{\href{http://ascl.net/#1}{\nolinkurl{http://ascl.net/#1}}}
\providecommand{\doarXiv}[1]{\href{https://arxiv.org/abs/#1}{\nolinkurl{https://arxiv.org/abs/#1}}}

\bibitem[{{Accurso} {et~al.}(2017){Accurso}, {Saintonge}, {Catinella},
  {Cortese}, {Dav{\'e}}, {Dunsheath}, {Genzel}, {Gracia-Carpio}, {Heckman},
  {Jimmy}, {Kramer}, {Li}, {Lutz}, {Schiminovich}, {Schuster}, {Sternberg},
  {Sturm}, {Tacconi}, {Tran}, \& {Wang}}]{accurso2017}
{Accurso}, G., {Saintonge}, A., {Catinella}, B., {et~al.} 2017, \mnras, 470,
  4750, \dodoi{10.1093/mnras/stx1556}

\bibitem[{{Ahoranta} {et~al.}(2016){Ahoranta}, {Finoguenov}, {Pinto},
  {Sanders}, {Kaastra}, {de Plaa}, \& {Fabian}}]{ahoranta2016}
{Ahoranta}, J., {Finoguenov}, A., {Pinto}, C., {et~al.} 2016, \aap, 592, A145,
  \dodoi{10.1051/0004-6361/201527523}

\bibitem[{{Aihara} {et~al.}(2019){Aihara}, {AlSayyad}, {Ando}, {Armstrong},
  {Bosch}, {Egami}, {Furusawa}, {Furusawa}, {Goulding}, {Harikane}, {Hikage},
  {Ho}, {Hsieh}, {Huang}, {Ikeda}, {Imanishi}, {Ito}, {Iwata}, {Jaelani},
  {Kakuma}, {Kawana}, {Kikuta}, {Kobayashi}, {Koike}, {Komiyama}, {Li},
  {Liang}, {Lin}, {Luo}, {Lupton}, {Lust}, {MacArthur}, {Matsuoka}, {Mineo},
  {Miyatake}, {Miyazaki}, {More}, {Murata}, {Namiki}, {Nishizawa}, {Oguri},
  {Okabe}, {Okamoto}, {Okura}, {Ono}, {Onodera}, {Onoue}, {Osato}, {Ouchi},
  {Shibuya}, {Strauss}, {Sugiyama}, {Suto}, {Takada}, {Takagi}, {Takata},
  {Takita}, {Tanaka}, {Terai}, {Toba}, {Uchiyama}, {Utsumi}, {Wang}, {Wang}, \&
  {Yamada}}]{aihara2019}
{Aihara}, H., {AlSayyad}, Y., {Ando}, M., {et~al.} 2019, \pasj, 71, 114,
  \dodoi{10.1093/pasj/psz103}

\bibitem[{{Alatalo} {et~al.}(2013){Alatalo}, {Davis}, {Bureau}, {Young},
  {Blitz}, {Crocker}, {Bayet}, {Bois}, {Bournaud}, {Cappellari}, {Davies}, {de
  Zeeuw}, {Duc}, {Emsellem}, {Khochfar}, {Krajnovi{\'c}}, {Kuntschner},
  {Lablanche}, {Morganti}, {McDermid}, {Naab}, {Oosterloo}, {Sarzi}, {Scott},
  {Serra}, \& {Weijmans}}]{alatalo2013}
{Alatalo}, K., {Davis}, T.~A., {Bureau}, M., {et~al.} 2013, \mnras, 432, 1796,
  \dodoi{10.1093/mnras/sts299}

\bibitem[{{Alatalo} {et~al.}(2015){Alatalo}, {Appleton}, {Lisenfeld},
  {Bitsakis}, {Lanz}, {Lacy}, {Charmandaris}, {Cluver}, {Dopita}, {Guillard},
  {Jarrett}, {Kewley}, {Nyland}, {Ogle}, {Rasmussen}, {Rich},
  {Verdes-Montenegro}, {Xu}, \& {Yun}}]{alatalo2015}
{Alatalo}, K., {Appleton}, P.~N., {Lisenfeld}, U., {et~al.} 2015, \apj, 812,
  117, \dodoi{10.1088/0004-637X/812/2/117}

\bibitem[{{Astropy Collaboration} {et~al.}(2013){Astropy Collaboration},
  {Robitaille}, {Tollerud}, {Greenfield}, {Droettboom}, {Bray}, {Aldcroft},
  {Davis}, {Ginsburg}, {Price-Whelan}, {Kerzendorf}, {Conley}, {Crighton},
  {Barbary}, {Muna}, {Ferguson}, {Grollier}, {Parikh}, {Nair}, {Unther},
  {Deil}, {Woillez}, {Conseil}, {Kramer}, {Turner}, {Singer}, {Fox}, {Weaver},
  {Zabalza}, {Edwards}, {Azalee Bostroem}, {Burke}, {Casey}, {Crawford},
  {Dencheva}, {Ely}, {Jenness}, {Labrie}, {Lim}, {Pierfederici}, {Pontzen},
  {Ptak}, {Refsdal}, {Servillat}, \& {Streicher}}]{astropy:2013}
{Astropy Collaboration}, {Robitaille}, T.~P., {Tollerud}, E.~J., {et~al.} 2013,
  \aap, 558, A33, \dodoi{10.1051/0004-6361/201322068}

\bibitem[{{Astropy Collaboration} {et~al.}(2018){Astropy Collaboration},
  {Price-Whelan}, {Sip{\H{o}}cz}, {G{\"u}nther}, {Lim}, {Crawford}, {Conseil},
  {Shupe}, {Craig}, {Dencheva}, {Ginsburg}, {Vand erPlas}, {Bradley},
  {P{\'e}rez-Su{\'a}rez}, {de Val-Borro}, {Aldcroft}, {Cruz}, {Robitaille},
  {Tollerud}, {Ardelean}, {Babej}, {Bach}, {Bachetti}, {Bakanov}, {Bamford},
  {Barentsen}, {Barmby}, {Baumbach}, {Berry}, {Biscani}, {Boquien}, {Bostroem},
  {Bouma}, {Brammer}, {Bray}, {Breytenbach}, {Buddelmeijer}, {Burke},
  {Calderone}, {Cano Rodr{\'\i}guez}, {Cara}, {Cardoso}, {Cheedella}, {Copin},
  {Corrales}, {Crichton}, {D'Avella}, {Deil}, {Depagne}, {Dietrich}, {Donath},
  {Droettboom}, {Earl}, {Erben}, {Fabbro}, {Ferreira}, {Finethy}, {Fox},
  {Garrison}, {Gibbons}, {Goldstein}, {Gommers}, {Greco}, {Greenfield},
  {Groener}, {Grollier}, {Hagen}, {Hirst}, {Homeier}, {Horton}, {Hosseinzadeh},
  {Hu}, {Hunkeler}, {Ivezi{\'c}}, {Jain}, {Jenness}, {Kanarek}, {Kendrew},
  {Kern}, {Kerzendorf}, {Khvalko}, {King}, {Kirkby}, {Kulkarni}, {Kumar},
  {Lee}, {Lenz}, {Littlefair}, {Ma}, {Macleod}, {Mastropietro}, {McCully},
  {Montagnac}, {Morris}, {Mueller}, {Mumford}, {Muna}, {Murphy}, {Nelson},
  {Nguyen}, {Ninan}, {N{\"o}the}, {Ogaz}, {Oh}, {Parejko}, {Parley}, {Pascual},
  {Patil}, {Patil}, {Plunkett}, {Prochaska}, {Rastogi}, {Reddy Janga},
  {Sabater}, {Sakurikar}, {Seifert}, {Sherbert}, {Sherwood-Taylor}, {Shih},
  {Sick}, {Silbiger}, {Singanamalla}, {Singer}, {Sladen}, {Sooley},
  {Sornarajah}, {Streicher}, {Teuben}, {Thomas}, {Tremblay}, {Turner},
  {Terr{\'o}n}, {van Kerkwijk}, {de la Vega}, {Watkins}, {Weaver}, {Whitmore},
  {Woillez}, {Zabalza}, \& {Astropy Contributors}}]{astropy:2018}
{Astropy Collaboration}, {Price-Whelan}, A.~M., {Sip{\H{o}}cz}, B.~M., {et~al.}
  2018, \aj, 156, 123, \dodoi{10.3847/1538-3881/aabc4f}

\bibitem[{{Baldi} {et~al.}(2009){Baldi}, {Forman}, {Jones}, {Kraft}, {Nulsen},
  {Churazov}, {David}, \& {Giacintucci}}]{baldi2009}
{Baldi}, A., {Forman}, W., {Jones}, C., {et~al.} 2009, \apj, 707, 1034,
  \dodoi{10.1088/0004-637X/707/2/1034}

\bibitem[{{Balogh} {et~al.}(2007){Balogh}, {Wilman}, {Henderson}, {Bower},
  {Gilbank}, {Whitaker}, {Morris}, {Hau}, {Mulchaey}, {Oemler}, \&
  {Carlberg}}]{balogh2007}
{Balogh}, M.~L., {Wilman}, D., {Henderson}, R. D.~E., {et~al.} 2007, \mnras,
  374, 1169, \dodoi{10.1111/j.1365-2966.2006.11235.x}

\bibitem[{{Barnes}(1999)}]{barnes1999}
{Barnes}, D.~G. 1999, \pasa, 16, 77, \dodoi{10.1071/AS99077}

\bibitem[{{Bettoni} {et~al.}(2010){Bettoni}, {Buson}, \&
  {Galletta}}]{bettoni2010}
{Bettoni}, D., {Buson}, L.~M., \& {Galletta}, G. 2010, \aap, 519, A72,
  \dodoi{10.1051/0004-6361/201014750}

\bibitem[{{Bitsakis} {et~al.}(2014){Bitsakis}, {Charmandaris}, {Appleton},
  {D{\'\i}az-Santos}, {Le Floc'h}, {da Cunha}, {Alatalo}, \&
  {Cluver}}]{bitsakis2014}
{Bitsakis}, T., {Charmandaris}, V., {Appleton}, P.~N., {et~al.} 2014, \aap,
  565, A25, \dodoi{10.1051/0004-6361/201323349}

\bibitem[{{Boger} \& {Sternberg}(2005)}]{boger2005}
{Boger}, G.~I., \& {Sternberg}, A. 2005, \apj, 632, 302, \dodoi{10.1086/432864}

\bibitem[{{Bolatto} {et~al.}(2013){Bolatto}, {Wolfire}, \&
  {Leroy}}]{bolatto2013}
{Bolatto}, A.~D., {Wolfire}, M., \& {Leroy}, A.~K. 2013, \araa, 51, 207,
  \dodoi{10.1146/annurev-astro-082812-140944}

\bibitem[{{Bolatto} {et~al.}(2011){Bolatto}, {Leroy}, {Jameson}, {Ostriker},
  {Gordon}, {Lawton}, {Stanimirovi{\'c}}, {Israel}, {Madden}, {Hony},
  {Sandstrom}, {Bot}, {Rubio}, {Winkler}, {Roman-Duval}, {van Loon},
  {Oliveira}, \& {Indebetouw}}]{bolatto2011}
{Bolatto}, A.~D., {Leroy}, A.~K., {Jameson}, K., {et~al.} 2011, \apj, 741, 12,
  \dodoi{10.1088/0004-637X/741/1/12}

\bibitem[{{Bonnarel} {et~al.}(2000){Bonnarel}, {Fernique}, {Bienaym{\'e}},
  {Egret}, {Genova}, {Louys}, {Ochsenbein}, {Wenger}, \&
  {Bartlett}}]{bonnarel2000}
{Bonnarel}, F., {Fernique}, P., {Bienaym{\'e}}, O., {et~al.} 2000, \aaps, 143,
  33, \dodoi{10.1051/aas:2000331}

\bibitem[{{Boselli} {et~al.}(2014{\natexlab{a}}){Boselli}, {Cortese}, \&
  {Boquien}}]{boselli2014a}
{Boselli}, A., {Cortese}, L., \& {Boquien}, M. 2014{\natexlab{a}}, \aap, 564,
  A65, \dodoi{10.1051/0004-6361/201322311}

\bibitem[{{Boselli} {et~al.}(2014{\natexlab{b}}){Boselli}, {Cortese},
  {Boquien}, {Boissier}, {Catinella}, {Gavazzi}, {Lagos}, \&
  {Saintonge}}]{boselli2014b}
{Boselli}, A., {Cortese}, L., {Boquien}, M., {et~al.} 2014{\natexlab{b}}, \aap,
  564, A67, \dodoi{10.1051/0004-6361/201322313}

\bibitem[{{Bournaud}(2010)}]{bournaud2010}
{Bournaud}, F. 2010, in Astronomical Society of the Pacific Conference Series,
  Vol. 423, Galaxy Wars: Stellar Populations and Star Formation in Interacting
  Galaxies, ed. B.~{Smith}, J.~{Higdon}, S.~{Higdon}, \& N.~{Bastian}, 177,
  \dodoi{10.48550/arXiv.0909.1812}

\bibitem[{{Brough} {et~al.}(2006){Brough}, {Forbes}, {Kilborn}, \&
  {Couch}}]{brough2006}
{Brough}, S., {Forbes}, D.~A., {Kilborn}, V.~A., \& {Couch}, W. 2006, \mnras,
  370, 1223, \dodoi{10.1111/j.1365-2966.2006.10542.x}

\bibitem[{{Brown} {et~al.}(2017){Brown}, {Catinella}, {Cortese}, {Lagos},
  {Dav{\'e}}, {Kilborn}, {Haynes}, {Giovanelli}, \&
  {Rafieferantsoa}}]{brown2017}
{Brown}, T., {Catinella}, B., {Cortese}, L., {et~al.} 2017, \mnras, 466, 1275,
  \dodoi{10.1093/mnras/stw2991}

\bibitem[{{Brown} {et~al.}(2021){Brown}, {Wilson}, {Zabel}, {Davis}, {Boselli},
  {Chung}, {Ellison}, {Lagos}, {Stevens}, {Cortese}, {Bah{\'e}}, {Bisaria},
  {Bolatto}, {Cashmore}, {Catinella}, {Chown}, {Diemer}, {Elahi}, {Hani},
  {Jim{\'e}nez-Donaire}, {Lee}, {Leidig}, {Mok}, {Olsen}, {Parker}, {Roberts},
  {Smith}, {Spekkens}, {Thorp}, {Tonnesen}, {Vienneau}, {Villanueva}, {Vogel},
  {Wadsley}, {Welker}, \& {Yoon}}]{brown2021}
{Brown}, T., {Wilson}, C.~D., {Zabel}, N., {et~al.} 2021, \apjs, 257, 21,
  \dodoi{10.3847/1538-4365/ac28f5}

\bibitem[{{Buta} \& {Combes}(1996)}]{buta1996}
{Buta}, R., \& {Combes}, F. 1996, \fcp, 17, 95

\bibitem[{{Calzetti}(2013)}]{calzetti2013}
{Calzetti}, D. 2013, {Star Formation Rate Indicators}, ed.
  J.~{Falc{\'o}n-Barroso} \& J.~H. {Knapen}, 419

\bibitem[{{Castignani} {et~al.}(2022){Castignani}, {Combes}, {Jablonka},
  {Finn}, {Rudnick}, {Vulcani}, {Desai}, {Zaritsky}, \&
  {Salom{\'e}}}]{castignani2021}
{Castignani}, G., {Combes}, F., {Jablonka}, P., {et~al.} 2022, \aap, 657, A9,
  \dodoi{10.1051/0004-6361/202040141}

\bibitem[{{Catinella} {et~al.}(2018){Catinella}, {Saintonge}, {Janowiecki},
  {Cortese}, {Dav{\'e}}, {Lemonias}, {Cooper}, {Schiminovich}, {Hummels},
  {Fabello}, {Ger{\'e}b}, {Kilborn}, \& {Wang}}]{catinella2018}
{Catinella}, B., {Saintonge}, A., {Janowiecki}, S., {et~al.} 2018, \mnras, 476,
  875, \dodoi{10.1093/mnras/sty089}

\bibitem[{{Chabrier}(2003)}]{chabrier2003}
{Chabrier}, G. 2003, \pasp, 115, 763, \dodoi{10.1086/376392}

\bibitem[{{Chung} {et~al.}(2009){Chung}, {van Gorkom}, {Kenney}, {Crowl}, \&
  {Vollmer}}]{chung2009}
{Chung}, A., {van Gorkom}, J.~H., {Kenney}, J. D.~P., {Crowl}, H., \&
  {Vollmer}, B. 2009, \aj, 138, 1741, \dodoi{10.1088/0004-6256/138/6/1741}

\bibitem[{{Chung} \& {Kim}(2014)}]{chung2014}
{Chung}, E.~J., \& {Kim}, S. 2014, \pasj, 66, 11, \dodoi{10.1093/pasj/pst011}

\bibitem[{{Conselice} {et~al.}(2000){Conselice}, {Bershady}, \&
  {Jangren}}]{conselice2000}
{Conselice}, C.~J., {Bershady}, M.~A., \& {Jangren}, A. 2000, \apj, 529, 886,
  \dodoi{10.1086/308300}

\bibitem[{{Cortese} {et~al.}(2020){Cortese}, {Catinella}, {Cook}, \&
  {Janowiecki}}]{cortese2020}
{Cortese}, L., {Catinella}, B., {Cook}, R.~H.~W., \& {Janowiecki}, S. 2020,
  \mnras, 494, L42, \dodoi{10.1093/mnrasl/slaa032}

\bibitem[{{Cortese} {et~al.}(2021){Cortese}, {Catinella}, \&
  {Smith}}]{cortese2021}
{Cortese}, L., {Catinella}, B., \& {Smith}, R. 2021, \pasa, 38, e035,
  \dodoi{10.1017/pasa.2021.18}

\bibitem[{{Cramer} {et~al.}(2020){Cramer}, {Kenney}, {Cortes}, {Cortes P.~C.},
  {Vlahakis}, {J{\'a}chym}, {Pompei}, \& {Rubio}}]{cramer2020}
{Cramer}, W.~J., {Kenney}, J.~D.~P., {Cortes}, J.~R., {et~al.} 2020, \apj, 901,
  95, \dodoi{10.3847/1538-4357/abaf54}

\bibitem[{{Cybulski} {et~al.}(2014){Cybulski}, {Yun}, {Fazio}, \&
  {Gutermuth}}]{cybulski2014}
{Cybulski}, R., {Yun}, M.~S., {Fazio}, G.~G., \& {Gutermuth}, R.~A. 2014,
  \mnras, 439, 3564, \dodoi{10.1093/mnras/stu200}

\bibitem[{{Dahlem} {et~al.}(1997){Dahlem}, {Petr}, {Lehnert}, {Heckman}, \&
  {Ehle}}]{dahlem1997}
{Dahlem}, M., {Petr}, M.~G., {Lehnert}, M.~D., {Heckman}, T.~M., \& {Ehle}, M.
  1997, \aap, 320, 731

\bibitem[{{De Lucia} {et~al.}(2012){De Lucia}, {Weinmann}, {Poggianti},
  {Arag{\'o}n-Salamanca}, \& {Zaritsky}}]{delucia2012}
{De Lucia}, G., {Weinmann}, S., {Poggianti}, B.~M., {Arag{\'o}n-Salamanca}, A.,
  \& {Zaritsky}, D. 2012, \mnras, 423, 1277,
  \dodoi{10.1111/j.1365-2966.2012.20983.x}

\bibitem[{{D{\'e}nes} {et~al.}(2016){D{\'e}nes}, {Kilborn}, {Koribalski}, \&
  {Wong}}]{denes2016}
{D{\'e}nes}, H., {Kilborn}, V.~A., {Koribalski}, B.~S., \& {Wong}, O.~I. 2016,
  \mnras, 455, 1294, \dodoi{10.1093/mnras/stv2391}

\bibitem[{{Doi} {et~al.}(2011){Doi}, {Nakanishi}, {Nagai}, {Kohno}, \&
  {Kameno}}]{doi2011}
{Doi}, A., {Nakanishi}, K., {Nagai}, H., {Kohno}, K., \& {Kameno}, S. 2011,
  \aj, 142, 167, \dodoi{10.1088/0004-6256/142/5/167}

\bibitem[{{Dressler}(1980)}]{dressler1980}
{Dressler}, A. 1980, \apj, 236, 351, \dodoi{10.1086/157753}

\bibitem[{{D{\v{z}}ud{\v{z}}ar} {et~al.}(2019){D{\v{z}}ud{\v{z}}ar}, {Kilborn},
  {Murugeshan}, {Meurer}, {Sweet}, \& {Putman}}]{dzudzar2019}
{D{\v{z}}ud{\v{z}}ar}, R., {Kilborn}, V., {Murugeshan}, C., {et~al.} 2019,
  \mnras, 490, L6, \dodoi{10.1093/mnrasl/slz139}

\bibitem[{{Eke} {et~al.}(2004){Eke}, {Baugh}, {Cole}, {Frenk}, {Norberg},
  {Peacock}, {Baldry}, {Bland-Hawthorn}, {Bridges}, {Cannon}, {Colless},
  {Collins}, {Couch}, {Dalton}, {de Propris}, {Driver}, {Efstathiou}, {Ellis},
  {Glazebrook}, {Jackson}, {Lahav}, {Lewis}, {Lumsden}, {Maddox}, {Madgwick},
  {Peterson}, {Sutherland}, \& {Taylor}}]{eke2004}
{Eke}, V.~R., {Baugh}, C.~M., {Cole}, S., {et~al.} 2004, \mnras, 348, 866,
  \dodoi{10.1111/j.1365-2966.2004.07408.x}

\bibitem[{{Fern{\'a}ndez Lorenzo} {et~al.}(2013){Fern{\'a}ndez Lorenzo},
  {Sulentic}, {Verdes-Montenegro}, \& {Argudo-Fern{\'a}ndez}}]{lorenzo2013}
{Fern{\'a}ndez Lorenzo}, M., {Sulentic}, J., {Verdes-Montenegro}, L., \&
  {Argudo-Fern{\'a}ndez}, M. 2013, \mnras, 434, 325,
  \dodoi{10.1093/mnras/stt1020}

\bibitem[{{For} {et~al.}(2019){For}, {Staveley-Smith}, {Westmeier}, {Whiting},
  {Oh}, {Koribalski}, {Wang}, {Wong}, {Bekiaris}, {Cortese}, {Elagali},
  {Kleiner}, {Lee-Waddell}, {Madrid}, {Popping}, {Rhee}, {Reynolds}, {Collier},
  {Phillips}, {Voronkov}, {M{\"u}ller}, \& {Jerjen}}]{for2019}
{For}, B.~Q., {Staveley-Smith}, L., {Westmeier}, T., {et~al.} 2019, \mnras,
  489, 5723, \dodoi{10.1093/mnras/stz2501}

\bibitem[{{For} {et~al.}(2021){For}, {Wang}, {Westmeier}, {Wong}, {Murugeshan},
  {Staveley-Smith}, {Courtois}, {Pomar{\`e}de}, {Spekkens}, {Catinella},
  {McQuinn}, {Elagali}, {Koribalski}, {Lee-Waddell}, {Madrid}, {Popping},
  {Reynolds}, {Rhee}, {Bekki}, {D{\`e}nes}, {Kamphuis}, \&
  {Verdes-Montenegro}}]{for2021}
{For}, B.~Q., {Wang}, J., {Westmeier}, T., {et~al.} 2021, \mnras, 507, 2300,
  \dodoi{10.1093/mnras/stab2257}

\bibitem[{{Forbes} {et~al.}(2020){Forbes}, {Dullo}, {Gannon}, {Couch},
  {Iodice}, {Spavone}, {Cantiello}, \& {Schipani}}]{forbes2020}
{Forbes}, D.~A., {Dullo}, B.~T., {Gannon}, J., {et~al.} 2020, \mnras, 494,
  5293, \dodoi{10.1093/mnras/staa1111}

\bibitem[{{Forbes} {et~al.}(1994){Forbes}, {Franx}, \&
  {Illingworth}}]{forbes1994}
{Forbes}, D.~A., {Franx}, M., \& {Illingworth}, G.~D. 1994, \apjl, 428, L49,
  \dodoi{10.1086/187390}

\bibitem[{{Forbes} {et~al.}(1995){Forbes}, {Reitzel}, \&
  {Williger}}]{forbes1995}
{Forbes}, D.~A., {Reitzel}, D.~B., \& {Williger}, G.~M. 1995, \aj, 109, 1576,
  \dodoi{10.1086/117386}

\bibitem[{{Forbes} {et~al.}(2006){Forbes}, {Ponman}, {Pearce}, {Osmond},
  {Kilborn}, {Brough}, {Raychaudhury}, {Mundell}, {Miles}, \&
  {Kern}}]{forbes2006}
{Forbes}, D.~A., {Ponman}, T., {Pearce}, F., {et~al.} 2006, \pasa, 23, 38,
  \dodoi{10.1071/AS06002}

\bibitem[{{Franx} \& {Illingworth}(1988)}]{franx1988}
{Franx}, M., \& {Illingworth}, G.~D. 1988, \apjl, 327, L55,
  \dodoi{10.1086/185139}

\bibitem[{{George} {et~al.}(2019){George}, {Poggianti}, {Bellhouse},
  {Radovich}, {Fritz}, {Paladino}, {Bettoni}, {Jaff{\'e}}, {Moretti},
  {Gullieuszik}, {Vulcani}, {Fasano}, {Stalin}, {Subramaniam}, \&
  {Tandon}}]{george2019}
{George}, K., {Poggianti}, B.~M., {Bellhouse}, C., {et~al.} 2019, \mnras, 487,
  3102, \dodoi{10.1093/mnras/stz1443}

\bibitem[{{Giacintucci} {et~al.}(2011){Giacintucci}, {O'Sullivan}, {Vrtilek},
  {David}, {Raychaudhury}, {Venturi}, {Athreya}, {Clarke}, {Murgia},
  {Mazzotta}, {Gitti}, {Ponman}, {Ishwara-Chandra}, {Jones}, \&
  {Forman}}]{giacintucci2011}
{Giacintucci}, S., {O'Sullivan}, E., {Vrtilek}, J., {et~al.} 2011, \apj, 732,
  95, \dodoi{10.1088/0004-637X/732/2/95}

\bibitem[{{Giese} {et~al.}(2016){Giese}, {van der Hulst}, {Serra}, \&
  {Oosterloo}}]{giese2016}
{Giese}, N., {van der Hulst}, T., {Serra}, P., \& {Oosterloo}, T. 2016, \mnras,
  461, 1656, \dodoi{10.1093/mnras/stw1426}

\bibitem[{{G{\'o}mez} {et~al.}(2003){G{\'o}mez}, {Nichol}, {Miller}, {Balogh},
  {Goto}, {Zabludoff}, {Romer}, {Bernardi}, {Sheth}, {Hopkins}, {Castander},
  {Connolly}, {Schneider}, {Brinkmann}, {Lamb}, {SubbaRao}, \&
  {York}}]{gomez2003}
{G{\'o}mez}, P.~L., {Nichol}, R.~C., {Miller}, C.~J., {et~al.} 2003, \apj, 584,
  210, \dodoi{10.1086/345593}

\bibitem[{{Goto} {et~al.}(2003){Goto}, {Yamauchi}, {Fujita}, {Okamura},
  {Sekiguchi}, {Smail}, {Bernardi}, \& {Gomez}}]{goto2003}
{Goto}, T., {Yamauchi}, C., {Fujita}, Y., {et~al.} 2003, \mnras, 346, 601,
  \dodoi{10.1046/j.1365-2966.2003.07114.x}

\bibitem[{{Guo} {et~al.}(2021){Guo}, {Jones}, {Wang}, \& {Lin}}]{guo2021}
{Guo}, H., {Jones}, M.~G., {Wang}, J., \& {Lin}, L. 2021, \apj, 918, 53,
  \dodoi{10.3847/1538-4357/ac062e}

\bibitem[{{Haines} {et~al.}(2015){Haines}, {Pereira}, {Smith}, {Egami},
  {Babul}, {Finoguenov}, {Ziparo}, {McGee}, {Rawle}, {Okabe}, \&
  {Moran}}]{haines2015}
{Haines}, C.~P., {Pereira}, M.~J., {Smith}, G.~P., {et~al.} 2015, \apj, 806,
  101, \dodoi{10.1088/0004-637X/806/1/101}

\bibitem[{Harris {et~al.}(2020)Harris, Millman, van~der Walt, Gommers,
  Virtanen, Cournapeau, Wieser, Taylor, Berg, Smith, Kern, Picus, Hoyer, van
  Kerkwijk, Brett, Haldane, del R{\'{i}}o, Wiebe, Peterson,
  G{\'{e}}rard-Marchant, Sheppard, Reddy, Weckesser, Abbasi, Gohlke, \&
  Oliphant}]{harris2020array}
Harris, C.~R., Millman, K.~J., van~der Walt, S.~J., {et~al.} 2020, Nature, 585,
  357, \dodoi{10.1038/s41586-020-2649-2}

\bibitem[{{Haynes} {et~al.}(2000){Haynes}, {Jore}, {Barrett}, {Broeils}, \&
  {Murray}}]{haynes2000}
{Haynes}, M.~P., {Jore}, K.~P., {Barrett}, E.~A., {Broeils}, A.~H., \&
  {Murray}, B.~M. 2000, \aj, 120, 703, \dodoi{10.1086/301457}

\bibitem[{{Hess} \& {Wilcots}(2013)}]{hess2013}
{Hess}, K.~M., \& {Wilcots}, E.~M. 2013, \aj, 146, 124,
  \dodoi{10.1088/0004-6256/146/5/124}

\bibitem[{{Ho} {et~al.}(1997){Ho}, {Filippenko}, \& {Sargent}}]{ho1997}
{Ho}, L.~C., {Filippenko}, A.~V., \& {Sargent}, W. L.~W. 1997, \apjs, 112, 315,
  \dodoi{10.1086/313041}

\bibitem[{{Hogg} {et~al.}(2004){Hogg}, {Blanton}, {Brinchmann}, {Eisenstein},
  {Schlegel}, {Gunn}, {McKay}, {Rix}, {Bahcall}, {Brinkmann}, \&
  {Meiksin}}]{hogg2004}
{Hogg}, D.~W., {Blanton}, M.~R., {Brinchmann}, J., {et~al.} 2004, \apjl, 601,
  L29, \dodoi{10.1086/381749}

\bibitem[{{Holwerda} {et~al.}(2011{\natexlab{a}}){Holwerda}, {Pirzkal}, {de
  Blok}, {Bouchard}, {Blyth}, {van der Heyden}, \& {Elson}}]{holwerda2011a}
{Holwerda}, B.~W., {Pirzkal}, N., {de Blok}, W.~J.~G., {et~al.}
  2011{\natexlab{a}}, \mnras, 416, 2415,
  \dodoi{10.1111/j.1365-2966.2011.17683.x}

\bibitem[{{Holwerda} {et~al.}(2011{\natexlab{b}}){Holwerda}, {Pirzkal}, {de
  Blok}, \& {van Driel}}]{holwerda2011b}
{Holwerda}, B.~W., {Pirzkal}, N., {de Blok}, W.~J.~G., \& {van Driel}, W.
  2011{\natexlab{b}}, \mnras, 416, 2447,
  \dodoi{10.1111/j.1365-2966.2011.18662.x}

\bibitem[{{Houghton}(2015)}]{houghton2015}
{Houghton}, R.~C.~W. 2015, \mnras, 451, 3427, \dodoi{10.1093/mnras/stv1113}

\bibitem[{Hunter(2007)}]{Hunter:2007}
Hunter, J.~D. 2007, Computing in Science \& Engineering, 9, 90,
  \dodoi{10.1109/MCSE.2007.55}

\bibitem[{{Iodice} {et~al.}(2020){Iodice}, {Spavone}, {Cattapan}, {Bannikova},
  {Forbes}, {Rampazzo}, {Ciroi}, {Corsini}, {D'Ago}, {Oosterloo}, {Schipani},
  \& {Capaccioli}}]{iodice2020}
{Iodice}, E., {Spavone}, M., {Cattapan}, A., {et~al.} 2020, \aap, 635, A3,
  \dodoi{10.1051/0004-6361/201936435}

\bibitem[{{J{\'a}chym} {et~al.}(2019){J{\'a}chym}, {Kenney}, {Sun}, {Combes},
  {Cortese}, {Scott}, {Sivanandam}, {Brinks}, {Roediger}, {Palou{\v{s}}}, \&
  {Fumagalli}}]{jachym2019}
{J{\'a}chym}, P., {Kenney}, J. D.~P., {Sun}, M., {et~al.} 2019, \apj, 883, 145,
  \dodoi{10.3847/1538-4357/ab3e6c}

\bibitem[{{Jaff{\'e}} {et~al.}(2016){Jaff{\'e}}, {Verheijen}, {Haines}, {Yoon},
  {Cybulski}, {Montero-Casta{\~n}o}, {Smith}, {Chung}, {Deshev},
  {Fern{\'a}ndez}, {van Gorkom}, {Poggianti}, {Yun}, {Finoguenov}, {Smith}, \&
  {Okabe}}]{jaffe2016}
{Jaff{\'e}}, Y.~L., {Verheijen}, M. A.~W., {Haines}, C.~P., {et~al.} 2016,
  \mnras, 461, 1202, \dodoi{10.1093/mnras/stw984}

\bibitem[{{Jarrett} {et~al.}(2017){Jarrett}, {Cluver}, {Magoulas}, {Bilicki},
  {Alpaslan}, {Bland-Hawthorn}, {Brough}, {Brown}, {Croom}, {Driver},
  {Holwerda}, {Hopkins}, {Loveday}, {Norberg}, {Peacock}, {Popescu}, {Sadler},
  {Taylor}, {Tuffs}, \& {Wang}}]{jarrett2017}
{Jarrett}, T.~H., {Cluver}, M.~E., {Magoulas}, C., {et~al.} 2017, \apj, 836,
  182, \dodoi{10.3847/1538-4357/836/2/182}

\bibitem[{{Jogee} {et~al.}(2005){Jogee}, {Scoville}, \& {Kenney}}]{jogee2005}
{Jogee}, S., {Scoville}, N., \& {Kenney}, J. D.~P. 2005, \apj, 630, 837,
  \dodoi{10.1086/432106}

\bibitem[{{Juneau} {et~al.}(2009){Juneau}, {Narayanan}, {Moustakas}, {Shirley},
  {Bussmann}, {Kennicutt}, \& {Vanden Bout}}]{juneau2009}
{Juneau}, S., {Narayanan}, D.~T., {Moustakas}, J., {et~al.} 2009, \apj, 707,
  1217, \dodoi{10.1088/0004-637X/707/2/1217}

\bibitem[{{Jung} {et~al.}(2018){Jung}, {Choi}, {Wong}, {Kimm}, {Chung}, \&
  {Yi}}]{jung2018}
{Jung}, S.~L., {Choi}, H., {Wong}, O.~I., {et~al.} 2018, \apj, 865, 156,
  \dodoi{10.3847/1538-4357/aadda2}

\bibitem[{{Kaneko} {et~al.}(2017){Kaneko}, {Kuno}, {Iono}, {Tamura}, {Tosaki},
  {Nakanishi}, \& {Sawada}}]{kaneko2017}
{Kaneko}, H., {Kuno}, N., {Iono}, D., {et~al.} 2017, \pasj, 69, 66,
  \dodoi{10.1093/pasj/psx041}

\bibitem[{{Kantharia} {et~al.}(2005){Kantharia}, {Ananthakrishnan},
  {Nityananda}, \& {Hota}}]{kantharia2005}
{Kantharia}, N.~G., {Ananthakrishnan}, S., {Nityananda}, R., \& {Hota}, A.
  2005, \aap, 435, 483, \dodoi{10.1051/0004-6361:20042261}

\bibitem[{{Kauffmann} {et~al.}(2004){Kauffmann}, {White}, {Heckman},
  {M{\'e}nard}, {Brinchmann}, {Charlot}, {Tremonti}, \&
  {Brinkmann}}]{Kauffmann2004}
{Kauffmann}, G., {White}, S. D.~M., {Heckman}, T.~M., {et~al.} 2004, \mnras,
  353, 713, \dodoi{10.1111/j.1365-2966.2004.08117.x}

\bibitem[{{Kilborn} {et~al.}(2009){Kilborn}, {Forbes}, {Barnes}, {Koribalski},
  {Brough}, \& {Kern}}]{kilborn2009}
{Kilborn}, V.~A., {Forbes}, D.~A., {Barnes}, D.~G., {et~al.} 2009, \mnras, 400,
  1962, \dodoi{10.1111/j.1365-2966.2009.15587.x}

\bibitem[{{Kilborn} {et~al.}(2005){Kilborn}, {Koribalski}, {Forbes}, {Barnes},
  \& {Musgrave}}]{kilborn2005}
{Kilborn}, V.~A., {Koribalski}, B.~S., {Forbes}, D.~A., {Barnes}, D.~G., \&
  {Musgrave}, R.~C. 2005, \mnras, 356, 77,
  \dodoi{10.1111/j.1365-2966.2004.08450.x}

\bibitem[{{Kleiner} {et~al.}(2021){Kleiner}, {Serra}, {Maccagni}, {Venhola},
  {Morokuma-Matsui}, {Peletier}, {Iodice}, {Raj}, {de Blok}, {Comrie},
  {J{\'o}zsa}, {Kamphuis}, {Loni}, {Loubser}, {Moln{\'a}r}, {Passmoor},
  {Ramatsoku}, {Sivitilli}, {Smirnov}, {Thorat}, \& {Vitello}}]{kleiner2021}
{Kleiner}, D., {Serra}, P., {Maccagni}, F.~M., {et~al.} 2021, \aap, 648, A32,
  \dodoi{10.1051/0004-6361/202039898}

\bibitem[{{Koribalski} {et~al.}(2020){Koribalski}, {Staveley-Smith},
  {Westmeier}, {Serra}, {Spekkens}, {Wong}, {Lee-Waddell}, {Lagos},
  {Obreschkow}, {Ryan-Weber}, {Zwaan}, {Kilborn}, {Bekiaris}, {Bekki},
  {Bigiel}, {Boselli}, {Bosma}, {Catinella}, {Chauhan}, {Cluver}, {Colless},
  {Courtois}, {Crain}, {de Blok}, {D{\'e}nes}, {Duffy}, {Elagali}, {Fluke},
  {For}, {Heald}, {Henning}, {Hess}, {Holwerda}, {Howlett}, {Jarrett}, {Jones},
  {Jones}, {J{\'o}zsa}, {Jurek}, {J{\"u}tte}, {Kamphuis}, {Karachentsev},
  {Kerp}, {Kleiner}, {Kraan-Korteweg}, {L{\'o}pez-S{\'a}nchez}, {Madrid},
  {Meyer}, {Mould}, {Murugeshan}, {Norris}, {Oh}, {Oosterloo}, {Popping},
  {Putman}, {Reynolds}, {Rhee}, {Robotham}, {Ryder}, {Schr{\"o}der}, {Shao},
  {Stevens}, {Taylor}, {van{\^A} der Hulst}, {Verdes-Montenegro}, {Wakker},
  {Wang}, {Whiting}, {Winkel}, \& {Wolf}}]{koribalski2020}
{Koribalski}, B.~S., {Staveley-Smith}, L., {Westmeier}, T., {et~al.} 2020,
  \apss, 365, 118, \dodoi{10.1007/s10509-020-03831-4}

\bibitem[{{Kroupa}(2001)}]{kroupa2001}
{Kroupa}, P. 2001, \mnras, 322, 231, \dodoi{10.1046/j.1365-8711.2001.04022.x}

\bibitem[{{Krumholz} {et~al.}(2009){Krumholz}, {McKee}, \&
  {Tumlinson}}]{krumholz2009}
{Krumholz}, M.~R., {McKee}, C.~F., \& {Tumlinson}, J. 2009, \apj, 693, 216,
  \dodoi{10.1088/0004-637X/693/1/216}

\bibitem[{{Lee} \& {Chung}(2018)}]{lee2018}
{Lee}, B., \& {Chung}, A. 2018, \apjl, 866, L10,
  \dodoi{10.3847/2041-8213/aae4d9}

\bibitem[{{Lee} {et~al.}(2017){Lee}, {Chung}, {Tonnesen}, {Kenney}, {Wong},
  {Vollmer}, {Petitpas}, {Crowl}, \& {van Gorkom}}]{lee2017}
{Lee}, B., {Chung}, A., {Tonnesen}, S., {et~al.} 2017, \mnras, 466, 1382,
  \dodoi{10.1093/mnras/stw3162}

\bibitem[{{Lee-Waddell} {et~al.}(2019){Lee-Waddell}, {Koribalski}, {Westmeier},
  {Elagali}, {For}, {Kleiner}, {Madrid}, {Popping}, {Reynolds}, {Rhee},
  {Serra}, {Shao}, {Staveley-Smith}, {Wang}, {Whiting}, {Wong}, {Allison},
  {Bhandari}, {Collier}, {Heald}, {Marvil}, \& {Ord}}]{leewaddell2019}
{Lee-Waddell}, K., {Koribalski}, B.~S., {Westmeier}, T., {et~al.} 2019, \mnras,
  487, 5248, \dodoi{10.1093/mnras/stz017}

\bibitem[{{Leroy} {et~al.}(2005){Leroy}, {Bolatto}, {Simon}, \&
  {Blitz}}]{leroy2005}
{Leroy}, A., {Bolatto}, A.~D., {Simon}, J.~D., \& {Blitz}, L. 2005, \apj, 625,
  763, \dodoi{10.1086/429578}

\bibitem[{{Leroy} {et~al.}(2021){Leroy}, {Hughes}, {Liu}, {Pety}, {Rosolowsky},
  {Saito}, {Schinnerer}, {Schruba}, {Usero}, {Faesi}, {Herrera}, {Chevance},
  {Hygate}, {Kepley}, {Koch}, {Querejeta}, {Sliwa}, {Will}, {Wilson}, {Anand},
  {Barnes}, {Belfiore}, {Be{\v{s}}li{\'c}}, {Bigiel}, {Blanc}, {Bolatto},
  {Boquien}, {Cao}, {Chandar}, {Chastenet}, {Chiang}, {Congiu}, {Dale},
  {Deger}, {den Brok}, {Eibensteiner}, {Emsellem},
  {Garc{\'\i}a-Rodr{\'\i}guez}, {Glover}, {Grasha}, {Groves}, {Henshaw},
  {Jim{\'e}nez Donaire}, {Kim}, {Klessen}, {Kreckel}, {Kruijssen}, {Larson},
  {Lee}, {Mayker}, {McElroy}, {Meidt}, {Mok}, {Pan}, {Puschnig}, {Razza},
  {S{\'a}nchez-Bl'azquez}, {Sandstrom}, {Santoro}, {Sardone}, {Scheuermann},
  {Sun}, {Thilker}, {Turner}, {Ubeda}, {Utomo}, {Watkins}, \&
  {Williams}}]{leroy2021}
{Leroy}, A.~K., {Hughes}, A., {Liu}, D., {et~al.} 2021, \apjs, 255, 19,
  \dodoi{10.3847/1538-4365/abec80}

\bibitem[{{Lewis} {et~al.}(2002){Lewis}, {Balogh}, {De Propris}, {Couch},
  {Bower}, {Offer}, {Bland-Hawthorn}, {Baldry}, {Baugh}, {Bridges}, {Cannon},
  {Cole}, {Colless}, {Collins}, {Cross}, {Dalton}, {Driver}, {Efstathiou},
  {Ellis}, {Frenk}, {Glazebrook}, {Hawkins}, {Jackson}, {Lahav}, {Lumsden},
  {Maddox}, {Madgwick}, {Norberg}, {Peacock}, {Percival}, {Peterson},
  {Sutherland}, \& {Taylor}}]{lewis2002}
{Lewis}, I., {Balogh}, M., {De Propris}, R., {et~al.} 2002, \mnras, 334, 673,
  \dodoi{10.1046/j.1365-8711.2002.05558.x}

\bibitem[{{Lin} {et~al.}(2023){Lin}, {Wang}, {Kilborn}, {Peng}, {Cortese},
  {Boselli}, {Liang}, {Lee}, {Yang}, {Catinella}, {Deg}, {D{\'e}nes},
  {Elagali}, {Kamphuis}, {Koribalski}, {Lee-Waddell}, {Rhee}, {Shao},
  {Spekkens}, {Staveley-Smith}, {Westmeier}, {Wong}, {Bekki}, {Bosma}, {Du},
  {Ho}, {Madrid}, {Verdes-Montenegro}, {Wang}, \& {Wang}}]{lin2023}
{Lin}, X., {Wang}, J., {Kilborn}, V., {et~al.} 2023, arXiv e-prints,
  arXiv:2304.09795, \dodoi{10.48550/arXiv.2304.09795}

\bibitem[{{Liz{\'e}e} {et~al.}(2021){Liz{\'e}e}, {Vollmer}, {Braine}, \&
  {Nehlig}}]{lizee2021}
{Liz{\'e}e}, T., {Vollmer}, B., {Braine}, J., \& {Nehlig}, F. 2021, \aap, 645,
  A111, \dodoi{10.1051/0004-6361/202038910}

\bibitem[{{Makarov} {et~al.}(2014){Makarov}, {Prugniel}, {Terekhova},
  {Courtois}, \& {Vauglin}}]{makarov2014}
{Makarov}, D., {Prugniel}, P., {Terekhova}, N., {Courtois}, H., \& {Vauglin},
  I. 2014, \aap, 570, A13, \dodoi{10.1051/0004-6361/201423496}

\bibitem[{{Martin} {et~al.}(2005){Martin}, {Fanson}, {Schiminovich},
  {Morrissey}, {Friedman}, {Barlow}, {Conrow}, {Grange}, {Jelinsky},
  {Milliard}, {Siegmund}, {Bianchi}, {Byun}, {Donas}, {Forster}, {Heckman},
  {Lee}, {Madore}, {Malina}, {Neff}, {Rich}, {Small}, {Surber}, {Szalay},
  {Welsh}, \& {Wyder}}]{martin2005}
{Martin}, D.~C., {Fanson}, J., {Schiminovich}, D., {et~al.} 2005, \apjl, 619,
  L1, \dodoi{10.1086/426387}

\bibitem[{{Martinez-Badenes} {et~al.}(2012){Martinez-Badenes}, {Lisenfeld},
  {Espada}, {Verdes-Montenegro}, {Garc{\'\i}a-Burillo}, {Leon}, {Sulentic}, \&
  {Yun}}]{martinez-badenes2012}
{Martinez-Badenes}, V., {Lisenfeld}, U., {Espada}, D., {et~al.} 2012, \aap,
  540, A96, \dodoi{10.1051/0004-6361/201117281}

\bibitem[{{Matsushita} {et~al.}(1998){Matsushita}, {Makishima}, {Ikebe},
  {Rokutanda}, {Yamasaki}, \& {Ohashi}}]{matsushita1998}
{Matsushita}, K., {Makishima}, K., {Ikebe}, Y., {et~al.} 1998, \apjl, 499, L13,
  \dodoi{10.1086/311339}

\bibitem[{{McAlpine} {et~al.}(2011){McAlpine}, {Satyapal}, {Gliozzi}, {Cheung},
  {Sambruna}, \& {Eracleous}}]{mcalpine2011}
{McAlpine}, W., {Satyapal}, S., {Gliozzi}, M., {et~al.} 2011, \apj, 728, 25,
  \dodoi{10.1088/0004-637X/728/1/25}

\bibitem[{{McGee} {et~al.}(2009){McGee}, {Balogh}, {Bower}, {Font}, \&
  {McCarthy}}]{mcgee2009}
{McGee}, S.~L., {Balogh}, M.~L., {Bower}, R.~G., {Font}, A.~S., \& {McCarthy},
  I.~G. 2009, \mnras, 400, 937, \dodoi{10.1111/j.1365-2966.2009.15507.x}

\bibitem[{{McMullin} {et~al.}(2007){McMullin}, {Waters}, {Schiebel}, {Young},
  \& {Golap}}]{mcmullin2007}
{McMullin}, J.~P., {Waters}, B., {Schiebel}, D., {Young}, W., \& {Golap}, K.
  2007, in Astronomical Society of the Pacific Conference Series, Vol. 376,
  Astronomical Data Analysis Software and Systems XVI, ed. R.~A. {Shaw},
  F.~{Hill}, \& D.~J. {Bell}, 127

\bibitem[{{Miles} {et~al.}(2004){Miles}, {Raychaudhury}, {Forbes},
  {Goudfrooij}, {Ponman}, \& {Kozhurina-Platais}}]{miles2004}
{Miles}, T.~A., {Raychaudhury}, S., {Forbes}, D.~A., {et~al.} 2004, \mnras,
  355, 785, \dodoi{10.1111/j.1365-2966.2004.08356.x}

\bibitem[{{Mok} {et~al.}(2017){Mok}, {Wilson}, {Knapen}, {S{\'a}nchez-Gallego},
  {Brinks}, \& {Rosolowsky}}]{mok2017}
{Mok}, A., {Wilson}, C.~D., {Knapen}, J.~H., {et~al.} 2017, \mnras, 467, 4282,
  \dodoi{10.1093/mnras/stx345}

\bibitem[{{Mok} {et~al.}(2016){Mok}, {Wilson}, {Golding}, {Warren}, {Israel},
  {Serjeant}, {Knapen}, {S{\'a}nchez-Gallego}, {Barmby}, {Bendo}, {Rosolowsky},
  \& {van der Werf}}]{mok2016}
{Mok}, A., {Wilson}, C.~D., {Golding}, J., {et~al.} 2016, \mnras, 456, 4384,
  \dodoi{10.1093/mnras/stv2958}

\bibitem[{{Moretti} {et~al.}(2020{\natexlab{a}}){Moretti}, {Paladino},
  {Poggianti}, {Serra}, {Roediger}, {Gullieuszik}, {Tomi{\v{c}}i{\'c}},
  {Radovich}, {Vulcani}, {Jaff{\'e}}, {Fritz}, {Bettoni}, {Ramatsoku}, \&
  {Wolter}}]{moretti2020a}
{Moretti}, A., {Paladino}, R., {Poggianti}, B.~M., {et~al.} 2020{\natexlab{a}},
  \apj, 889, 9, \dodoi{10.3847/1538-4357/ab616a}

\bibitem[{{Moretti} {et~al.}(2020{\natexlab{b}}){Moretti}, {Paladino},
  {Poggianti}, {Serra}, {Ramatsoku}, {Franchetto}, {Deb}, {Gullieuszik},
  {Tomi{\v{c}}i{\'c}}, {Mingozzi}, {Vulcani}, {Radovich}, {Bettoni}, \&
  {Fritz}}]{moretti2020b}
---. 2020{\natexlab{b}}, \apjl, 897, L30, \dodoi{10.3847/2041-8213/ab9f3b}

\bibitem[{{Morokuma-Matsui} {et~al.}(2021){Morokuma-Matsui}, {Kodama},
  {Morokuma}, {Nakanishi}, {Koyama}, {Yamashita}, {Koyama}, \&
  {Okamoto}}]{morokuma-matsui2021}
{Morokuma-Matsui}, K., {Kodama}, T., {Morokuma}, T., {et~al.} 2021, \apj, 914,
  145, \dodoi{10.3847/1538-4357/abedb6}

\bibitem[{{Mun} {et~al.}(2021){Mun}, {Hwang}, {Lee}, {Chung}, {Yoon}, \&
  {Lee}}]{mun2021}
{Mun}, J.~Y., {Hwang}, H.~S., {Lee}, M.~G., {et~al.} 2021, Journal of Korean
  Astronomical Society, 54, 17, \dodoi{10.5303/JKAS.2021.54.1.17}

\bibitem[{{Nolthenius}(1993)}]{nolthenius1993}
{Nolthenius}, R. 1993, \apjs, 85, 1, \dodoi{10.1086/191753}

\bibitem[{{Oosterloo} {et~al.}(2018){Oosterloo}, {Zhang}, {Lucero}, \&
  {Carignan}}]{oosterloo2018}
{Oosterloo}, T.~A., {Zhang}, M.~L., {Lucero}, D.~M., \& {Carignan}, C. 2018,
  arXiv e-prints, arXiv:1803.08263.
\newblock \doarXiv{1803.08263}

\bibitem[{{Osmond} \& {Ponman}(2004)}]{osmond2004}
{Osmond}, J. P.~F., \& {Ponman}, T.~J. 2004, \mnras, 350, 1511,
  \dodoi{10.1111/j.1365-2966.2004.07742.x}

\bibitem[{{O'Sullivan} {et~al.}(2018){O'Sullivan}, {Combes}, {Salom{\'e}},
  {David}, {Babul}, {Vrtilek}, {Lim}, {Olivares}, {Raychaudhury}, \&
  {Schellenberger}}]{osullivan2018}
{O'Sullivan}, E., {Combes}, F., {Salom{\'e}}, P., {et~al.} 2018, \aap, 618,
  A126, \dodoi{10.1051/0004-6361/201833580}

\bibitem[{{Pan} {et~al.}(2018){Pan}, {Lin}, {Hsieh}, {Xiao}, {Gao}, {Ellison},
  {Scudder}, {Barrera-Ballesteros}, {Yuan}, {Saintonge}, {Wilson}, {Hwang}, {De
  Looze}, {Gao}, {Ho}, {Brinks}, {Mok}, {Brown}, {Davis}, {Williams}, {Chung},
  {Parsons}, {Bureau}, {Sargent}, {Chung}, {Kim}, {Liu}, {Micha{\l}owski}, \&
  {Tosaki}}]{pan2018}
{Pan}, H.-A., {Lin}, L., {Hsieh}, B.-C., {et~al.} 2018, \apj, 868, 132,
  \dodoi{10.3847/1538-4357/aaeb92}

\bibitem[{{Persic} {et~al.}(2004){Persic}, {Cappi}, {Rephaeli}, {Bassani},
  {Della Ceca}, {Franceschini}, {Hunt}, {Malaguti}, \& {Palazzi}}]{persic2004}
{Persic}, M., {Cappi}, M., {Rephaeli}, Y., {et~al.} 2004, \aap, 427, 35,
  \dodoi{10.1051/0004-6361:20041128}

\bibitem[{{Pisano}(2004)}]{pisano2004}
{Pisano}, D.~J. 2004, \pasa, 21, 390, \dodoi{10.1071/AS04055}

\bibitem[{{Prichard} {et~al.}(2019){Prichard}, {Vaughan}, \&
  {Davies}}]{prichard2019}
{Prichard}, L.~J., {Vaughan}, S.~P., \& {Davies}, R.~L. 2019, \mnras, 488,
  1679, \dodoi{10.1093/mnras/stz1191}

\bibitem[{{Rasmussen} {et~al.}(2006){Rasmussen}, {Ponman}, \&
  {Mulchaey}}]{rasmussen2006}
{Rasmussen}, J., {Ponman}, T.~J., \& {Mulchaey}, J.~S. 2006, \mnras, 370, 453,
  \dodoi{10.1111/j.1365-2966.2006.10492.x}

\bibitem[{{Renaud} {et~al.}(2019){Renaud}, {Bournaud}, {Agertz}, {Kraljic},
  {Schinnerer}, {Bolatto}, {Daddi}, \& {Hughes}}]{renaud2019}
{Renaud}, F., {Bournaud}, F., {Agertz}, O., {et~al.} 2019, \aap, 625, A65,
  \dodoi{10.1051/0004-6361/201935222}

\bibitem[{{Reynolds} {et~al.}(2020){Reynolds}, {Westmeier}, {Staveley-Smith},
  {Chauhan}, \& {Lagos}}]{reynolds2020}
{Reynolds}, T.~N., {Westmeier}, T., {Staveley-Smith}, L., {Chauhan}, G., \&
  {Lagos}, C.~D.~P. 2020, \mnras, 493, 5089, \dodoi{10.1093/mnras/staa597}

\bibitem[{{Roberts} \& {Parker}(2017)}]{robert2017}
{Roberts}, I.~D., \& {Parker}, L.~C. 2017, \mnras, 467, 3268,
  \dodoi{10.1093/mnras/stx317}

\bibitem[{{Roberts} {et~al.}(2021){Roberts}, {van Weeren}, {McGee}, {Botteon},
  {Ignesti}, \& {Rottgering}}]{roberts2021}
{Roberts}, I.~D., {van Weeren}, R.~J., {McGee}, S.~L., {et~al.} 2021, \aap,
  652, A153, \dodoi{10.1051/0004-6361/202141118}

\bibitem[{{Robotham} {et~al.}(2011){Robotham}, {Norberg}, {Driver}, {Baldry},
  {Bamford}, {Hopkins}, {Liske}, {Loveday}, {Merson}, {Peacock}, {Brough},
  {Cameron}, {Conselice}, {Croom}, {Frenk}, {Gunawardhana}, {Hill}, {Jones},
  {Kelvin}, {Kuijken}, {Nichol}, {Parkinson}, {Pimbblet}, {Phillipps},
  {Popescu}, {Prescott}, {Sharp}, {Sutherland}, {Taylor}, {Thomas}, {Tuffs},
  {van Kampen}, \& {Wijesinghe}}]{robotham2011}
{Robotham}, A.~S.~G., {Norberg}, P., {Driver}, S.~P., {et~al.} 2011, \mnras,
  416, 2640, \dodoi{10.1111/j.1365-2966.2011.19217.x}

\bibitem[{{Rodriguez-Franco} {et~al.}(1998){Rodriguez-Franco},
  {Martin-Pintado}, \& {Fuente}}]{rodriguez-franco1998}
{Rodriguez-Franco}, A., {Martin-Pintado}, J., \& {Fuente}, A. 1998, \aap, 329,
  1097

\bibitem[{{Roychowdhury} {et~al.}(2022){Roychowdhury}, {Meyer}, {Rhee},
  {Zwaan}, {Chauhan}, {Davies}, {Bellstedt}, {Driver}, {Lagos}, {Robotham},
  {Bland-Hawthorn}, {Dodson}, {Holwerda}, {Hopkins}, {Lara-L{\'o}pez},
  {L{\'o}pez-S{\'a}nchez}, {Obreschkow}, {Rozgonyi}, {Whiting}, \&
  {Wright}}]{roychowdhury2022}
{Roychowdhury}, S., {Meyer}, M.~J., {Rhee}, J., {et~al.} 2022, \apj, 927, 20,
  \dodoi{10.3847/1538-4357/ac49ea}

\bibitem[{{Ryder} {et~al.}(1997){Ryder}, {Purcell}, {Davis}, \&
  {Andersen}}]{ryder1997}
{Ryder}, S.~D., {Purcell}, G., {Davis}, D., \& {Andersen}, V. 1997, \pasa, 14,
  81, \dodoi{10.1071/AS97081}

\bibitem[{{Saintonge} {et~al.}(2016){Saintonge}, {Catinella}, {Cortese},
  {Genzel}, {Giovanelli}, {Haynes}, {Janowiecki}, {Kramer}, {Lutz},
  {Schiminovich}, {Tacconi}, {Wuyts}, \& {Accurso}}]{saintonge2016}
{Saintonge}, A., {Catinella}, B., {Cortese}, L., {et~al.} 2016, \mnras, 462,
  1749, \dodoi{10.1093/mnras/stw1715}

\bibitem[{{Saintonge} {et~al.}(2017){Saintonge}, {Catinella}, {Tacconi},
  {Kauffmann}, {Genzel}, {Cortese}, {Dav{\'e}}, {Fletcher},
  {Graci{\'a}-Carpio}, {Kramer}, {Heckman}, {Janowiecki}, {Lutz}, {Rosario},
  {Schiminovich}, {Schuster}, {Wang}, {Wuyts}, {Borthakur}, {Lamperti}, \&
  {Roberts-Borsani}}]{saintonge2017}
{Saintonge}, A., {Catinella}, B., {Tacconi}, L.~J., {et~al.} 2017, \apjs, 233,
  22, \dodoi{10.3847/1538-4365/aa97e0}

\bibitem[{{Salim} {et~al.}(2007){Salim}, {Rich}, {Charlot}, {Brinchmann},
  {Johnson}, {Schiminovich}, {Seibert}, {Mallery}, {Heckman}, {Forster},
  {Friedman}, {Martin}, {Morrissey}, {Neff}, {Small}, {Wyder}, {Bianchi},
  {Donas}, {Lee}, {Madore}, {Milliard}, {Szalay}, {Welsh}, \& {Yi}}]{salim2007}
{Salim}, S., {Rich}, R.~M., {Charlot}, S., {et~al.} 2007, \apjs, 173, 267,
  \dodoi{10.1086/519218}

\bibitem[{{Salo} {et~al.}(2015){Salo}, {Laurikainen}, {Laine}, {Comer{\'o}n},
  {Gadotti}, {Buta}, {Sheth}, {Zaritsky}, {Ho}, {Knapen}, {Athanassoula},
  {Bosma}, {Laine}, {Cisternas}, {Kim}, {Mu{\~n}oz-Mateos}, {Regan}, {Hinz},
  {Gil de Paz}, {Menendez-Delmestre}, {Mizusawa}, {Erroz-Ferrer}, {Meidt}, \&
  {Querejeta}}]{salo2015}
{Salo}, H., {Laurikainen}, E., {Laine}, J., {et~al.} 2015, \apjs, 219, 4,
  \dodoi{10.1088/0067-0049/219/1/4}

\bibitem[{{S{\'a}nchez} {et~al.}(2017){S{\'a}nchez}, {Barrera-Ballesteros},
  {S{\'a}nchez-Menguiano}, {Walcher}, {Marino}, {Galbany}, {Bland-Hawthorn},
  {Cano-D{\'\i}az}, {Garc{\'\i}a-Benito}, {L{\'o}pez-Cob{\'a}}, {Zibetti},
  {Vilchez}, {Igl{\'e}sias-P{\'a}ramo}, {Kehrig}, {L{\'o}pez S{\'a}nchez},
  {Duarte Puertas}, \& {Ziegler}}]{sanchez2017}
{S{\'a}nchez}, S.~F., {Barrera-Ballesteros}, J.~K., {S{\'a}nchez-Menguiano},
  L., {et~al.} 2017, \mnras, 469, 2121, \dodoi{10.1093/mnras/stx808}

\bibitem[{{Saponara} {et~al.}(2018){Saponara}, {Koribalski}, {Benaglia}, \&
  {Fern{\'a}ndez L{\'o}pez}}]{saponara2018}
{Saponara}, J., {Koribalski}, B.~S., {Benaglia}, P., \& {Fern{\'a}ndez
  L{\'o}pez}, M. 2018, \mnras, 473, 3358, \dodoi{10.1093/mnras/stx2475}

\bibitem[{{Satyapal} {et~al.}(2008){Satyapal}, {Vega}, {Dudik}, {Abel}, \&
  {Heckman}}]{satyapal2008}
{Satyapal}, S., {Vega}, D., {Dudik}, R.~P., {Abel}, N.~P., \& {Heckman}, T.
  2008, \apj, 677, 926, \dodoi{10.1086/529014}

\bibitem[{{Schaye}(2004)}]{schaye2004}
{Schaye}, J. 2004, \apj, 609, 667, \dodoi{10.1086/421232}

\bibitem[{{Schruba} {et~al.}(2011){Schruba}, {Leroy}, {Walter}, {Bigiel},
  {Brinks}, {de Blok}, {Dumas}, {Kramer}, {Rosolowsky}, {Sand strom},
  {Schuster}, {Usero}, {Weiss}, \& {Wiesemeyer}}]{schruba2011}
{Schruba}, A., {Leroy}, A.~K., {Walter}, F., {et~al.} 2011, \aj, 142, 37,
  \dodoi{10.1088/0004-6256/142/2/37}

\bibitem[{{Serra} {et~al.}(2015{\natexlab{a}}){Serra}, {Koribalski}, {Kilborn},
  {Allison}, {Amy}, {Ball}, {Bannister}, {Bell}, {Bock}, {Bolton}, {Bowen},
  {Boyle}, {Broadhurst}, {Brodrick}, {Brothers}, {Bunton}, {Chapman}, {Cheng},
  {Chippendale}, {Chung}, {Cooray}, {Cornwell}, {DeBoer}, {Diamond}, {Forsyth},
  {Gough}, {Gupta}, {Hampson}, {Harvey-Smith}, {Hay}, {Hayman}, {Heywood},
  {Hotan}, {Hoyle}, {Humphreys}, {Indermuehle}, {Jacka}, {Jackson}, {Jackson},
  {Jeganathan}, {Johnston}, {Joseph}, {Kamphuis}, {Leach}, {Lenc}, {Lensson},
  {Mackay}, {Marquarding}, {Marvil}, {McClure-Griffiths}, {McConnell}, {Meyer},
  {Mirtschin}, {Neuhold}, {Ng}, {Norris}, {O'Sullivan}, {Pathikulangara},
  {Pearce}, {Phillips}, {Popping}, {Qiao}, {Reynolds}, {Roberts}, {Sault},
  {Schinckel}, {Shaw}, {Shimwell}, {Staveley-Smith}, {Storey}, {Sweetnam},
  {Troup}, {Tzioumis}, {Voronkov}, {Westmeier}, {Whiting}, {Wilson}, {Wong}, \&
  {Wu}}]{serra2015}
{Serra}, P., {Koribalski}, B., {Kilborn}, V., {et~al.} 2015{\natexlab{a}},
  \mnras, 452, 2680, \dodoi{10.1093/mnras/stv1326}

\bibitem[{{Serra} {et~al.}(2015{\natexlab{b}}){Serra}, {Westmeier}, {Giese},
  {Jurek}, {Fl{\"o}er}, {Popping}, {Winkel}, {van der Hulst}, {Meyer},
  {Koribalski}, {Staveley-Smith}, \& {Courtois}}]{serra2015a}
{Serra}, P., {Westmeier}, T., {Giese}, N., {et~al.} 2015{\natexlab{b}}, \mnras,
  448, 1922, \dodoi{10.1093/mnras/stv079}

\bibitem[{{Seth} \& {Raychaudhury}(2020)}]{seth2020}
{Seth}, R., \& {Raychaudhury}, S. 2020, \mnras, 497, 466,
  \dodoi{10.1093/mnras/staa1779}

\bibitem[{{Shibatsuka} {et~al.}(2003){Shibatsuka}, {Matsushita}, {Kohno}, \&
  {Kawabe}}]{shibatsuka2003}
{Shibatsuka}, T., {Matsushita}, S., {Kohno}, K., \& {Kawabe}, R. 2003, \pasj,
  55, 87, \dodoi{10.1093/pasj/55.1.87}

\bibitem[{{Sofue} {et~al.}(2003){Sofue}, {Koda}, {Nakanishi}, {Onodera},
  {Kohno}, {Tomita}, \& {Okumura}}]{sofue2003}
{Sofue}, Y., {Koda}, J., {Nakanishi}, H., {et~al.} 2003, \pasj, 55, 17,
  \dodoi{10.1093/pasj/55.1.17}

\bibitem[{{Solomon} \& {Vanden Bout}(2005)}]{solomon2005}
{Solomon}, P.~M., \& {Vanden Bout}, P.~A. 2005, \araa, 43, 677,
  \dodoi{10.1146/annurev.astro.43.051804.102221}

\bibitem[{{Sorai} {et~al.}(2019){Sorai}, {Kuno}, {Muraoka}, {Miyamoto},
  {Kaneko}, {Nakanishi}, {Nakai}, {Yanagitani}, {Tanaka}, {Sato}, {Salak},
  {Umei}, {Morokuma-Matsui}, {Matsumoto}, {Ueno}, {Pan}, {Noma}, {Takeuchi},
  {Yoda}, {Kuroda}, {Yasuda}, {Yajima}, {Oi}, {Shibata}, {Seta}, {Watanabe},
  {Kita}, {Komatsuzaki}, {Kajikawa}, {Yashima}, {Cooray}, {Baji}, {Segawa},
  {Tashiro}, {Takeda}, {Kishida}, {Hatakeyama}, {Tomiyasu}, \&
  {Saita}}]{sorai2019}
{Sorai}, K., {Kuno}, N., {Muraoka}, K., {et~al.} 2019, \pasj, 71, S14,
  \dodoi{10.1093/pasj/psz115}

\bibitem[{{Speagle} {et~al.}(2014){Speagle}, {Steinhardt}, {Capak}, \&
  {Silverman}}]{speagle2014}
{Speagle}, J.~S., {Steinhardt}, C.~L., {Capak}, P.~L., \& {Silverman}, J.~D.
  2014, \apjs, 214, 15, \dodoi{10.1088/0067-0049/214/2/15}

\bibitem[{{Strong} \& {Mattox}(1996)}]{strong1996}
{Strong}, A.~W., \& {Mattox}, J.~R. 1996, \aap, 308, L21

\bibitem[{{Tucker} {et~al.}(2000){Tucker}, {Oemler}, {Hashimoto}, {Shectman},
  {Kirshner}, {Lin}, {Landy}, {Schechter}, \& {Allam}}]{tucker2000}
{Tucker}, D.~L., {Oemler}, Augustus, J., {Hashimoto}, Y., {et~al.} 2000, \apjs,
  130, 237, \dodoi{10.1086/317348}

\bibitem[{{Tully} \& {Shaya}(1984)}]{tully1984}
{Tully}, R.~B., \& {Shaya}, E.~J. 1984, \apj, 281, 31, \dodoi{10.1086/162073}

\bibitem[{Virtanen {et~al.}(2020)Virtanen, Gommers, Oliphant, Haberland, Reddy,
  Cournapeau, Burovski, Peterson, Weckesser, Bright, {van der Walt}, Brett,
  Wilson, Millman, Mayorov, Nelson, Jones, Kern, Larson, Carey, Polat, Feng,
  Moore, {VanderPlas}, Laxalde, Perktold, Cimrman, Henriksen, Quintero, Harris,
  Archibald, Ribeiro, Pedregosa, {van Mulbregt}, \& {SciPy 1.0
  Contributors}}]{2020SciPy-NMeth}
Virtanen, P., Gommers, R., Oliphant, T.~E., {et~al.} 2020, Nature Methods, 17,
  261, \dodoi{10.1038/s41592-019-0686-2}

\bibitem[{{Vulcani} {et~al.}(2018){Vulcani}, {Poggianti}, {Jaff{\'e}},
  {Moretti}, {Fritz}, {Gullieuszik}, {Bettoni}, {Fasano}, {Tonnesen}, \&
  {McGee}}]{vulcani2018}
{Vulcani}, B., {Poggianti}, B.~M., {Jaff{\'e}}, Y.~L., {et~al.} 2018, \mnras,
  480, 3152, \dodoi{10.1093/mnras/sty2095}

\bibitem[{{Walter} {et~al.}(2008){Walter}, {Brinks}, {de Blok}, {Bigiel},
  {Kennicutt}, {Thornley}, \& {Leroy}}]{walter2008}
{Walter}, F., {Brinks}, E., {de Blok}, W.~J.~G., {et~al.} 2008, \aj, 136, 2563,
  \dodoi{10.1088/0004-6256/136/6/2563}

\bibitem[{{Walter} {et~al.}(2004){Walter}, {Dahlem}, \&
  {Lisenfeld}}]{walter2004}
{Walter}, F., {Dahlem}, M., \& {Lisenfeld}, U. 2004, \apj, 606, 258,
  \dodoi{10.1086/382774}

\bibitem[{{Wang} {et~al.}(2020){Wang}, {Xu}, {Lee}, {Du}, {Overzier}, \&
  {Shao}}]{wang2020}
{Wang}, J., {Xu}, W., {Lee}, B., {et~al.} 2020, \apj, 903, 103,
  \dodoi{10.3847/1538-4357/abb9aa}

\bibitem[{{Wang} {et~al.}(2017){Wang}, {Koribalski}, {Jarrett}, {Kamphuis},
  {Li}, {Ho}, {Westmeier}, {Shao}, {Lagos}, {Wong}, {Serra}, {Staveley-Smith},
  {J{\'o}zsa}, {van der Hulst}, \& {L{\'o}pez-S{\'a}nchez}}]{wang2017}
{Wang}, J., {Koribalski}, B.~S., {Jarrett}, T.~H., {et~al.} 2017, \mnras, 472,
  3029, \dodoi{10.1093/mnras/stx2073}

\bibitem[{{Wang} {et~al.}(2021){Wang}, {Staveley-Smith}, {Westmeier},
  {Catinella}, {Shao}, {Reynolds}, {For}, {Lee}, {Liang}, {Wang}, {Elagali},
  {D{\'e}nes}, {Kleiner}, {Koribalski}, {Lee-Waddell}, {Oh}, {Rhee}, {Serra},
  {Spekkens}, {Wong}, {Bekki}, {Bigiel}, {Courtois}, {Hess}, {Holwerda},
  {McQuinn}, {Pandey-Pommier}, {van der Hulst}, \&
  {Verdes-Montenegro}}]{wang2021}
{Wang}, J., {Staveley-Smith}, L., {Westmeier}, T., {et~al.} 2021, \apj, 915,
  70, \dodoi{10.3847/1538-4357/abfc52}

\bibitem[{{Wang} {et~al.}(2022){Wang}, {Wang}, {For}, {Lee}, {Reynolds}, {Lin},
  {Staveley-Smith}, {Shao}, {Wong}, {Catinella}, {Serra}, {Verdes-Montenegro},
  {Westmeier}, {Lee-Waddell}, {Koribalski}, {Murugeshan}, {Elagali}, {Kleiner},
  {Rhee}, {Bigiel}, {Bosma}, {Holwerda}, {Oh}, \& {Spekkens}}]{s.wang2021}
{Wang}, S., {Wang}, J., {For}, B.-Q., {et~al.} 2022, \apj, 927, 66,
  \dodoi{10.3847/1538-4357/ac4270}

\bibitem[{{Westmeier} {et~al.}(2011){Westmeier}, {Braun}, \&
  {Koribalski}}]{westmeier2011}
{Westmeier}, T., {Braun}, R., \& {Koribalski}, B.~S. 2011, \mnras, 410, 2217,
  \dodoi{10.1111/j.1365-2966.2010.17596.x}

\bibitem[{{Westmeier} {et~al.}(2021){Westmeier}, {Kitaeff}, {Pallot}, {Serra},
  {van der Hulst}, {Jurek}, {Elagali}, {For}, {Kleiner}, {Koribalski},
  {Lee-Waddell}, {Mould}, {Reynolds}, {Rhee}, \&
  {Staveley-Smith}}]{westmeier2021}
{Westmeier}, T., {Kitaeff}, S., {Pallot}, D., {et~al.} 2021, \mnras, 506, 3962,
  \dodoi{10.1093/mnras/stab1881}

\bibitem[{{Wolter} {et~al.}(2015){Wolter}, {Esposito}, {Mapelli}, {Pizzolato},
  \& {Ripamonti}}]{wolter2015}
{Wolter}, A., {Esposito}, P., {Mapelli}, M., {Pizzolato}, F., \& {Ripamonti},
  E. 2015, \mnras, 448, 781, \dodoi{10.1093/mnras/stv054}

\bibitem[{{Wright} {et~al.}(2010){Wright}, {Eisenhardt}, {Mainzer}, {Ressler},
  {Cutri}, {Jarrett}, {Kirkpatrick}, {Padgett}, {McMillan}, {Skrutskie},
  {Stanford}, {Cohen}, {Walker}, {Mather}, {Leisawitz}, {Gautier}, {McLean},
  {Benford}, {Lonsdale}, {Blain}, {Mendez}, {Irace}, {Duval}, {Liu}, {Royer},
  {Heinrichsen}, {Howard}, {Shannon}, {Kendall}, {Walsh}, {Larsen}, {Cardon},
  {Schick}, {Schwalm}, {Abid}, {Fabinsky}, {Naes}, \& {Tsai}}]{wright2010}
{Wright}, E.~L., {Eisenhardt}, P. R.~M., {Mainzer}, A.~K., {et~al.} 2010, \aj,
  140, 1868, \dodoi{10.1088/0004-6256/140/6/1868}

\bibitem[{{Yang} {et~al.}(2007){Yang}, {Mo}, {van den Bosch}, {Pasquali}, {Li},
  \& {Barden}}]{yang2007}
{Yang}, X., {Mo}, H.~J., {van den Bosch}, F.~C., {et~al.} 2007, \apj, 671, 153,
  \dodoi{10.1086/522027}

\bibitem[{{Yoon} {et~al.}(2017){Yoon}, {Chung}, {Smith}, \&
  {Jaff{\'e}}}]{yoon2017}
{Yoon}, H., {Chung}, A., {Smith}, R., \& {Jaff{\'e}}, Y.~L. 2017, \apj, 838,
  81, \dodoi{10.3847/1538-4357/aa6579}

\bibitem[{{Yun} {et~al.}(1994){Yun}, {Ho}, \& {Lo}}]{yun1994}
{Yun}, M.~S., {Ho}, P.~T.~P., \& {Lo}, K.~Y. 1994, \nat, 372, 530,
  \dodoi{10.1038/372530a0}

\bibitem[{{Zabel} {et~al.}(2019){Zabel}, {Davis}, {Smith}, {Maddox}, {Bendo},
  {Peletier}, {Iodice}, {Venhola}, {Baes}, {Davies}, {de Looze}, {Gomez},
  {Grossi}, {Kenney}, {Serra}, {van de Voort}, {Vlahakis}, \&
  {Young}}]{zabel2019}
{Zabel}, N., {Davis}, T.~A., {Smith}, M. W.~L., {et~al.} 2019, \mnras, 483,
  2251, \dodoi{10.1093/mnras/sty3234}

\bibitem[{{Zabludoff} \& {Mulchaey}(1998)}]{zabludoff1998}
{Zabludoff}, A.~I., \& {Mulchaey}, J.~S. 1998, \apj, 496, 39,
  \dodoi{10.1086/305355}

\bibitem[{{Zahid} {et~al.}(2012){Zahid}, {Dima}, {Kewley}, {Erb}, \&
  {Dav{\'e}}}]{zahid2012}
{Zahid}, H.~J., {Dima}, G.~I., {Kewley}, L.~J., {Erb}, D.~K., \& {Dav{\'e}}, R.
  2012, \apj, 757, 54, \dodoi{10.1088/0004-637X/757/1/54}

\end{thebibliography}




\end{document}